\documentclass[sigconf,10pt]{acmart}
\renewcommand\footnotetextcopyrightpermission[1]{} 

\usepackage[english]{babel}
\usepackage{blindtext}
\usepackage{times}

\setcopyright{none}
\usepackage{tabularx,tablefootnote,wrapfig,multirow}
\usepackage{hyphenat,xspace,color,colortbl,enumitem}
\usepackage{booktabs,multirow,microtype}
\usepackage{bm}
\usepackage[normalem]{ulem}
\usepackage{newtxmath,bm,courier,textcomp}
\usepackage{enumitem,fancyref}
\usepackage{xfrac,sparklines,bigstrut,physics}
\usepackage[labelformat=simple,skip=0pt]{subcaption}
\usepackage{amsmath,amsfonts,amsbsy}
\usepackage[scaled=0.90]{helvet}
\newcommand{\systemname}{X-ResQ}
\newcommand{\systemnames}{X-ResQ's}

\definecolor{Gray}{gray}{0.9}
\definecolor{LightGreen}{rgb}{0.88,1,0.88}
\definecolor{DarkGreen}{rgb}{0.0,0.4,0.13}
\definecolor{LightOrange}{rgb}{1,0.85,0.8}
\definecolor{LightYellow}{rgb}{1,1.00,0.5}
\definecolor{LightRed}{rgb}{1,0.80,0.80}

\usepackage{soul}

\newcommand{\RN}[1]{%
  \textup{\uppercase\expandafter{\romannumeral#1}}%
}

\newcommand{\parahead}[1]{\vspace{2pt plus 0pt minus 2pt}\noindent{\bfseries #1}}
\newcommand{\parabreak}{\vspace*{1.00ex minus 0.25ex}\noindent}

\renewcommand{\paragraph}[1]{\parahead{#1}}

\usepackage{fontawesome}
\usepackage{tikz}
\usepackage{amsmath}

\usepackage{filecontents}

\begin{document}
% \date{}
\pagestyle{plain}

\title{X-ResQ: Reverse Annealing for Quantum MIMO Detection with Flexible Parallelism}

% \title{X-ResQ: Multi-Seed Ensemble Reverse Annealing for Quantum MIMO Detection with Scalable and Elastic Parallelism}

% \title{X-ResQ: Cross Reverse Annealing for Massively Parallel Quantum-Accelerated NextG MIMO Processing}

% \title{QuAMax 2.0: Enabling 
% High-Bit-Rate Quantum-Massive MIMO Processing\\via Decomposition Search and Reverse Quantum Annealing}
% \author{Paper \#71: 12 pages, plus References and Appendices (21 pages total)}  

\author{Minsung Kim$^{\star,\diamond}$, Abhishek Kumar Singh$^{\diamond}$, Davide Venturelli$^{\ast}$, John Kaewell$^{\dagger}$, Kyle Jamieson$^{\diamond}$}

\affiliation{%
 \institution{\small $^{\star}$Yale University, $^{\diamond}$Princeton University, $^{\dagger}$InterDigital, $^{\ast}$USRA Research Institute for Advanced Computer Science}
%  \institution{ }
}

\begin{abstract}
Quantum Annealing (QA)-accelerated MIMO detection is an emerging research approach in the context of NextG wireless networks. The opportunity is to enable large MIMO systems and thus improve wireless performance. The approach aims to leverage QA to expedite the computation required for theoretically optimal but computationally-demanding Maximum Likelihood detection to overcome the limitations of the currently deployed linear detectors. This paper presents \textbf{\systemname{}}, a QA-based MIMO detector system featuring fine-grained quantum task parallelism that is uniquely enabled by the Reverse Annealing (RA) protocol. Unlike prior designs, \systemname{} has many desirable system properties for a parallel QA detector and has effectively improved detection performance as more qubits are assigned. 
% Its parallelization strategy is designed 
% to make full use of the advantages of RA 
% by thorough RA initialization studies.
% Trading off with more qubit usage, \systemname{} alleviates two important challenges in QA-based MIMO detection: (1) enabling high-order modulations (2) mitigating error floor phenomenon. 
In our evaluations on a state-of-the-art quantum annealer, fully parallel \systemname{} achieves near-optimal throughput (over 10 bits/s/Hz) for $4\times6$ MIMO with 16-QAM using six levels of parallelism with 240 qubits and $220~\mu$s QA compute time, achieving 2.5--5$\times$ gains compared against other tested detectors.
% whereas other parallel QA detectors show limitations.
% in either performance or flexibility. 
% In our evaluations on a state-of-the-art quantum annealer, \systemname{} achieves near $10^{-4}$ BER for $4\times6$ MIMO with 16-QAM at SNR 20 dB using six levels of parallelism (XXX qubits) and $110~\mu$s QA compute time, resulting in near-optimal spectral efficiency for 1,500-byte packets (over 10 bits/s/Hz), whereas other parallel QA detectors show limitations in performance and flexibility. 
% enables near-optimal spectral efficiency with 16-QAM within a few hundred microseconds of QA compute time with reasonable qubit counts, whereas other parallel QA detectors show limitations in performance and flexibility. 
% near $10^{-4}$ BER performance for $4\times6$ MIMO with 16-QAM at SNR 20 dB using six levels of parallelism (XXX qubits) and $110~\mu$s QA compute time, resulting in near-optimal spectral efficiency performance for 1,500-byte packets (over 10 bits/s/Hz). 
% To our best knowledge, \systemname{} is the first QA-based MIMO detector that enables near-optimal performance with 16-QAM within a few hundreds of microseconds of QA compute time. 
% To test our design's scalability, 
For more comprehensive evaluations, 
we implement and evaluate X-ResQ in the non-quantum digital setting. This non-quantum X-ResQ demonstration showcases the potential to realize
ultra-large $1024\times1024$ MIMO, 
significantly outperforming other MIMO detectors, including the state-of-the-art RA detector 
classically
implemented in the same way.
\end{abstract}

\maketitle

\pagestyle{plain}
% \setcopyright{none}
% \settopmatter{printacmref=false, printccs=false, printfolios=false}

% Quantum Annealing (QA) is a new type of optimization algorithm that potentially accelerates the process of solving difficult combinatorial problems over classical computing methods, using quantum effects (\emph{e.g.,} quantum tunneling). 
% However, quantum optimization search for practical use includes various design considerations. 
% In this paper, we present \textbf{\systemname{}}, a flexibly parallel QA optimization design, uniquely enabled by \emph{quantum reverse annealing}. \systemname{} makes use of computing resources for multi-point initialized reverse annealing, while typical parallel QA designs are based on problem decomposition. 

% discuss parallelization approaches of QA and 

% The proposed technique has unique properties.
% , including one that all parallel processing can independently converge to the optimal solution. 

% The proposed technique is inspired by our hypothesis that there exists an unknown factor, which determines reverse quantum search, other than commonly-accepted factors (\emph{i.e.,} hamming distance and cost of the initializing state). In the paper, we empirically prove the hypothesis on a real-world quantum device. 
% Unlike other potential designs of parallel QA, this proposed technique has unique properties.
% , including one that all parallel processing can independently converge to the optimal solution. 

\vspace{-0.2cm}
\section{Introduction}
\label{s:Introduction}

To satisfy ever-increasing demand on mobile traffic, the \emph{multi-user Multiple-Input Multiple-Output} (MU-MIMO) technique with \emph{spatial multiplexing} becomes an essential building block in modern wireless standards. To enable parallel data streams and thus achieve capacity gains in MU-MIMO systems, a signal processing technique called \emph{MU-MIMO detection} is required to demultiplex mutually-interfering streams at the base station receiver into each user signal. Optimal \emph{Maximum Likelihood} (ML) detectors ensure the best possible performance. Nonetheless, current MIMO systems make use of sub-optimal detectors, sacrificing tremendous potential gains. This is because the computational complexity of the ML detectors increases at an exponential rate with more users. To resolve the computational bottleneck in ML MIMO detection, researchers have started to explore non-traditional heuristic optimization processing~\cite{rimimo,kim2019leveraging,kim2021physics,norimoto2023quantum,gulbahar2023maximum}.
% with emerging computing devices such as quantum computers and Ising machines~\cite{aramon2019physics}.
\emph{Physics-Inspired Computing} (PIC) is a computing paradigm defined by computational principles that imitate laws of physics. For example, PIC algorithms such as \emph{Quantum Annealing} (QA)~\cite{boixo-45467,kadowaki1998quantum} and \emph{Simulated Annealing} (SA)~\cite{metropolis1949monte} mimic the convergence nature of thermal annealing to find the global optimum of discrete optimization problems.

\parahead{Need of Parallelization Strategies in MIMO Detection.} Unlike most optimization problems, there exist extremely tight \emph{computation time deadlines} in MIMO detection, at most a few hundred microseconds, regardless of the difficulties of the problems. However, for difficult MIMO scenarios (\emph{e.g.,} large user counts), 
% simple linear detectors perform poorly, so 
the use of (near-)optimal detectors is encouraging~\cite{Geosphere,xu2022low}, but their required computations cannot be completed within the deadlines. Regarding this, parallel architecture-based detectors with \emph{detection task parallelism} are promising in that large computation amounts can be split into multiple compute tasks to reduce overall compute times by solving them in parallel at the expense of more computation resources~\cite{flexcore-nsdi17,nikitopoulos2022massively}. 
% The usage of more resources for the task parallelism implies much more resources for the entire system because of the coexisting task and data parallelism in MIMO detection. However,
While transistors' clock speed has stopped increasing rapidly with Moore's Law, the number of embedded transistors per chip keeps increasing nearly exponentially. With this trend, hardware that contains massive \emph{processing elements} (PEs) including GPUs and FPGAs are actively explored in this direction with a variety of parallel MIMO detection algorithms~\cite{roger2012fully,FCSD,Wenk06,guo2006algorithm,flexcore-nsdi17,he2019algorithm,nikitopoulos2018massively}. In this regard, ParaMax~\cite{kim2021physics}, a recent classical PIC MIMO detector, parallelizes its PIC heuristics to collect more candidate solutions simultaneously (\emph{i.e.,} sample parallelism) to reduce its overall latency without loss of performance. 
% For this reason, leveraging detection task parallelism is promising to enable near-optimal MIMO detection performance.

% For this reason, leveraging detection task parallelism in challenging MIMO scenarios (\emph{e.g.,} large user counts) has been studied on classical platforms~\cite{barbero2006rapid,roger2012fully,flexcore-nsdi17,nikitopoulos2018massively,nikitopoulos2021non,he2019algorithm}.  

We believe the same motivation holds for QA MIMO detectors, and thus designs of highly efficient and scalable parallelization strategies for QA MIMO detection need to be investigated in light of the expected large-scale qubit processors. Since 2011, every new generation of D-Wave quantum annealers has presented exponentially-increasing qubit counts as well as improved hardware connectivity: 128 qubits in 2011 (D-Wave One System) and 5000+ qubits in 2020 (Advantage System). Based on the extrapolation of historical records, over ten thousand qubits are expected in 2030 and hundreds of thousands in 2035 on a single QA machine~\cite{kasi2021challenge} (likely in an interconnected multi-core setting), which are sufficient enough to cover both data and task parallelism in the physical layer processing. Therefore, parallelization strategies in QA MIMO detectors will become more essential. This paper investigates \emph{quantum MIMO detection (optimization) task parallelism}. 
% In prior work, QA parallelization is discussed to some extent, but 
Prior QA MIMO detector design QuAMax~\cite{kim2019leveraging} discusses sample parallelism like ParaMax, but the strategy is not effective when optimization problems become difficult (\S\ref{s:xresq_vs_iotresq}).
% since it solves the same problem per task.
IoT-ResQ~\cite{kim2022warm} applies decomposition-based parallelism to make problems easier with further parallelization by decomposing a problem into simpler subproblems,
% that are solvable in parallel, 
but it cannot support flexible parallelism due to its coarse granularity. 
\parabreak{}In this paper, we introduce \emph{\systemname{}} (\S\ref{s:design}), a parallel QA MIMO detector system that supports flexible quantum task parallelism that is uniquely enabled by \emph{Reverse Annealing} (RA) (\S\ref{s:primer_ra}), a method called \emph{multi-seed parallel ensemble RA}. 
Unlike standard QA, RA starts its optimization operation from a classical state where its QA search tends to be localized around that state. \systemname{} applies multiple independent RA runs initiated from different initial classical states for its QA parallelization. 
Unlike prior designs, \systemname{} has many desirable features for a parallel QA system including fine-grained parallelism
(Table~\ref{t:qa_detectors_properties}), and has effectively improved performance as more qubits are assigned, providing an efficient trade-off between qubits and compute time. 
We describe its design principles based on 
somewhat unexpected experimental RA results~(\S\ref{s:desing_principle}). 
% , whereas prior QA detectors~\cite{kim2019leveraging,kim2022warm} show limitations in either performance or flexibility. 
% while reducing the required QA time for optimal detection (\emph{i.e.,} trade-off between qubits and compute time).
% (\S\ref{s:xresq_vs_iotresq}),
% including elastic parallelism and the capability to facilitate quantum optimization search per additional parallel task. 
% and also
% of \systemname{}, 
% point out limitations of other parallel QA approaches including widely-used decomposition methods. 
We also present a split-detection scheme (\S\ref{s:mmse_quadrant}) to mitigate the error floor phenomenon at high SNRs leveraging QA's exceptional performance with B/QPSK and analyze its effect both theoretically and experimentally.
% At the expense of more qubit usage, X-ResQ alleviates two important challenges in QA MIMO detection: (1) enabling high-order modulations (2)
% To mitigate QA MIMO detection's error floor phenomenon at high SNRs with high-order modulations, we introduce a split-detection scheme ($\S$\ref{s:mmse_quadrant}) and analyze it both experimentally and theoretically. 

We implement and evaluate \systemname{} on the state-of-the-art D-Wave quantum annealer. In our evaluations,
% on a state-of-the-art quantum annealer, 
\systemname{} achieves near $10^{-4}$ uncoded BER performance for $4\times6$ MIMO with 16-QAM at SNR 20 dB using six levels of parallelism (240 qubits) and $220~\mu$s QA compute time, resulting in near-optimal throughput for 1,500-byte packets (over 10 bits/s/Hz). The other parallel QA detectors tested on the same machine
with the same qubits either still perform quite poorly or cannot be programmed on the hardware due to inflexible parallelism. 
% To our best knowledge, \systemname{} is the first QA-based MIMO detector that enables near-optimal performance with 16-QAM within a few hundred microseconds of QA time, which
% The enabled sizes of MIMO are relatively small compared
% to the ones by QA detectors with low-order
% modulations and even by the conventional non-linear MIMO
% detectors. However, we believe this is an important step in
% the direction in that enabling high-order modulations with
% QA has been constantly identified as an important challenge.
Despite the relatively small sizes of enabled MIMO, we believe this is an important step in the direction in that enabling high-order modulations for QA MIMO detection has been constantly identified as an important challenge.
% To our best knowledge, \systemname{} is the first QA-based MIMO detector that enables near-optimal performance with 16-QAM within a few hundreds of microseconds of QA time, which 
% we believe is an important step in the direction despite the relatively small sizes of enabled MIMO. 
For more comprehensive evaluations beyond QA hardware, we classically implement \systemname{} using a generic PIC algorithm, \emph{parallel tempering} (\S\ref{s:implementation}). We observe the classical \systemname{} can potentially enable previously impossible ultra-large $1024\times1024$ MIMO, achieving BER below $10^{-7}$ with BPSK. 
% For example, the classical-version X-ResQ can achieve BER below $10^{-7}$ for $1024\times1024$ MIMO with BPSK at SNR 14~dB.
For $256\times256$ MIMO with QPSK at SNR 14~dB, \systemname{} obtains around $10^{-7}$ BER, showing four to six orders of magnitude better BER than the other tested detectors, with $\approx333~\mu$s.
\section{Background}
\label{s:background}

This section introduces background knowledge to understand the paper. Section~\ref{s:primer_mimo} explains MIMO detection and Section~\ref{s:QA_RA} introduces QA and Section~\ref{s:primer_ra} describes its Reverse Annealing protocol.
% \subsection{Primer: MIMO Detection}
% \label{s:primer_mimo}

% \textcolor{blue}{ABHISHEK -> MMSE EXPLANATION ON TOP OF ZF}

% While non-linear detectors more robust to $\mathbf{n}$, $\mathbf{H}$ are actively being studied~\cite{Geosphere,flexcore-nsdi17,albreem2019massive,FCSD,nikitopoulos2018massively}, they are still considered impractical for the similar reasons to the ML detector. This is the well-known trade-off between throughput and complexity in MIMO detection, which is one of the main challenge factors towards \emph{ultra-high-speed} wireless networks envisioned in the 5G/6G roadmap. The precoding is a dual problem of the MIMO detection in downlink (from BS to users), but the uplink is more challenging in general as shown in Table~{\ref{t:SNIR_dist}}. 

% \vspace{-0.15cm}
\vspace{-0.1cm}
\subsection{MIMO Detection}
\label{s:primer_mimo}

\begin{figure}
      \centering
    
    \begin{subfigure}[b]{\linewidth}
    \centering

    \includegraphics[width=0.85\linewidth]{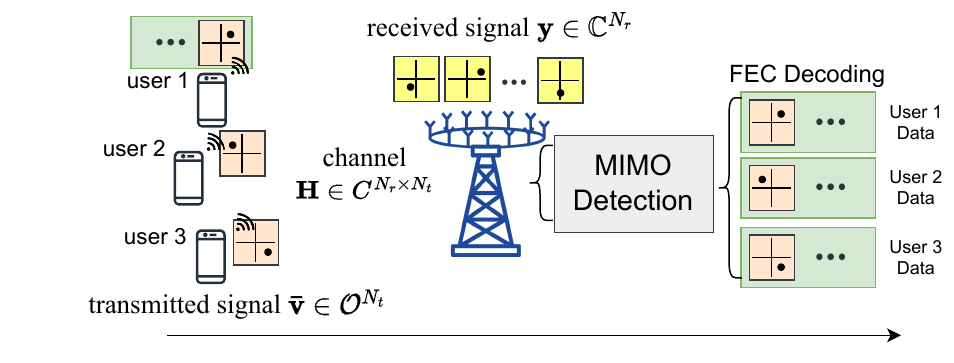}
    \caption{Uplink MIMO and overall detection and decoding structure.}
    \label{f:mimo_detection}
    \end{subfigure}
    % \qquad
    % \hfill
    \begin{subfigure}[b]
    {\linewidth}
    \centering    \includegraphics[width=0.75\linewidth]{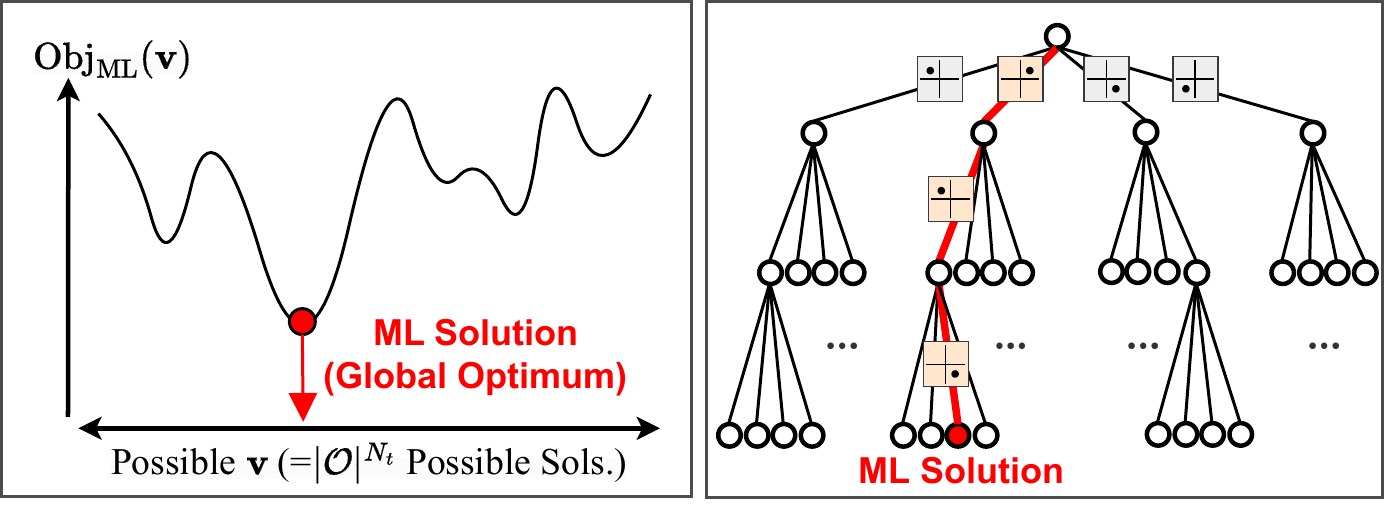}
    \caption{ML solution (\emph{left}) and equivalent path in a tree search (\emph{right}).}
    \label{f:MLD}
    \end{subfigure} 
    \caption{\textbf{\small MIMO model and optimal ML solution.}} 
\label{f:mimo_model}
\end{figure}

Figure~{\ref{f:mimo_detection}} shows the uplink MU-MIMO model and overall detection and decoding stricture, where $N_t$ single-antenna mobile users are simultaneously sending their signals to a base station (BS) with $N_r$ antennas ($N_r \geq N_t$). Each user sends out a wireless symbol $v \in \mathcal{O}$ that represents $M$ bits (out of Forward Error Correction (FEC)-coded bits) based on a constellation $\mathcal{O}$ with $|\mathcal{O}| = 2^M$ being the modulation size. 
% (\emph{e.g.,} $\mathcal{O}=\{-1-1j, 1-1j, -1+1j, +1+1j\}$ in the case of QPSK, $|\mathcal{O}|=4$). 
Then, the received signal at the BS can be expressed as 
$\mathbf{y} = \mathbf{H}\mathbf{\bar{v}}+\mathbf{n} \in \mathbb{C}^{N_r}$, where $\mathbf{\bar{v}}\in \mathcal{O}^{N_t}$ is the transmitted signal vector,  $\mathbf{H} \in \mathbb{C}^{N_r \times N_t}$ is wireless channel matrix, and  $\mathbf{n} \in \mathbb{C}^{N_r}$ is a noise vector. We denote this model as $N_t \times N_r$ MIMO and all OFDM subcarriers
% \footnote{OFDM is a technique to subdivide a large frequency band into many narrow subcarriers to avoid frequency selective fading~\cite{goldsmith2005wireless}.} 
hold the same model independently.

\parahead{MIMO Detection}. MU-MIMO detection is a signal processing through which the receiver generates an estimated signal $\mathbf{\hat{v}}$ (without $\mathbf{n}$,  $\mathbf{\hat{v}}=\mathbf{\bar{v}}$) based on the received signal and estimated channel~\cite{castaneda2016data}.\footnote{In downlink (from BS to users), MIMO precoding is a dual problem of
MIMO detection.}
% \footnote{In downlink (from BS to users), MIMO precoding is a dual problem of MIMO detection. In general, MIMO detection is more challenging than MIMO precoding due to low average SNRs in uplink scenarios.} 
After the detection, the detected symbol vector 
 $\mathbf{\hat{v}}$
is demapped into corresponding $M$ bits, and then blocks of those bits across subcarriers are decoded by a channel decoder to reconstruct \emph{error-free} data bits (per user), according to the applied FEC scheme (\emph{e.g.,} LDPC, convolutional coding).

\parahead{Optimal ML Detectors.} Optimal \emph{maximum likelihood} (ML) detectors (\emph{e.g.,} Sphere Decoders~\cite{Agrell02,Damen03,SD}) can ensure the \emph{best-possible} detection performance.
The ML method detects $\mathbf{\hat{v}}$ among all possible $\mathbf{v} \in \mathcal{O}^{N_t}$ that minimizes the objective function $\text{Obj}_{\text{ML}}(\mathbf{v})=||\mathbf{y}-\mathbf{H}\mathbf{v}||^2$ (\emph{i.e.,} $\mathbf{\hat{v}_{ML}}$ is global optimum). 
% \begin{equation}   
% \begin{small}
% \label{eqn:ml}
% \hat{\mathbf{v}} = \argmin{\mathbf{v} \in \mathcal{O}^{N_t}}
% \left\lVert\mathbf{y} - \mathbf{Hv}\right\rVert^2.
% \end{small}
% \end{equation}
% Through the QR decomposition of $\mathbf{H}$, 
The ML detection problem can be translated into an equivalent tree search problem that aims to find the best path (from the root node to a leaf node) with the minimum cumulative metric among $\mathcal{|O|}^{N_t}$ possible paths in a perfect tree as shown in Figure~\ref{f:MLD}. Each level corresponds to a user, while each branch decision is a symbol decision per user.
% , which is \emph{interdependent} among users. 
While optimal, an ML detector is not practically feasible today, because of its exponentially increasing search size with higher $N_t$ and/or $\mathcal{|O|}$ and
limited available processing time in wireless standards ($\approx$ 0.5 -- 4 ms)~\cite{BigStation,gong2023scalable,ding2020agora}.

\vspace{-0.1cm}
\subsection{Quantum Annealing}
\label{s:QA_RA}
% \vspace{-0.1cm}

Quantum annealers are analog devices designed to find the (near-)optimal solution of combinatorial optimization problems using quantum mechanical fluctuations with tremendous speedup potentials compared to classical computational resources and methods~\cite{denchev-44814,king2023quantum,boixo-45467}. For the state-of-the-art D-Wave quantum annealers, input optimization problems are \emph{Ising spin models} whose Ising energy (cost) functions $E$ are:
\begin{small}
\begin{equation}\label{eq:ising-ham}
E(\mathbf{s}) =  \sum_i f_i s_i + \sum_{i,j} g_{ij} s_i s_j
\end{equation}
\end{small}
\noindent where real-value $f_i$ and $g_{ij}$ represent the optimization problem we want to solve. The machines aim to find a spin \emph{configuration} (or state) $\mathbf{s} = \{s_1,s_2,\cdots,s_{N_V}\}$ that makes the Ising energy $E(\mathbf{s})$ minimized, where $\mathbf{s}$ consists of $N_V$ spins, with each spin variable $s_i$ being either 1 or -1. The configuration that corresponds to the minimum Ising energy is called a \emph{ground state} (global optimum). For an Ising model of ML MIMO detection, $N_V=N_t\log_2{|\mathcal{O}|}$ spins are required (\S\ref{s:parallel_QA}). 

\parahead{QA Heuristics.} The time evolution of \emph{Hamiltonian} $\mathcal{H}$ in a system enables QA heuristics to solve the input optimization problem, where (simplified) $\mathcal{H}$ can be defined as:
\begin{small}
\begin{align}
\mathcal{H}(\tau)
=A(\tau)\cdot\mathcal{H}_{\text{superposition}}+B(\tau)\cdot\mathcal{H}_{\text{problem}\,E},
% \notag\\
% =A(\tau)\left(\sum_i \sigma_i^x\right) +B(\tau)\left(\sum_i f_{i} \sigma_i^z + \sum_{i<j} g_{ij} \sigma_i^z\sigma_j^z\right),
% =A(\tau)\cdot\mathcal{H}_{\text{superposition}}+B(\tau)\cdot E(\mathbf{s})
\label{eq:hamiltonian}
\end{align}
\end{small}
where $\tau$ is a time-dependent function ($\tau \in [0,1]$). 
% while $\sigma^{x,z}$ are the Pauli matrices that represent the interaction of the spins, following the standard notation of quantum mechanics. 
% (\emph{i.e.,} how to do anneal scheduling) 
% determines QA  algorithms. Figure~\ref{f:anneal_scheudle} shows two anneal signals as a function of $\tau$, where
$A(\tau)$ represents the transverse energy that controls the strength of the \emph{quantum
fluctuations} (\emph{i.e.,} quantum-coherent exploration of the search
space), while $B(\tau)$ is the energy that is meant to correlate these fluctuations to the energy function of the problem (Eq.~\ref{eq:ising-ham}). How to control $\tau$ as a function of wall-clock \emph{anneal duration} ($T_a$) 
% (\emph{i.e.,} how to do anneal scheduling) 
determines QA  algorithms. 

In the standard \emph{Forward Annealing} or FA, the system initially applies strong quantum fluctuations and prepares a \emph{quantum superposition state} that holds every possible configuration information (\emph{i.e.,} $A(\tau) \gg B(\tau)$ at $\tau=0$). Then it drives the changes until the effect of the initial $\mathcal{H}_{\text{superposition}}$ diminishes (\emph{i.e.,} $A(\tau) \ll B(\tau)$ at $\tau=1$),
% \footnote{The state at $\tau=0$ indicates a fully quantum state, while the state at $\tau=1$ indicates a fully classical state.} During $T_a$, 
% (user parameter range: $T_a \in$ [$0.5,2,000\,\mu$s]), 
% $\tau$ increases incrementally from 0 to 1, 
hoping that the driven non-equilibrium quantum thermodynamics process (which follows phenomenology from the adiabatic and quasistatic evolution~\cite{DWPauseMarshall}) will lead to a low-energy
state (the ground state in ideal cases).
% , despite the noisy, imperfect implementation of the current hardware. 
At the end of the anneal protocol, 
% the $\sigma$ operators in Eq. \ref{eq:hamiltonian} can be identified to spin values, and
the system can interpret the result as a spin configuration $\mathbf{s}$, being a \emph{sample}. Since QA is a probabilistic technique, the end result is non-deterministic. Thus, multiple anneal iterations ($N_a$ or \emph{anneal counts}) are typically applied, leading to $N_a$ samples. Among them, the best sample $\mathbf{s}$ (\emph{i.e.,} min $E(\mathbf{s})$) is filtered as the final optimization solution. 

% More precise explanations of QA algorithms and machine programming used in X-ResQ are in Appendix~\ref{s:QA_apppendix}.

% Annealing functions A(s)
% , B(s)
% . Annealing begins at s=0
%  with A(s)≫B(s)
%  and ends at s=1
%  with A(s)≪B(s)
% .

% \footnote{More precisely, $A(\tau)\left(\sum_i \sigma_i^x\right) +B(\tau)\left(\sum_i {f}_{i} \sigma_i^z + \sum_{i<j} {g}_{i,j} \sigma_i^z\sigma_j^z\right)$, where $\sigma_i^{x}$ and $\sigma_i^{z}$ are Pauli matrices acting on $i^{th}$ qubits.} 

% where $\sigma_i^{x}$ and $\sigma_i^{z}$ are Pauli matrices acting on $i^{th}$ qubits. Figure

% \footnote{$\mathcal{H}(\tau)=A(\tau)\left(\sum_i \sigma_i^x\right) +B(\tau)\left(\sum_i \mathbf{M}_{(i,i)} \sigma_i^z + \sum_{i<j} \mathbf{M}_{(i,j)} \sigma_i^z\sigma_j^z\right)$, where $\sigma_i^{x}$ and $\sigma_i^{z}$ are Pauli matrices acting on $i^{th}$ qubits.}

% \subsubsection{QA Schedules and Algorithms}
% \label{s:QA_algorithms}

\begin{figure}
\centering
    \includegraphics[width=0.7\linewidth]{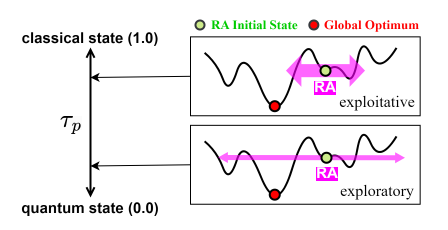}
    % {example-image-a}
    \vspace{-0.2cm}
    \caption{\small
    \textbf{Optimization search 
 of Reverse Annealing (RA) initiated from a classical state (cf. from a superposition state in FA).
    % \vspace{-0.2cm}
    }} 
\label{f:IllustrativeRA}
\end{figure}

% \parahead{QA Algorithm: Forward Annealing (FA).}

% \vspace{-0.18cm}
\subsection{Reverse Annealing (RA)}
\label{s:primer_ra}

\emph{Reverse Annealing} or RA~\cite{ohkuwa2018reverse} (\systemnames{} optimization core) is a variation of QA whose initialization process is reversed ($R$) compared to the standard FA.
For RA, the annealer initially programs ``a classical state'' on the hardware ($\tau=1$) and then introduces quantum fluctuations 
by reducing $\tau$ to the \emph{switching point} $\tau_p$ ($0 \leq \tau_p \leq 1$ \emph{e.g.,} $\tau_p=0.4$~\cite{kim2020towards} or 0.6~\cite{pelofske2023initial}). When the process reaches $\tau_p$, a superposition state is prepared with relatively weak fluctuations, which enables it to still hold the initial classical state information.  After the optional pause ($P$), the rest of $T_a$ follows the incremental increase of $\tau$, like FA ($F$). Accordingly,
during $T_a$ in the RA protocol, the state of $\mathcal{H}(\tau)$ is changed as following, compared to FA:
\begin{small}
\begin{itemize}
    \item FA:  quantum  $(\tau=0)$ $\xrightarrow{F}$ classical $(\tau=1)$
    % $\tau=0 \xrightarrow{F} \tau=1$

    \item RA: initial classical $(\tau=1) \xrightarrow{R}$ initial intermediate $(\tau=\tau_p)\xrightarrow{P}$ stable intermediate $ (\tau=\tau_p)\xrightarrow{F}$ final classical $(\tau=1)$.
    % ($\tau=1) \xrightarrow{R} \tau=\tau_p\xrightarrow{P} \tau=\tau_p\xrightarrow{F} \tau=1$

\end{itemize}
\end{small}

% \noindent \textcolor{red}{The later intermediate state in RA is more at equilibrium.}

This RA protocol promises to improve QA's optimization performance by limiting quantum fluctuations around the initial state~\cite{crosson2021prospects}. In doing so, RA works like a \emph{refined local search}, where optimization search tends to be localized around the initial
state based on the quasi-(non)local search~\cite{king2019quantum}, 
% When $\tau_p$ is too far from the initial $\tau=1$ (\emph{i.e.,} $\tau_p$ close to 0), the information related to the initial state would be wiped out with strong fluctuations (\emph{i.e.,} no effects of using the initial state). 
% On the other way, when $\tau_p$ is too close to the initial $\tau=1$, the strength of quantum fluctuations is not enough to trigger sufficient superposition (\emph{i.e.,} no effects of quantum search).
% In other words, 
providing a chance of 
% classical-quantum hybridization, 
exploring a trade-off in searching between exploration and exploitation as shown in Figure~\ref{f:IllustrativeRA}.\footnote{When $\tau_p$ is too far from the initial $\tau=1$ (\emph{i.e.,} $\tau_p$ close to 0), the information related to the initial state would be wiped out with strong fluctuations (\emph{i.e.,} no effects of using the initial state). On the other hand, when $\tau_p$ is too close to the initial $\tau=1$, the strength of quantum fluctuations is not enough to trigger sufficient superposition (\emph{i.e.,} no effects of quantum search).} 
In a rough way, FA can be considered a \emph{wide} and \emph{shallow} search (entire search space w/ scattered search power), whereas RA is a \emph{narrow} and \emph{deep} search (focused search space w/ concentrated search power). 
Therefore, providing a good initial classical state to RA is a key factor for optimization performance, which will be further discussed later in the paper (\S\ref{s:desing_principle}). It has been reported that RA could outperform FA for various applications~\cite{ReverseVenturelli,kim2020towards,golden2021reverse,king2019quantum,pelofske2023initial}. 
% \vspace{-0.3cm}

\newcolumntype{Y}{>{\centering\arraybackslash}X}
\begin{table}[htbp]
\begin{tiny}
% \begin{footnotesize}
\centering
\caption{\small Experimental Massive ($N_t \times N_r$) MIMO systems.}
% \vspace{-0.3cm}
\begin{tabularx}{\linewidth}{*{5}{Y}}
\toprule
\textbf{\scriptsize{MIMO System}} & \textbf{\scriptsize{MIMO Detector}} & \textbf{\scriptsize{MIMO Config.}} & \textbf{\scriptsize{$N_r/N_t$}} \scriptsize{(BS~ant.~/~user)} & \textbf{\scriptsize{System Imple.}}  \\ \midrule
%Access time & 26.647~ms\\
%Access overhead & 5.834~ms\\
\textbf{\footnotesize{Agora}~\cite{ding2020agora}} &  \scriptsize{linear ZF} & \scriptsize{$16\times64$} & \scriptsize{4} &
\scriptsize{software} \\
\textbf{\footnotesize{Argos}~\cite{argos-mobicom12}} & \scriptsize{linear ZF} & \scriptsize{$16\times64$} & \scriptsize{4} & \scriptsize{hardware} \\
% \textbf{\footnotesize{ArgosV2}~\cite{shepard2013argosv2}} & \scriptsize{linear ZF} & \scriptsize{$32\times96$} & \scriptsize{3} & \scriptsize{hardware} \\
% \textbf{\footnotesize{BigStation}~\cite{BigStation}} & \scriptsize{ZF} & \scriptsize{$9\times12$} & \scriptsize{1.4} & \scriptsize{7000+} \\
\textbf{\footnotesize{BigStation}~\cite{BigStation}} & \scriptsize{linear ZF} & \scriptsize{$16\times128$} & \scriptsize{8} & \scriptsize{software} \\
\textbf{\footnotesize{Hydra}~\cite{gong2023scalable}} & \scriptsize{linear ZF} & \scriptsize{$32\times150$} & \scriptsize{4.69} & \scriptsize{software} \\
\textbf{\footnotesize{LuMaMi}~\cite{malkowsky2017world}} & \scriptsize{linear ZF} & \scriptsize{$12\times100$} & \scriptsize{8.33} & \scriptsize{hardware} \\
\midrule
% \textbf{\footnotesize{IoT-ResQ}} & \scriptsize{MMSE-RA} & \scriptsize{$1024\times1024$} & \scriptsize{1} & \scriptsize{BPSK} \\
% \textbf{\footnotesize{ParaMax}} & \scriptsize{PT} & \scriptsize{$512\times512$} & \scriptsize{1 (w/ BPSK)} & \scriptsize{BER} \\
% \textbf{\footnotesize{IoT-ResQ}} & \scriptsize{FSD+RA} & \scriptsize{$512\times512$} & \scriptsize{1 (w/ BPSK)} & \scriptsize{BER} \\
% \textbf{\footnotesize{X-ResQ}} & \scriptsize{MMSE+RA} & \scriptsize{$1024\times1024$} & \scriptsize{1 (w/ BPSK)} & \scriptsize{BER} \\

% \textbf{\scriptsize{Hydra}} & \scriptsize{6} & \scriptsize{15} & \scriptsize{20} & \scriptsize{20} \\
% Connectivity & $\simeq$ 9000~$\mu$s\\
 % & $\simeq$ 300~$\mu$s\\
% Post processing & 445~$\mu$s \\
% \midrule
% $(N_t)$\\
% \bottomrule
% \toprule
% \multicolumn{6}{c}{\textbf{Dense Urban}}\\ 
% \multicolumn{3}{c}{\textbf{Downlink}} & \multicolumn{3}{c}{\textbf{Uplink}} \\\midrule
% \text{poor} & \text{medium} & good & \text{poor} & \text{medium} & good\\
% 5~dB & 12.5~dB  & 17.5~dB  & 5~dB  & 7.5~dB & 9~dB\\
% \bottomrule
\end{tabularx}
% \vspace{-0.4cm}
\label{t:nr_nt_ratio}
\end{tiny}
\end{table}

\section{Motivation and Related Work}
\label{s:related_work}
\subsection{MIMO Systems with Linear Detector}
\label{s:massive_mimo}
\vspace{-0.1cm}
% \parahead{Linear Detectors and Massive MIMO Systems.} 
Linear detectors, such as \emph{Zero-Forcing} (ZF) and \emph{Minimum Mean Square Error} (MMSE) methods, feature simple processing, thus being commonly used in experimental wireless systems as well as practical cellular systems.
% However, linear detectors perform well only when high $N_r/N_t$ is available.
% They result in poor detection performance, when the concurrently-served user count ($N_t$) approaches the given BS antenna count ($N_r$) due to the large noise amplification (Appendix~\ref{s:primer_linear}).
% since 
However, when the concurrently-served user count ($N_t$) approaches the given BS antenna count ($N_r$), 
% (\emph{i.e.,} smaller $N_r/N_t$), 
their detection suffers from large \emph{noise amplification} (Appendix~\ref{s:primer_linear}), resulting in poor detection performance~\cite{Geosphere}.  
% Since linear methods such as MMSE and ZF feature simple processing, they are widely used in experimental wireless systems as well as practical cellular systems. 
% However, linear detectors perform well only when high $N_r/N_t$ ratio is available (\emph{i.e.,} massive MIMO like $16\times64$ MIMO). They cannot guarantee sufficient performance due to their sub-optimality, when the concurrently-served user count ($N_t$) approaches the given BS antenna count ($N_r$), where the channel $\mathbf{H}$ becomes ill-conditioned, resulting in large noise amplification and thus poor linear detection performance.
% Massive MIMO systems featuring a large number of $N_r$ with linear detectors such as ZF and MMSE are the mainstream systems that are deployed in practice including 5G cellular systems.
% When Massive MIMO serves relatively small user counts ($N_r \gg N_t$, commonly with $N_r/N_t \geq 6$), the channel $\mathbf{H}$ becomes ``well-conditioned'' and thus the amplification factor in $\mathbf{n'}$ becomes small, making a linear detector perform well~\cite{Geosphere}. 
For this reason, many systems use a \emph{massive MIMO} regime where the BS with massive 64--128 antennas supports only a few users at a time (\emph{i.e.,} $N_r\gg N_t$);
% For example, Agora and Argos  
% Those systems blindly apply the error correction to the detected bits, but the performance of the linear detectors will become an eventual block factor to . 
% Agora~\cite{ding2020agora} supports $16\times64$ MIMO, $N_r/N_t=4$, Argos~\cite{Argos, argos-asilomar17} $16--32\times64--96$ MIMO, $N_r/N_t \approx 3--4$), BigStation $16\times 128$ 
% Argos~\cite{Argos, argos-asilomar17} $8\times96$ MIMO (\emph{i.e.,} $N_r/N_t = 12$), BigStation~\cite{Argos, argos-asilomar17} $8\times96$ MIMO (\emph{i.e.,} $N_r/N_t = 12$), Hydra~\cite{gong2023scalable} $32\times150$ MIMO (\emph{i.e.,} $N_r/N_t \approx 4.69$), and LuMaMi~\cite{malkowsky2017world} $10\times100$ MIMO (\emph{i.e.,} $N_r/N_t = 10$), while all of them using linear detectors. 
Table~\ref{t:nr_nt_ratio} shows examples of state-of-the-art Massive MIMO systems with their $N_r/N_t$.
% When Massive MIMO serves relatively small user counts ($N_r \gg N_t$, commonly with $N_r/N_t \geq 6$), the channel $\mathbf{H}$ becomes ``well-conditioned'' and thus the amplification factor in $\mathbf{n'}$ becomes small, making a linear detector perform well~\cite{Geosphere}. 
% For this reason, many MU-MIMO systems such as 
% Agora~\cite{ding2020agora}, Argos~\cite{argos-mobicom12,argos-asilomar17}, Bigstation~\cite{BigStation}, Hydra~\cite{gong2023scalable}, and LuMaMi~\cite{vieira2014flexible} exploit massive MIMO with linear detectors with 64--128 receiver antennas ($N_r$), supporting $\approx$ 8--10 users ($N_t$) at a time. 
% Blindly increasing the number of users will eventually result in very poor communication performance due to the limited detection capability of linear detectors.
% However, to support more users without detection performance degradation, even more BS-antennas than additional users need to be deployed. 
% (\emph{e.g.,} to support $N_t\approx 100$ users, $N_r\approx1000$ is required). 
% This implies the required high $N_r/N_t$ ratio will eventually become a bottleneck
% to spatial multiplexing gains since the deployment of larger numbers of antennas will become impractical.
% , particularly in WLAN. Accordingly, 
% for any given Massive MIMO antenna counts, limited users can be supported at a time~\cite{kim2021physics}.
% While many computing innovations have been made in prior work, 
However, the gain in spectral efficiency and MIMO capacity is proportional to $N_t$, and blindly increasing $N_t$ under the given $N_r$ (relying on FEC) is not effective, since \emph{the fundamental BER performance} of linear detectors will be an eventual bottleneck, resulting in significant errors and thus tremendous time and signaling overheads. To keep scaling up $N_t$ towards $N_r/N_t=1$, a regime called \emph{Large MIMO} (cf. Massive MIMO) for the ideal gains, (near-)optimal ML performance is necessitated~\cite{flexcore-nsdi17}, which is the case for (small $N_r$) WLAN scenarios as well. However, enabling ML detectors for large-scale MIMO has a complexity issue (see \S\ref{s:primer_mimo}). 
% \vspace{-0.3cm}
% , which QA MIMO detectors aim to resolve.
% Thus, to increase $N_t$ 
% \emph{i.e.}, $N_r/N_t \rightarrow  1$ called \emph{Large MIMO} (cf. Massive MIMO), in those systems. 
% For example, Agora tested for $32\times32$ MIMO with QPSK modulation at SNR xx dB leads to XXX throughputs even with FEC decoding.  
% there is a
% strong incentive to scale up MIMO sizes.

% \textcolor{red}{
% \parahead{E2E Discussion.}
% Current MIMO systems blindly report the system performance without considering MIMO detection's performance. They rely largely on channel decoding. We believe sophisticated MIMO detection can open up a new research topic: sophisticated MIMO detection and simplified error correction decoding, instead of the current simple MIMO detection with sophisticated decoding. NEED TO UPDATE}

% (Appendix~\ref{s:primer_linear} for details). 
% In other words, linear detectors require high $N_r/N_t$ ratio to perform well (\emph{e.g.,} $16\times 64$ MIMO).

\vspace{-0.2cm}
\subsection{PIC for Wireless Processing}
\label{s:PIC_MIMO_detectors}
From the perspective of non-traditional PIC algorithms, various specialized devices have also emerged such as digital annealers~\cite{aramon2019physics}, analog quantum annealers~\cite{denchev-44814,king2023quantum}, bifurcation machines~\cite{tatsumura2021scaling}, 
% optical processors for photonic computing~\cite{yang2023fiber},
digital quantum gate-model computers~\cite{monroe2013scaling,arute2019quantum}, memristor and spintronic Ising machines~\cite{sutton2017intrinsic,cai2020power,grollier2020neuromorphic}, and oscillator-based, photonic, and optical coherent Ising machines~\cite{dopo,mcmahon2016fully,oim,inagaki2016coherent,yang2023fiber,pierangeli2019large}, targeting the ability to accelerate computing of solving combinatorial optimization problems, leading to new areas to explore.
% in many different fields of research.
In wireless networks, despite its infancy, the PIC approach has already shown great potential in many different modules such as MIMO precoding~\cite{kasi2021quantum,winter2024lattice}, MIMO beam selection and satellite beam placement~\cite{dinh2023efficient,huang2023quantum}, error control coding~\cite{kasi2023quantum,kasi2020towards,sarma2022quantum}, intelligent meta-surfaces and discrete phase shifting~\cite{ross2021engineering,lim2023quantum,kim2023physics},
% beamforming of antenna arrays~\cite{kim2023physics}, 
and scheduling~\cite{vista2023hybrid}. Particularly for MIMO detection, extensive research efforts 
% on non-traditional hardware and algorithms 
have been made to expedite the computation required for optimal ML MIMO detection processing~\cite{rimimo,cui2022quantum,dimimo,cui2022general,kim2021physics,norimoto2023quantum,marosits2021exploring,sreedhara2023mu,takabe2023deep}.

\vspace{-0.15cm}
\subsection{Challenges in QA MIMO Detction}
\label{s:common_challenges}

Through prior studies~\cite{kim2019leveraging,marosits2021exploring,kim2022warm,ducoing2022quantum,tabi2021evaluation}, the promise of QA MIMO detectors has been experimentally demonstrated with great potential of accelerating ML processing. However, their challenges have been identified as well, which are rather different from the ones in conventional MIMO detectors.\footnote{These challenges are present due to multiple reasons having to do with fundamental spin-glass physics bottlenecks~\cite{knysh2016zero}, finite temperature fluctuations~\cite{albash2017temperature}, and the effect of analog parameter misspecification~\cite{pearson2019analog}).} 
% (\emph{i.e.,} intrinsic control error~\cite{pearson2019analog}).} 

% \parahead{Severe Degradation with High-Order Modulations.}
\parahead{(1)} Severe performance degradation with high-order modulations such as 16/64-QAM is commonly observed in PIC/QA-based MIMO detection. With low-order modulations such as BPSK and QPSK, large MIMO (like $60\times60$) with near-optimal performance 
% (\emph{e.g.,} over 100 times of spatial gains w/ BPSK) 
has been successfully reported within a few hundred microseconds of QA compute time, whereas 16-QAM MIMO problems even with only a few users (like $3\times3$ MIMO) are quite challenging, requiring several milliseconds. 
% \systemname{} enables the improved 16-QAM performance by using multi-seed parallel ensemble RA ($\S$\ref{s:design_overview}) at the cost of more qubit usage (note that FA with the same qubit usage still performs quite poorly).
% Given the search sizes of the ML problems, the observed abrupt performance degradation with 16/64-QAM is surprising.
% With this abnormal phenomenon, enabling high-performance high-order modulations becomes a well-known open challenge in QA-MIMO detection. Since high-order modulations are being used in 4G and 5G, and will be essential for NextG wireless systems as well, this challenge must be resolved for future general use of QA MIMO detection. 
% To our best knowledge, \systemname{} is the first QA MIMO detector that enables near-optimal detection with 16-QAM within a few hundred microseconds, while the sizes of enabled MIMO are quite small compared to ones with low-order modulations.

% \parahead{Error Floor at High SNRs.}
\parahead{(2)}
A BER \emph{error floor} is observed at high SNRs; BER flattens despite increasing SNRs.
% (sometimes even for low-order modulations). 
% The error floor (commonly encountered in sub-optimal methods) is generally acceptable since systems have FEC decoding module after MIMO detection module (Figure~\ref{f:mimo_detection}) through which a block of coded bits with errors becomes error-free data bits. However, 
When detection BER at the point of the error floor is not low enough, FEC decoding could result in errors. Then, data retransmission is required (overheads).

% the senders are required to  the data.
% need to re-transmit the  (ACK) to the senders for the retransmission, 
% resulting in tremendous time and signaling overheads. 
% In this work,

% \begin{figure*}
%     \centering    \includegraphics[width=0.7\linewidth]{Figures/IoT_ResQ.pdf}
%     \caption{\small\textbf{System architecture of IoT-ResQ~\cite{kim2022warm} consisting of FSD and RA (decomposition-based parallel RA). Since IoT-ResQ relies on the decomposition approach, only a single sub-problem retains the global optimum (sub-Ising model \#2). 
%     % IoT-ResQ can support only limited parallelism. Further, IoT-ResQ has high overheads due to non-linear classical detector FSD, requiring parallelism in both classical and quantum processing. 
%     When FA is directly applied to sub-Ising problems (instead of applying RA initialized by GS solutions), we call it QFSD (Quantum FSD).     \vspace{-0.2cm}
%     }} 
%     \label{f:IoT_ResQ}
% \end{figure*}

% , introducing the rudimentary structure of QA-based MIMO detector designs as shown in Figure XXX. Unlike the standard QA algorithm, QuAMax applies anneal pause in the middle of FA. 
%  However, this approach 

% \parahead{RA-based Detectors.} RA was studied for MIMO detection~\cite{kim2020towards,kim2022warm}. IoT-ResQ makes use of RA for MIMO detection in the context of IoT connectivity.  

\vspace{-0.1cm}
\parabreak To mitigate these challenges, \systemname{} tweaks QA optimization using a \emph{parallelization} strategy uniquely enabled by RA.

\begin{figure}
\centering
    \includegraphics[width=0.86\linewidth]{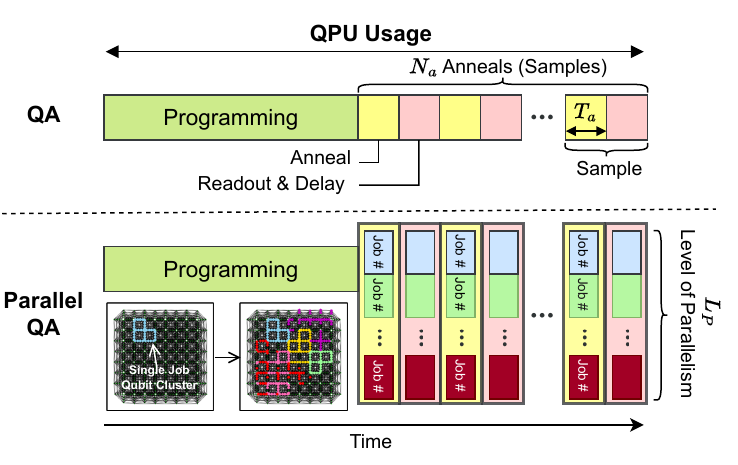}
    \vspace{-0.2cm}
    \caption{\small
    \textbf{Machine QPU operation for QA and parallel QA.
    }} 
\label{f:parallel_QA}
\end{figure}

\vspace{-0.1cm}
\subsection{Parallel QA Optimization}
\label{s:QA_parallel_strategy}

\begin{figure*}
    \centering    \includegraphics[width=0.75\linewidth]{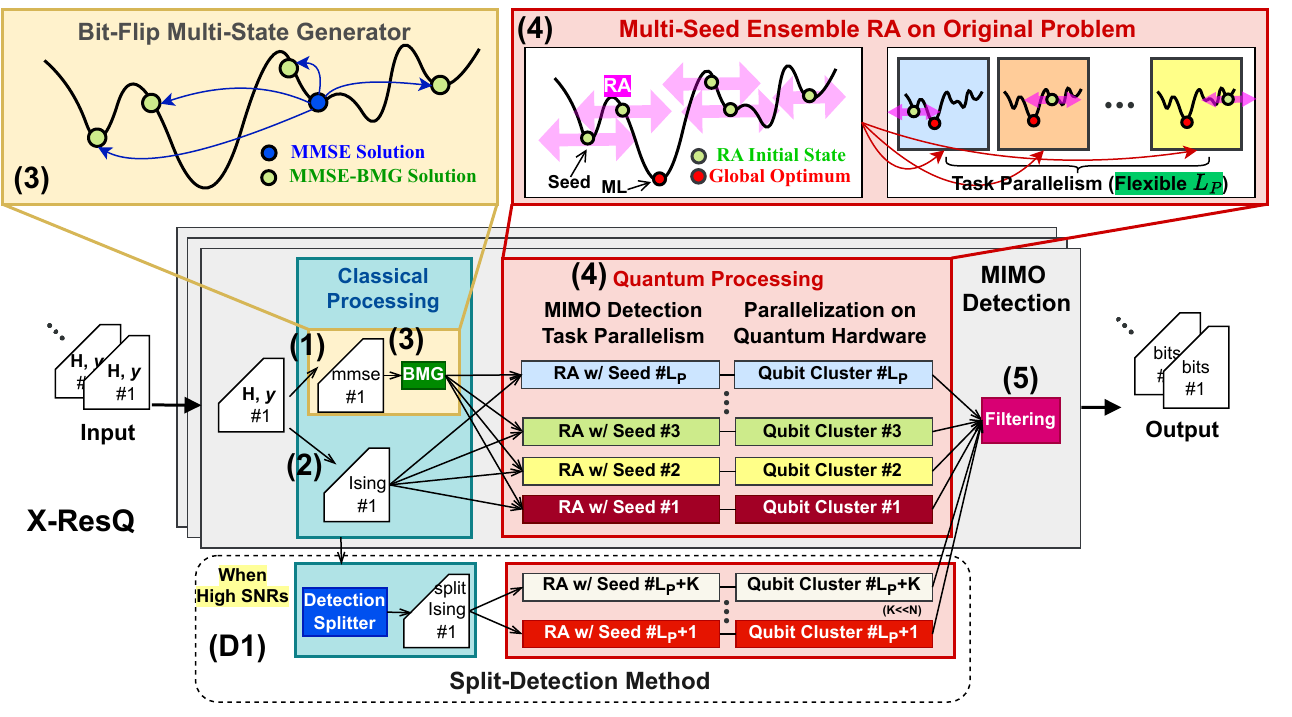}\vspace{-0.2cm}
    \caption{\small\textbf{System Architecture of X-ResQ (multi-seed parallel ensemble RA). \systemname{} is based on the currently-deployed MMSE detector, 
    % (cf. FSD in IoT-ResQ), 
    requiring only simple preprocessing. Unlike IoT-ResQ, 
    % with its multi-seed ensemble RA (optimization core),
    \systemname{} can support flexible parallelism with fine granularity for any $L_P$, where all the tasks can converge to the ML solution, thus increasing the ML fidelity through sample parallelism.  \vspace{-0.2cm}
    % Further, it can mitigate the error-floor phenomenon at high SNRs with its split-detection method.
    }} 
    \label{f:design_overview}
\end{figure*}

Parallelization of QA is an efficient approach to further accelerate QA optimization or boost the overall QA system performance by using more qubits~\cite{pelofske2022parallel,pelofske2022solving,huang2023quantumparallel}.
% , just like parallelism on classical platforms. 
Figure~\ref{f:parallel_QA} shows the machine operation for (non-)parallel QA, consisting of hardware programming, annealing, and readout~\cite{DwaveQPU}. In parallel QA, each anneal handles $L_P$ independent (same or different) jobs/tasks simultaneously on different qubit clusters, where $L_P$ is the \emph{level of parallelism}.
% (\S\ref{s:parallel_QA}.
With \emph{fully parallel} QA that does not require any iterative QPU operations (cf. anneal iterations), the system can avoid multiple occurrences of inevitable overheads related to the considerable programming time.\footnote{The programming time on the current machine is several milliseconds, while the readout with delay is around 25 -- 150~$\mu$s per sample. However, they are being reduced exponentially every generation, and specific techniques to make them within a few (tens of) microseconds have been already identified~\cite{grover2020fast,weber2021high,reed2010fast,walter2017rapid}. We discuss \systemnames{} potential practicality in \S\ref{s:discussion}.} Thus, parallel QA will become essential for latency-intensive applications like MIMO detection; for example, multi-stage approaches~\cite{singh2023uplink,karimi2017boosting,kim2021physics} do not work for QA MIMO detectors due to the aforementioned tight deadline. While data parallelism is quite straightforward (\emph{e.g.,} across subcarriers)~\cite{ducoing2022quantum}, in this paper we are interested in \emph{task parallelism} to further accelerate detection, which has been barely studied so far in the context of PIC methods like QA. 

A straightforward parallel QA method is \emph{sample parallelism} that programs/embeds the ``same'' problem multiple times on QA hardware~\cite{pelofske2022parallel,huang2023quantumparallel}. This approach aims to collect more samples per anneal (total sample counts = $L_P\cdot N_a$), thus achieving the (linear) speedup across used qubits to obtain the target sample count. 
QuAMax~\cite{kim2019leveraging} (and classical ParaMax~\cite{kim2021physics}) applies this technique with FA to solve Ising spin models
of ML MIMO problems based on the available qubit counts (without actual implementations).
Parallel ensemble QA~\cite{ayanzadeh2022equal} has also been studied, where multiple Ising models are generated by adding controlled noises to the original one to overcome the system bias, but the method also uses FA and is not tested by actual parallel QA runs. 
% However, it uses FA, and high overheads (compared to \systemname{}) are required due to the generation of the noises and the multiple Ising models.
\emph{Decomposition-based parallelism} is 
% one of the most conventional 
a task parallelization strategy commonly used in both classical~\cite{FCSD} and QA~\cite{pelofske2022solving} optimization. Its typical idea is to decompose an original problem into multiple ``easier'' subproblems (by prefixing some hard variables via \emph{full search}) and process the resulting subproblems in parallel. 

IoT-ResQ~\cite{kim2022warm} is a decomposition-based parallel QA MIMO detector that consists of Fixed-complexity Sphere Decoder (FSD)~\cite{FCSD} and RA.
% uses this decomposition approach for QA MIMO detection,
For detection, IoT-ResQ first transforms an ML problem into an equivalent tree search (Figure~\ref{f:MLD}) and conducts the full search for the initial $N_{fs}$ levels in the tree (\emph{i.e.,} symbol full expansion for $N_{fs}$ users), resulting in $|\mathcal{O}|^{N_{fs}}$ subproblems (subtrees with reduced $N_t-N_{fs}$ levels).
Then, decomposed subproblems are solved by a greedy search (so far, FSD), followed by RA optimization initialized by the corresponding greedy search solutions, all in parallel. The overall architecture of IoT-ResQ is available in Appendix~\ref{s:iot-resq}. We discuss limitations of QuAMax and IoT-ResQ in \S\ref{s:xresq_vs_iotresq}.

\vspace{-0.1cm}
\section{Design}
\label{s:design}

In this section, we describe \systemname{}. Section~\ref{s:design_overview} introduces \systemnames{} overall architecture and operation. In Section~\ref{s:xresq_vs_iotresq}, we explain its properties, highlighting the differences against those of other QA MIMO detectors. Section~\ref{s:desing_principle} discusses the design rationale, and Section~\ref{s:parallel_QA} explains preprocessing for multi-seed parallel RA. Section~\ref{s:mmse_quadrant} introduces the split-detection method to mitigate the error floor at high SNRs.

\parahead{Quantum-based C-RAN.} Envisioned scenarios for \systemname{} (and other QA MIMO detectors) assume
\emph{Centralized Radio Access Networks} (C-RAN) with QA processors co-placed in a cloud or edge \emph{data center} along with existing classical processors, separating their computing roles~\cite{kim2021heuristic}.
The quantum-based C-RAN resolves many practical issues (\emph{e.g.,} on cryogenic temperatures for operations and machine access overheads).
More discussions on QA feasibility in wireless cellular networks are available in Section~\ref{s:discussion}.

% \systemname{} supports flexible and scalable parallelism without any tremendous changes in the current MIMO systems and QA hardware.

% In this section, we describe \systemname{} starting from introducing the motivation of the system (based on related prior work) to the explanation of the design components. 

% \begin{figure}
%     \centering
%     \includegraphics[width=0.85\linewidth]{Figures/CRA.png}
%     \caption{\textbf{Cross Reverse Annealing}
%     % is equivalent to direct 16-QAM process, but not directly applied to MU-MIMO.
%     } 
% \label{f:multi_qpsk}
% \end{figure}

% \subsection{Motivation}
% \label{s:xresq_motivation}

\vspace{-0.1cm}
\subsection{\systemname{} Architecture Overview}
\label{s:design_overview}

% In this subsection, we introduce \systemnames{} overall architecture, compared to IoT-ResQ (recent RA-based MIMO detectors).
\systemname{} is a parallel QA-accelerated MIMO detection system that utilizes multi-seed ensemble RA as its task parallelism. Its parallelization strategy is 
% uniquely enabled by the RA protocol (\S\ref{s:primer_ra}), 
designed to maximize RA's advantages and thus further expedite quantum ML detection efficiently with more qubits. 
Figure~\ref{f:design_overview} shows the overall system architecture and operation of \systemname{}.
For MIMO detection, \systemnames{} structure has the following stages:
% However, FSD cannot support flexible parallelism; it can support only $
% |\mathcal{O}|^{N_{fs}}$ levels of parallelism, where $N_{fs}$ ($0 \leq N_{fs}\leq N_t$, typically 1 or 2) is controllable user counts whose symbols are fully expanded for the decomposition.
% \parahead{\systemname{}.} 

\noindent \textbf{(1) Initial MMSE Detection.} \systemname{} initially applies MMSE detection based on the received signal and estimated channel. This stage has very low complexity as the MMSE detector computes its equalizer only once per channel coherence time (remaining same for several milliseconds) and the detection operation for instances (\emph{i.e.,} channel uses) involves just matrix-vector multiplications.

\noindent \textbf{(2) ML Ising Model Generation for Parallel Ensemble QA.}
At the same time, the system prepares an Ising model consisting of multiple ML Ising forms for parallel QA ($\S$\ref{s:parallel_QA}). During the channel coherence time, only $f_i$ (in Eq.~\ref{eq:ising-ham}) needs to be updated per channel use based on generalized forms~\cite{kim2019leveraging}, requiring also only insignificant matrix-vector multiplications. 
% Thus, the computational amount in this step is  as well.
% As noted before,  requires representing the problem as an Ising problem. 
% In this stage, the X-ResQ prepares the ML Ising form. 

\noindent \textbf{(3) Multi-Seed Generation.}
Based on the MMSE solution, a \emph{bit-flip multi-state generator} (BMG) brings out multiple $L_P$ different solutions/states by flipping a single bit on it ($\S$\ref{s:parallel_QA}).

\noindent \textbf{(4) Multi-Seed Parallel Ensemble RA.} Then, the system makes use of them as multi-seed initialization states, each of which becomes an initial state for each RA run. \systemname{} applies RA to all $L_P$ runs in parallel (all with the original Ising form). We call this structure and operation, \emph{multi-seed ensemble RA}. Unlike prior designs, with this simple parallelization strategy, \systemname{} features many desirable properties for parallel QA systems, as described in the next subsection. 
% In Section~\ref{s:desing_principle}, we explain its design rationale.
% (the core parallel QA design in \systemname{}).

\noindent \textbf{(D1) Split-Detection (for high SNRs).}
When SNRs are high,
% starting to lead to the error floor, 
\systemname{} generates an additional Ising form based on the \textit{split-detection approach} described in $\S$\ref{s:mmse_quadrant}, which is designed to mitigate QA detectors' error floor at high SNRs ($\S$\ref{s:common_challenges}). This new Ising form consists of two parts: \emph{quadrant search} and \emph{position search} in the quadrant. While these two are jointly considered in the original Ising form, they are considered \emph{separately} in \systemnames{} split-detection Ising form by transforming the original problem into simpler B/QPSK problems. 
% The form is the same size ($N_V$) as the original form, is independent of the generated forms for the baseline \systemname{}, and can be solved also in parallel together with them.  
% Theoretical noise analysis is available in Appendix~\ref{s:noise_analysis}.

% Relatively small number of resources will be assigned to this form. 
% due to the chance of the qubit waste for the same reason as a decomposition approach.

% \fromAKS{
% \parahead{(3) Quantum Processing:} X-ResQ uses RA to solve the Ising instance(s). These RA runs will generate several candidate solutions, one or more of which can be the optimal solution. 
% }

\noindent \textbf{(5) Filtering as Post-Processing:} \systemname{} accumulates all $L_P\cdot N_a$ samples generated by all RA runs, 
% ($L_P$ is the applied levels of parallelism and $N_a$ is anneal counts),
and filters the best one that has the minimum energy (Eq.\ref{eq:ising-ham}) as the final solution, which will be translated into detected bits.\vspace{-0.18cm}
% Finally, X-ResQ will output the solution that has the lowest energy. 
% \parahead{(1)}
% \systemname{} first applies MMSE detection. Meanwhile, the system prepares the ML Ising form. Based on the MMSE solution, a bit-flip multi-state generator (BMG) brings out multiple $N$ different solutions/states by flipping a single bit on it ($\S$\ref{s:parallel_QA}). Then, the system makes use of them as multi-seed initialization states, each of which becomes an initial state for each RA run. \systemname{} applies RA to all $N$ runs in parallel with the original Ising form. We call this structure and operation, \emph{multi-seed parallel ensemble RA}. 
% % Here, the Ising form is the same as the original problem. 

% \vspace{-0.1cm}
% \parabreak{}

% \subsection{\systemname{} versus IoT-ResQ}

% \newcolumntype{Y}{>{\centering\arraybackslash}X}
\begin{table}[htbp]
\begin{tiny}
% \begin{footnotesize}
\centering
\caption{\small Comparison of QA MIMO detectors (Section~\ref{s:xresq_vs_iotresq})}
\vspace{-0.1cm}
\begin{tabularx}{\linewidth}{*{6}{Y}}
\toprule
\textbf{\tiny{QA MIMO Detectors}} & \textbf{\tiny{QA Algorithm}} & \textbf{\tiny{Flexible Parallelism}} & \textbf{\tiny{Classical Pre-Process}}  & \textbf{\tiny{Sample Parallelism}} & \textbf{\tiny{Easier Subproblems}}  \\ \midrule
%Access time & 26.647~ms\\
%Access overhead & 5.834~ms\\
\textbf{\scriptsize{QuAMax}} &  \scriptsize{\cellcolor{LightRed}FA} & \scriptsize{\cellcolor{LightGreen}yes} & \scriptsize{\cellcolor{LightGreen} light} &
\scriptsize{\cellcolor{LightGreen}yes} & \scriptsize{\cellcolor{LightRed}no} \\
\textbf{\scriptsize{IoT-ResQ}} & \scriptsize{\cellcolor{LightGreen}RA} & \scriptsize{\cellcolor{LightRed}no} & \scriptsize{\cellcolor{LightRed}heavy} & \scriptsize{\cellcolor{LightRed}no} & \scriptsize{\cellcolor{LightGreen} yes} \\
\textbf{\scriptsize{X-ResQ}} & \scriptsize{\cellcolor{LightGreen}RA} & \scriptsize{\cellcolor{LightGreen}yes} & \scriptsize{\cellcolor{LightGreen} light} & \scriptsize{\cellcolor{LightGreen}yes} & \scriptsize{\cellcolor{LightGreen}yes} 
\end{tabularx}
\label{t:qa_detectors_properties}
\end{tiny}
\end{table} 
\vspace{-0.4cm}
\subsection{\systemname{} System Characteristics}
\label{s:xresq_vs_iotresq}

% In this subsection, we explain the properties of \systemname{}, highlighting the differences against those of other QA MIMO detectors, QuAMax and IoT-ResQ. 
% Despite its simple design (\S\ref{s:design_overview}), 
\systemname{} features many desirable properties for a parallel QA MIMO detector, also satisfying all the requirements for \emph{ideal} parallel MIMO detectors~\cite{nikitopoulos2022massively}.\footnote{\systemname{} 
% is implementable only on specialized quantum hardware due to QA, implying that it 
does \emph{not} satisfy ``being transparent to the implementation technology'' (fifth requirement in~\cite{nikitopoulos2022massively}) due to QA. However, note that we also demonstrate non-quantum \systemname{} (\S\ref{s:implementation}) that is based on a generic PIC algorithm that can be implemented on any platform.}
The comparison of QA MIMO detectors (QuAMax, IoT-ResQ, \systemname{}) is summarized in Table~\ref{t:qa_detectors_properties}, and we discuss the characteristics one by one in this subsection.

\parahead{QA Algorithm.} In the context of MIMO detection, unlike FA (used in QuAMax), RA (\S\ref{s:primer_ra}) can provide opportunities for classical-quantum \emph{hybrid} optimization, where RA heuristically corrects the initial classical detector's (non-ML) solutions into the optimal ML solutions. This hybrid structure (also known as \emph{warm-started} quantum optimization)\footnote{This hybridization also applies to gate-model quantum processors~\cite{egger2021warm}.} provides not only improved QA detection performance over FA~\cite{kim2022warm,kim2020towards}, but also a chance of \emph{opportunistic} quantum optimization. 
% without having to significantly change the current system structures.
Specifically, in the case of \systemname{} comprised of linear MMSE and RA, RA quantum optimization can be skipped for MIMO scenarios where MMSE performs well, and be applied only when MMSE often fails to find the ML solution (\emph{e.g.,} user/traffic peak times). 
For these reasons, we argue that RA is a more pragmatic QA algorithm 
% (or wireless networks generally) 
than FA in MIMO detection. Furthermore, \systemnames{} design does not have to significantly change the current wireless systems, since it relies on the currently deployed MMSE detector as an initial classical detector (cf. non-linear FSD in IoT-ResQ).

\parahead{Flexible Parallelism with Fine-Granularity.}
% can support any levels of parallelism ($L_P$) 
Available levels of parallelism ($L_P$) in \systemname{} are \emph{any} number (until hardware limits). Thanks to this flexibility, 
% depending on qubit counts on the hardware, 
\systemname{} can achieve detection performance gains \emph{elastically} with fine-grained parallel QA, thus requiring fewer $L_P$ (\emph{i.e.}, fewer qubits) to obtain (near-)optimal performance.
IoT-ResQ cannot support this flexible parallelism; because of the symbol full expansion (\S\ref{s:QA_parallel_strategy}), available $L_P$ is limited to the power of the modulation size (\emph{i.e.,} $L_P = \mathcal{|O|}^1,\mathcal{|O|}^2,\cdots$). For example, with 16-QAM, IoT-ResQ cannot achieve any gains between  $L_P= 16^{1}$ and $16^{2}$, which implies coarse-grained parallelism with inflexible and inefficient qubit usage (see Table~\ref{t:qubitcounts} for comparisons).

\begin{table}[t]
\centering
\caption{\small The required number of qubits for 16-user fully parallel QA MIMO detection operations with QPSK ($|\mathcal{O}|=4$) on Zephyr-topology QA hardware (\S\ref{s:implementation}) are shown.}\vspace{-0.1cm}
% Available qubit count divided by required qubit count can provide approximate available parallelism.
% Both the corresponding maximum supportable user numbers (without parallelism) at a time for various modulations and 
% The corresponding maximum parallelisms on a single QPU are provided.
% \vspace{-0.1cm}

\begin{small}
\begin{tabularx}{\linewidth}{*{5}{X}}
% \multicolumn{6}{c}{\textbf{Dense Urban}}\\
% \toprule
% \midrule
{\textbf{\footnotesize \,\,\,\,\,\,\,\,\,\,\,\,\,\,\,\,\,\,\,\,\,\,\,\,\,\,\,\,\,\,\,\,\,\,\,\,\,\,\,\,\,\,\,\,Available $L_P$}} & \multicolumn{4}{c}{\textbf{\footnotesize Required Qubits}}\\
\midrule
{\tiny\textbf{\,\,\,\,\,\,\,\,\,\,\,IoT-ResQ}\,\,\,\,\,\,\,\,\,\,\,\,  4 / 16 / 64 / 256 / 1024 / $\cdots$}& \multicolumn{4}{c}{{\tiny480 / 1,792 / 6,656 / 18,432 / 67,584 / $\cdots$} }\\
{\tiny \textbf{X-ResQ, QuAMax} \,\,\,\,\,  1 / 2 / 3 / 4 / 5 / $\cdots$}& \multicolumn{4}{c}{{\tiny128 / 256 / 384 / 512 / 640 / $\cdots$ \,\,\,\,\,  } }\\
\bottomrule
\end{tabularx} 
\end{small}
\label{t:qubitcounts}
\end{table}

% and generally in wireless networks.
% , and
% this paper investigates a QA parallelization strategy that is uniquely enabled by RA for enhanced efficiency of QA MIMO detection.

\parahead{Classical QA Preprocessing.} 
Classical preprocessing is light in \systemname{} because of the linear MMSE. 
Furthermore, MMSE detection and Ising formulation can be processed in parallel, and the time required for the bit flip-based BMG is negligible. 
Also, X-ResQ requires only quantum parallelism, while IoT-ResQ requires parallelism in both classical and quantum processing parts. This implies
classical resources can be used for data parallelism (\emph{e.g.}, subcarriers) in \systemname{}.
% Furthermore, while IoT-ResQ requires task parallelism both in the classical and quantum processing parts, \systemname{} requires only quantum task parallelism. This implies classical resources can be used for data parallelism (\emph{e.g.,} subcarriers).

\parahead{Sample Parallelism.} Due to the use of the original Ising form for parallel tasks, all the jobs can converge to the ML solution in \systemname{} (\emph{i.e.,} sample parallelism in \S\ref{s:QA_parallel_strategy}). 
In terms of resource usage, this is a desirable feature in that
there are no qubits used on searches that never
reach the ML solution. Furthermore,
given that QA is a probabilistic technique, sample parallelism not only accelerates optimization convergence but also increases the fidelity of getting the ML solution by collecting more samples: the optimum probability is $1-(1-P_G)^{\text{sample count}}$, where $P_G$ is \emph{the probability of finding the ML solution per sample}. In the case of IoT-ResQ, sample parallelism is not supported, since
all parallel QA jobs solve different subproblems because of the decomposition nature, where among all the subproblems, \emph{only one} entails the ML solution (\emph{i.e.,} $P_G=0$ for the others).
% symbol prefixing during the full search. 
While this is inevitable for the approach, 
% in terms of resource usage 
% it is not quite efficient in that 
% qubits are expensive and precious resources,
% among all sub-problems, \emph{only one} entails the global optimum,
its optimum probability could be low, despite slightly increased $\hat{P}_G$ in the correct subproblem\footnote{To satisfy $1-(1-P_G)^{N_a\cdot|\mathcal{O}|} < 1-(1-\hat{P}_G)^{N_a}$, the gain ($\hat{P}_G / P_G$) has to be high enough (\emph{e.g.,} over 8 for $P_G$ =0.1 w/ 16-QAM).
% high to outperform the sample parallelism in terms of the optimum probability (\emph{i.e.,} ). 
However, the (FSD-based) decomposition method has not been quite effective for QA MIMO detection~\cite{kim2023finer}, and surprisingly we experimentally found that it often results in adverse effects, particularly for $N\times N$ Large MIMO (see Appendix~\ref{s:decom_adverse_effect}).} with the reduced search space ($2^{N_V} \rightarrow2^{N_V-N_{fs}log_2(|\mathcal{O}|)}$).

% ($\hat{P}_G$ from the subproblem of $2^{ (N_V-N_{fs}log_2(|\mathcal{O}|))}$ space, cf. $P_G$ from $2^{N_V}$).
% (\emph{i.e.,} $P_G$ obtained from the original problem of $2^{N_V}$ search space size vs. $\hat{P}_G$ from the subproblem of $2^{ (N_V-N_{fs}log_2(|\mathcal{O}|))}$). 
% , though each subproblem has slightly higher $P_G$\footnote{For $(1-P_G)^{16} > (1-P_G')$  with $L_P=16$, $P_G'$ needs to $P_G' > P_G$} 

% \footnote{Surprisingly, we experimentally found that decomposed subproblems could be even harder problems than the original problem for QA, which does not occur in the case of deterministic classical solvers like GS in FSD.}

% QuAMax~\cite{kim2019leveraging} made use of the technique with FA in order to solve Ising spin models of ML problems. 

% \parahead{Elastic Parallelism.}

% \parahead{Elastic Parallelism.}

% \parahead{Elastic Parallelism.}

\begin{figure}
\centering
    \includegraphics[width=0.83\linewidth]{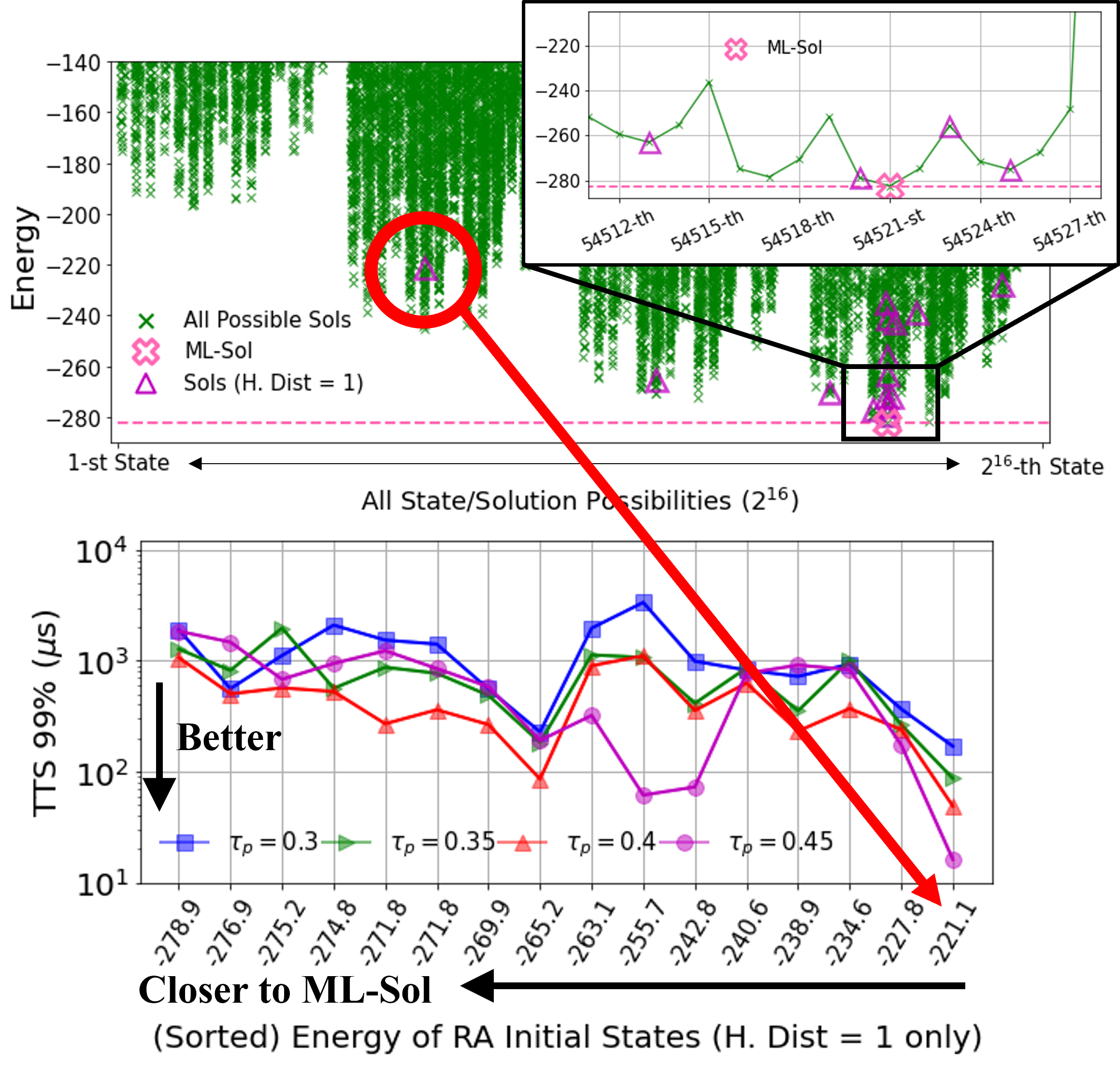}    \vspace{-0.1cm}
    \caption{\small
    \textbf{TTS Analysis of RA that is initialized from H. Dist=1 initial states using a $4\times4$ MIMO detection instance ($N_V=16$) at SNR 20~dB. 
    It demonstrates initial states that have lower Ising energies do not necessarily result in better RA results.     \vspace{-0.1cm}}}
    % ; the highest-$E$ state (circle) turns out to be the best initial state in this example.  
    % (Figure~\ref{f:TTS_SEED} for \systemname{} is related, where parallel RA is considered).
    % }} 
\label{f:tthamming_1}
\end{figure}

\parahead{Making Problems Easier with Parallelization.} 
% The main objective of detection task parallelism is to improve the detection performance as more resources (qubits) are assigned.
% (\emph{i.e.,} higher $P_G$ is expected as $L_P$ increases).
% IoT-ResQ can facilitate solving optimization problems by generating multiple easier subproblems (w/ reduced search space) that are solvable in parallel at the expense of qubits. 
As discussed, IoT-ResQ makes a detection problem easier by decomposing it into simpler subproblems 
% (\emph{i.e.,} $\hat{P}_G > P_G$) 
% that are solvable in parallel, at the expense of qubits 
(\emph{i.e.,} impact on $P_G$ and thus TTS (\S\ref{s:desing_principle}) across $L_P$). QuAMax's sample parallelism does not have the capability
% facilitate the optimization search itself 
because of the same Ising form used for parallel tasks (\emph{i.e.,} same $P_G$ across tasks); when detection scenarios become challenging where most of the anneal trials fail ($P_G\approx0$), it achieves nearly no effects. On the other hand,
\systemname{} 
% (\emph{i.e.,} multi-seed ensemble RA) 
has an impact on $P_G$ across $L_P$, even with the same Ising form.
% Each task is used for RA initiated from a different state, and this design allows the facilitation ability 
In \systemname{}, RA is initiated by a different state per task. In doing so, a similar effect of solving a different-difficulty optimization problem is caused,
% , which leads to a similar effect of solving different problems, 
since RA's optimization search tends to be localized around the
initial state (quasi-nonlocal search~\cite{king2019quantum}), thus achieving better performance with higher $L_P$.
% For comparison, a parallel QA detector only with sample parallelism (through either FA or RA with the same initial state) does not facilitate the optimization search. When detection scenarios become challenging (like 16-QAM MIMO),  most of the anneal trials fail it and thus the detector achieves nearly no effects. 
% However, because this approach does not facilitate the optimization search itself, when detection scenarios become challenging (like 16-QAM MIMO) where most of the anneal trials fail, it achieves nearly no effects. 
% uniquely enabled by \systemnames{} parallelization strategy. 
We discuss this further in the next subsection with the rationale of \systemnames{} strategy, showing its experimental validation. 
\vspace{-0.1cm}
\subsection{Rationale of Parallelization Strategy}
\label{s:desing_principle}

% \systemnames{} multi-seed parallel ensemble RA is based on a \emph{hypothesis} on initial state qualities that there exists a currently-unknown factor on RA initial states that determines RA's MIMO optimization performance other than commonly-considered factors (\emph{e.g.,} hamming
% distance or energy of the initial state). In other words, we argue that we do not know what states are good initial states for RA detection optimization. \systemname{} is designed to make use of this ambiguity as a tool to provide a benefit to the optimization performance. In this subsection, we clarify our hypothesis and empirically verify it.

In this subsection, we explain the rationale for the \systemnames{} parallelization strategy (\emph{i.e.,} multi-seed ensemble RA).

% \textcolor{red}{
\parahead{QA Benchmark Metric: Time-to-Solution.} We use a commonly accepted metric for QA performance microbenchmarks, \emph{time-to-solution} (TTS)~\cite{ronnow2014defining}. TTS indicates the estimated required time ($\mu$s) 
 % median (or mean)across instances 
to find the global optimum in an optimization problem (\emph{i.e.,} ML solution in MIMO detection) with 
% $C_T$\%:
% \vspace{-0.1cm}
% \begin{small}
% \begin{equation}\label{eq:n-rep}
%     TTS(C_T\%) = \text{Anneal duration } T_a\cdot \frac{\log(1-C_T\%/100)}{\log(1 - P_G)}, 
% \end{equation}
% \label{eqn:tts}
% \end{small}
% \vspace{-0.1cm}
% \noindent where $C_T$\% is 
\emph{target confidence} (typically 99\%): $\text{TTS}(99\%) = T_a\cdot \log(1-0.99)/\log(1 - P_G)$, where $P_G$ is empirically obtained (out of 5,000 samples in this paper). 
% Since there is neither theoretical guidance nor precise simulators to acquire $P_G$, it is empirically obtained by the QA run based on the \emph{occurrence of hitting the ML solution} out of sufficiently large $N_a$ (5,000 in this paper). 
% For QA algorithms of the same anneal duration, higher $P_{G}$ results in lower TTS, implying better optimization performance. 
Lower TTS typically implies better QA optimization performance. 
% The TTS benchmark between FA and RA is in~\cite{kim2022warm}.

Recall that RA optimization performance depends on its classical initial state ($\S$\ref{s:primer_ra}).
Thus, we delve into RA to identify high-quality initial states that trigger better RA optimization (lower TTS).
% and to establish an efficient parallelization strategy on how to make use of extra qubits for parallel QA. For RA, it is challenging to define high-quality initial states, since there are too many related interdependent factors as well as imperfect hardware implementation. 
There have been some attempts to correlate the quality to initial states' Hamming distance (H. Dist) or energies against the global optimum~\cite{golden2021reverse,passarelli2023counterdiabatic,kim2020towards}. Since H. Dist is unknown information in MIMO detection, we analyze the impact of energies of initial states.\footnote{In \cite{kim2022warm}, the correlation between H. Dist and Ising energies of possible states is observed to some extent, but it is only limited to low-order modulations.}

\begin{figure}\centering
    \includegraphics[width=0.76\linewidth]{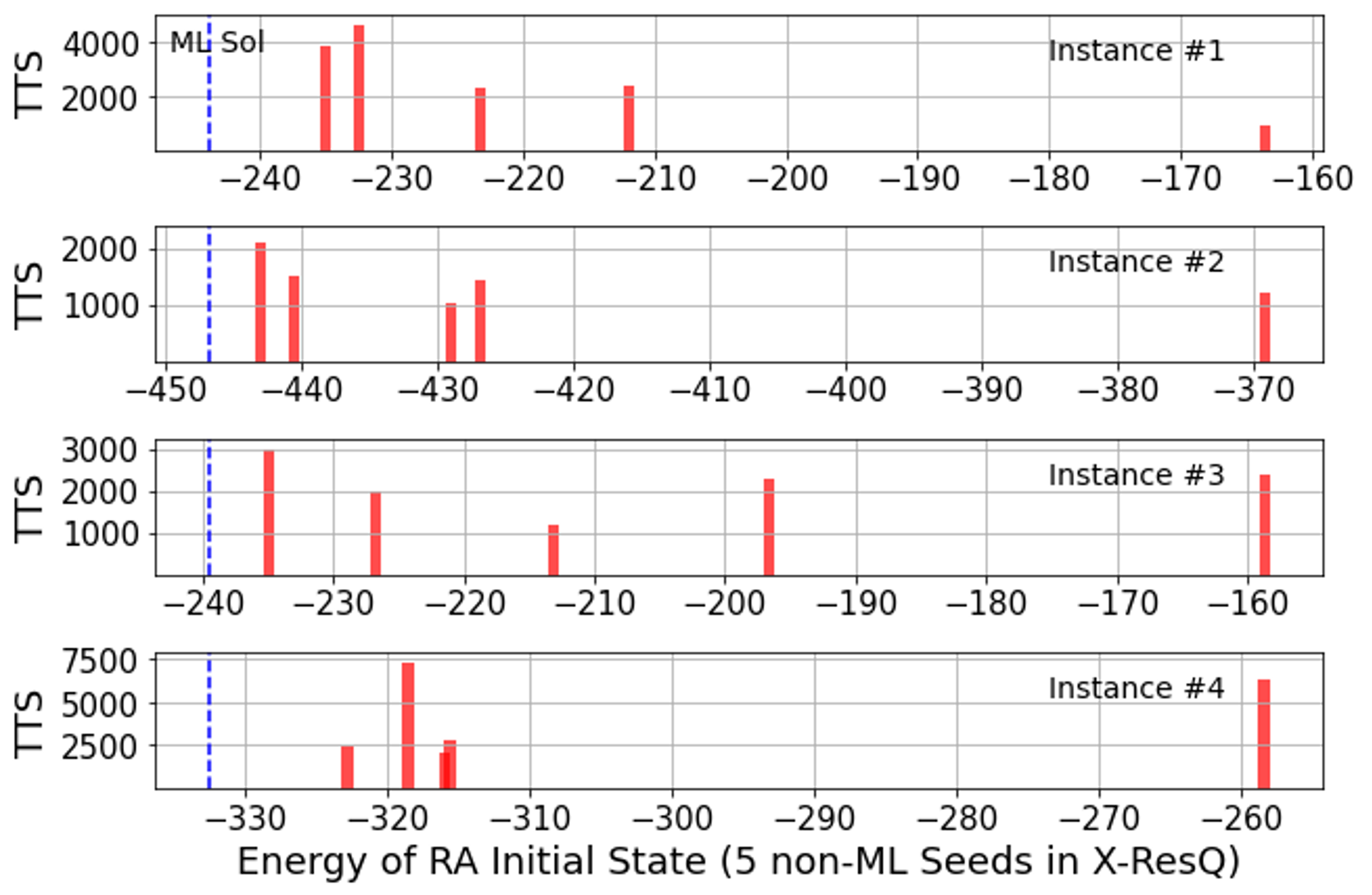}    \vspace{-0.1cm}
    \caption{\small
    \textbf{TTS 99\% ($\mu$s) of fully parallel \systemname{} ($L_P=5)$ for 4$\times$4 MIMO detection at SNR 20~dB. Each task out of the five tasks is parsed for TTS analysis from parallel \systemname{} results. A similar insight as Figure~\ref{f:tthamming_1} is observed for fully parallel runs.}} 
\label{f:TTS_SEED}
\end{figure}

Figure~\ref{f:tthamming_1} shows all possible states and the 16 possible states that have H. Dist 1 against the ML solution with their corresponding energies for a single, illustrative 16 spin-variable detection instance at SNR 20 dB (\emph{upper}). Then, we plot their TTS performance of RA optimization (\emph{lower}) with these H.Dist-1 states being an RA initial state, for different switching points $\tau_p$ ($\S$\ref{s:primer_ra}). \textbf{We observe that the states of lower energies do not necessarily result in better RA optimization, regardless of $\tau_p$ choices} (for QPSK as well). Except for $\tau_p=0.45$, two states are considered good initial states (one at $E=-265.2$ and the other at $E=-221.1$). Interestingly, the latter is the state with the highest energy among them. We verify this unexpected outcome with \emph{fully parallel} \systemname{} runs as well (with states generated by BMG). Figure~\ref{f:TTS_SEED} plots TTS of \systemnames{} RA with four random instances, for each of which we parse the parallel run results into each task and analyze its TTS. It is also observed that initial states that are closer to the global optimum (ML Sol) do not necessarily result in better RA performance. For all the instances, \emph{unpredictable} states turn out to result in the best TTS (among them), sometimes showing even large TTS gaps against the others.

\parabreak{}These surprising results led us to three following conclusions that \systemnames{} design derives from: \textbf{(1)} There are currently unknown (but existing) factors that determine the quality of initial states in RA. \textbf{(2)} Leveraging multiple initial states could improve the overall RA performance due to this ambiguity. 
% since the states with lower $E$ cannot guarantee better RA performance. 
\textbf{(3)} Thus, for parallel tasks, assigning qubits to multiple independent RA runs that are initialized by different states could be more efficient than focusing all resources on an RA run with a single initial state (unless we can precisely define the factors and hence better initial states). Intuitively, by avoiding using a single state, \systemname{} can prevent its optimization search from being stuck at the same local minima that is hard to escape from and make better use of the effect of RA's quasi-(non)local search with distributed initial points in parallel. 

\vspace{-0.1cm}
\subsection{Generating Ising Models and Initialization States for Parallel Ensemble RA}
\label{s:parallel_QA}

For QA MIMO detection, Ising models of the ML problems need to be prepared. Then, a QA solver applies a QA algorithm to solve the generated Ising models whose ground states correspond to optimal ML solutions. 
% With $N_a$ anneal runs and $N$ levels of parallelism, $N_a\times N$ solution candidates will be collected and the best one with the minimum $E(\mathbf{s})$ is selected as the final solution. 
% Since $\mathbf{s}$ is mapped to wireless symbols one by one (\emph{e.g.,} $\mathbf{s}=\{-1,1,1,-1\} \leftrightarrow \mathbf{v}=\{-1+1j,1-1j\}$) without any ancillary variables, it can be directly translated into the corresponding bits (\emph{i.e.,} MIMO detected bits). 

\parahead{Ising Spin Models of ML Problems.} 
ML problems can be transformed into equivalent Ising models with \emph{linear mapping} between possible wireless symbols and spin variables~\cite{kim2019leveraging}. Per instance, $N_V=N_t\log_2(\mathcal{|O|})$ spins are required, and the resulting ML Ising models are (nearly) fully-connected (\emph{i.e.,} non-zero $g_{ij}$ in Eq.~\ref{eq:ising-ham}). 
% Atypically, the Ising models do not include any penalty terms, and each $v \in \mathcal{O}$ is mapped `one by one' into $\mathbf{s}\in \{-1,+1\}^{log_2{\mathcal{|O|}}}$, requiring $N_V=N_tlog_2{\mathcal{|O|}}$ spins to express $\mathbf{v}\in\mathcal{O}^{N_t}$ without any ancillary variables. 
% With `one by one' mapping between $\mathbf{s}\in \{-1,+1\}^{log_2{\mathcal{|O|}}}$ and $\mathbf{v}\in\mathcal{O}^{N_t}$, 
The objective function of ML (\emph{i.e.,} $\text{Obj}_{\text{ML}}(\mathbf{v})$ on p.~\pageref{f:mimo_model}) is expressed as an Ising energy function (Eq.~\ref{eq:ising-ham}) by one-by-one mapping, 
% This implies  can be translated into equivalent Ising energy landscape
where $E(\mathbf{s})$ corresponds to $\text{Obj}_{\text{ML}}(\mathbf{v})$ with $\mathbf{s} \leftrightarrow \mathbf{v}$ (Figure~\ref{f:MLD}).
% , and the ML solution corresponds to the ground state of the problem $E$. 
Based on generalized forms, all $f$ and $g$ values can be obtained independently in parallel for any MIMO sizes and modulations. 
% Linear coefficient $f$ values are decided by $\mathbf{H}$ and $\mathbf{y}$, while quadratic $g$ values depend only on $\mathbf{H}$. For a frame (\emph{i.e.,} channel coherence time), $\mathbf{H}$ remains unchanged, so only $f$ is updated per channel use.
  
% , the required resources for the formulation are insignificant.

% The generalized forms  are introduced. 
% In $E$, linear coefficient $f$  values are decided by $\mathbf{H}$ and $\mathbf{y}$, while quadratic $g$ values depend only on $\mathbf{H}$. For a frame (\emph{i.e.,} channel coherence time), $\mathbf{H}$ remains unchanged, so only $f$ coefficients are updated per channel use.  

\parahead{Combined Model for Parallel QA.} In QA, only a single Hamiltonian $\mathcal{H}$ (Eq.~\ref{eq:hamiltonian}) can be run (with a single Quantum Machine Instruction or QMI)~\cite{DwaveQPU}. Thus, for parallel QA, we need to prepare $\mathcal{H}$ with a single $\mathcal{H}_{\text{problem}\,E}$ that contains all parallelization information. In \systemname{}, the same Ising model $E(\bf{s})$ is used for parallel QA. Simply adding the same one to the end of one another leads to a single larger Ising model.
% that combines the multiple numbers of the same Ising problem. 
As an example, we consider three levels of parallelism for a $N_V$-variable Ising model: $\text{\emph{combined }} E = E(\{s_1,\cdots,s_{N_V}\}) + E(\{s_{N_V+1},\cdots,s_{N_V+N_V}\}) + E(\{s_{2N_V+1},\cdots,s_{2N_V+N_V}\})$, requiring $3N_V$ variables, 
% to express it.
% In the combined model, there are no connections among them and 
whose ground state is the same as the three-time repeated ground state of the original $E$. Despite the combined Ising spin model, different qubit clusters (near or far) on QA hardware can be assigned to each of them.

 % For parallel QA processing, different qubit clusters can be assigned to different jobs as resource allocation.
% (Figure~\ref{f:quantum_parallel}).

\parahead{Bit-Flip Multi-State Generator
(BMG).} 
% This is a simple multi-state generator based on the MMSE solution. 
Based on the MMSE solution, BMG generates multiple states as seeds for RA initial states by flipping one bit (in default). 
% When the required seed counts are over $N_V$ (large-scale parallelism), BMG flips two (or more) bits. 
While there are many possible variations in selecting which bit to flip, we use a random (but non-overlapped) index for the minimum overhead. Note that this randomness is a reasonable choice according to our design principle ($\S$\ref{s:desing_principle}). The generated multiple states are prepared as a single appended array, since a single initialization state can be run for RA, like the Hamiltonian.
    
% \vspace{-0.5cm}
\vspace{-0.1cm}
\subsection{Split-Detection Method}
% \subsection{Improving performance at High SNRs: Split-Detection Approach}
\label{s:mmse_quadrant} 

% Recall that a common observation in QA MIMO detectors is notable performance degradation with high-order modulations such as 16/64-QAM and error floor at high SNRs ($\S$\ref{s:common_challenges}).
\systemnames{} split detection scheme is designed to mitigate the error floor phenomenon with high-order modulations at high SNRs by exploiting the exceptional performance of QA MIMO detectors with low-order modulations such as BPSK and QPSK ($\S$\ref{s:common_challenges}). The method transforms the original ML Ising problem with high-order modulations into multiple simpler B/QPSK problems based on the initial MMSE solution.
% , while still supporting the one-shot parallel QA process. 

% Let us try to visualize split detection for 16-QAM modulation. 
For 16-QAM ML Ising models, 
% the optimization variable ($\S$\ref{s:primer_mimo}) can expressed as $\mathbf{v} = 2\mathbf{{q}}_2 + \mathbf{{q}}_1$, where $\mathbf{{q}}_i \in \{-1-j, -1+j, 1-j, 1+j\}^{N_t}$ consists of QPSK symbols. Note that $\mathbf{{q}}_2$ represents the \emph{quadrant}, while $\mathbf{{q}}_1$ represents the \emph{position of the symbol} within the quadrant, and that $\mathbf{q}$ can be expressed with two spin variables ($\mathbf{q_i}=s_{2i-1}+js_{2i}$).
four spin variables are required to represent an ML variable: two spins are related to the quadrant decision, while the other two spins are related to the position decision. 
% For $n$-th user symbol, odd-numbered ($4n-3$-,$4n-1$-th) spins are related to the quadrant decision in the Ising representation, while even-numbered ($4n-2$-,$4n$-th) spins are related to the position decision. 
% (as illustrated in Figure~\ref{f:sym_quad}). 
Using these separate detection roles of the 
% odd/even-numbered 
spin variables along with the MMSE solution, in the split-detection we generate an Ising form of $N_V$ spins (\emph{same size} as the original form) that contains two independent QPSK Ising ML problems. In one problem, we decide the quadrant (based on the position of the MMSE solution), and in the other, we decide the position (based on the quadrant of the MMSE solution). In the original Ising form, these two are \emph{jointly} considered, but in the split detection method, they are considered \emph{separately} as two independent \emph{simpler} problems.
% (one for quadrant search and the other for position search).  
% To precisely explain the principles of the approach in detail,   
% perform an analytical estimation of the effective noise of the split-detection method (using multiple applications Cauchy-Schwartz inequality to find an upper bound on the effective noise) as theoretical analysis. 
% in our Appendix~\ref{s:noise_analysis}.
The scheme can be easily generalized for any modulation size, requiring the bare minimum additional forms and overheads (\emph{e.g.,} only two for 16-QAM, three for 64-QAM, and so on). Further, multi-seed parallel ensemble RA can be applied to them together with the baseline \systemnames{} Ising representation ($\S$\ref{s:parallel_QA}) without requiring any iterative QPU runs. 

Through multiple applications of Cauchy-Schwartz inequality to find an upper bound on the effective noise, we also theoretically prove that the method can effectively improve the performance at high SNRs. 
% where the error floor occurs with QA MIMO detectors.
% \footnote{The analytical proof is available in our Appendix~\ref{s:noise_analysis}, where we also compare the noise against the one in a na\"ive alternative split scheme.} 
However, note that the scheme does not work well at low SNRs where both MMSE's quadrant and position are often wrong. The analytical proof of the method with MMSE analysis is available in Appendix~\ref{s:noise_analysis}.

\vspace{-0.12cm}
\section{Implementation}
\label{s:implementation}
\vspace{-0.1cm}
\parahead{Quantum Annealers.} We implement \systemname{} on the state-of-the-art D-Wave Advantage Annealers. 
% Both Advantage\_6.1 (pegasus-topology) and
The \emph{prototype} of Advantage2 System with Zephyr topology with 20 couplers per qubit~\cite{DwaveZephyr} (1.1ver. released in 2022) is used because of the observed QA improvement for 16-QAM performance over the previous Pegasus-topology systems~\cite{kim2022warm}.
% with increased connectivity~\cite{kim2022warm}. 
% Advantage\_6.1 has 5,760 qubits with 15 couplers per qubit (\emph{i.e.,} connectivity).  
Despite the most advanced topology, the prototype machine has only 576 qubits (cf. previous Pegasus Advantage\_6.1 has 5,760 qubits with 15 couplers), and has been very recently upgraded to 1,200 qubits (2.2ver. released in 2024). A 7,000 qubits generation is planned to be released by the end of 2025. 
% Some QPSK experiments that require more qubits than 576 even for a non-parallel run are conducted on the Advantage\_6.1. 
All the QA experiments for comparisons among different QA detectors are conducted on the exactly same machines and Ising forms; We also implement fully parallel QuAMax using the same physical qubits for the same $L_P$ as \systemname{}. More details on QA hardware programming and implementations including embedding benchmarks are in Appendix~\ref{s:cumul_QA}.
Note that \systemname{} is one of the first reports of fully parallel RA.

\parahead{Classical Implementation with Parallel Tempering.} To test large-scale parallelism and more comprehensive MIMO scenarios beyond QA hardware, we also implement classical variants of \systemname{} and IoT-ResQ using \emph{Parallel Tempering} (PT)~\cite{swendsen1986replica}, a generic PIC algorithm that the state-of-the-art classical PIC MIMO detector, ParaMax~\cite{kim2021physics}, utilizes. Inspired by quantum RA ($\S\ref{s:primer_ra}$), we initiate PT from a given state (instead of a random state) for classical IoT-ResQ and X-ResQ, and apply different parallelization and initialization strategies corresponding to ParaMax, IoT-ResQ, and X-ResQ. 
They feature the same compute time for their PIC optimization processing due to the same PT engine, except for their preprocessing including their initial detectors (\emph{i.e.,} FSD in IoT-ResQ and MMSE in X-ResQ).
Unlike the QA implementations, each parallel task collects only one sample for the minimum latency here, so $L_P$ is equal to collected sample counts.
% We use Numba-based PySA [XXX] for the implementation (cf. C++ in ~\cite{kim2021physics}). 
% For parallelism benchmarks, we use cumulative sequential runs to estimate the performance, since unlike QA experiments, classical computing platforms do not suffer from critical precision or cross-talk issues and thus the same detection BER performance as fully parallel runs can be safely assumed.
% for benchmark purposes. 
Classical experiments are executed
on an Intel i9-9820X.
% with 20 cores.

% Discussion on mimicking a parallel QA run via cumulative sequential QA runs is available in Appendix~\ref{s:cumul_QA}.

% Due to the limited qubit counts on the prototype hardware, 
 
\vspace{-0.1cm}
% \vspace{-0.15cm}
\section{Evaluation}
\label{s:evaluation}
% \vspace{-0.05cm}
\vspace{-0.1cm}
In this section, we evaluate \systemnames{} MIMO detection performance. 
In Section~\ref{s:ber_eval}, the baseline \systemname{} is tested and compared against QuAMax (FA), MMSE (linear), and ML detector (optimal). 
% Then, \systemname{} with the split detection method at high SNRs is evaluated.
Pure QA compute times for detection are calculated as anneal counts multiplied by each anneal duration (QuAMax: $N_a\times 2~\mu$s, X-ResQ: $N_a\times 2.2~\mu$s). In  Section~\ref{s:classical_eval}, we conduct more comprehensive evaluations with various classical and PIC-based detectors. X-ResQ and IoT-ResQ are implemented classically using a PT technique ($\S$\ref{s:implementation}), allowing for experiments with large-scale parallelism and various MIMO scenarios that current QA hardware cannot support. 
% By default, 16-QAM modulation is used, unless otherwise specified. 

\parahead{Wireless Channel and Ising spin models.} Our default channel setting is (i.i.d) Gaussian channels. We also use 
% Argos's real-trace channels and measured noises~\cite{Argos} and 
NVIDIA's open-source library Sionna~\cite{sionna}. Sionna generates OFDMA channel traces (Ray tracing channel) corresponding to 3GPP Urban Macrocell (UMa) channel model 
% for $N_t \times N_r$ MIMO 
(link-level simulations instead of a
stochastic model). We use a single-sector topology with 
users moving around at 3 m/s and 
a base station operating at 2.4 GHz
% , with 128 subcarriers and 15 KHz subcarrier spacing. 
with horizontally polarized antennas arranged in a linear array. 
% Each user is equipped with one horizontally polarized antenna. 
% All antennas have omnidirectional antenna patterns. 
Using the channels with random data and noises, we formulate up to 100,000 ML Ising models per scenario ($N_t$, $N_r$, SNRs, and modulations); relatively fewer (thousands of) instances are tested for QA experiments.

% \parahead{Set-up.} 
% For heuristic optimization evaluations, we run a few tens of instances but with large anneal iteration counts ($N_a=5,000$) for statistical significance ($\S$\ref{s:tts_eval}). 

% Only 16-QAM results are reported since the prototype hardware cannot allow large-size MIMO (\emph{e.g.,} 48-users scenarios) which is required for low-order modulation QA-MIMO evaluations.
% In our BER evalutions, 
% \vspace{-0.3cm}

% \subsection{QA Optimization Benchmark}
% \label{s:tts_eval}

% Recall that the tested MIMO sizes are relatively small due to the qubit counts on the prototype, which could be enabled by classical non-linear detectors. However, enabling even the small MIMO sizes with high-order modulations through QA (or PIC approaches in general) is one of the most critical challenges left in the line of research~\cite{kim2019leveraging,kim2022warm,ducoing2022quantum,tabi2021evaluation,rimimo,dimimo,singh2023uplink,cui2022general,kim2021physics,norimoto2023quantum}. 

% \vspace{-0.33cm}
\vspace{-0.2cm}
\subsection{Detection Performance of \systemname{}}
% \subsection{Baseline \systemname{}}
\label{s:ber_eval}
\vspace{-0.1cm}
 
% For performance comparisons, we test multiple detectors for various detection scenarios. 
% Here, we do not report IoT-ResQ performance for the following reasons: (1) Even for the minimum parallelism, 16 levels of parallelism are required with 16-QAM, where even $4\times4$ MIMO problems cannot be programmed on the prototype machine. This indeed shows the limitations of decomposition-based IoT-ResQ caused by inflexible parallelism. (2) IoT-ResQ's performance is near-optimal in nearly all the tested scenarios in the subsection, not because of the QA part, but because of the initial classical non-linear FSD. Instead, we compare the parallelization strategies of \systemname{} and IoT-ResQ in Section~\ref{s:classical_eval}.

% \begin{figure}
% \centering
%     \includegraphics[width=0.88\linewidth]{example-image-a}
%     % \vspace{-0.1cm}
%     \caption{\small
%     \textbf{BER performance of fully-parallel \systemname{} as a function of QA compute time at SNR 16 dB for 4-user MIMO varying BS antenna counts $N_r$. 
%     % \vspace{-0.2cm}
%     }} 
% \label{f:BER_parallelism}
% \end{figure}

\begin{figure}
\centering
  
    % \vspace{-0.1cm}

      \begin{subfigure}[b]{\linewidth}
    \centering  \includegraphics[width=0.76\linewidth]{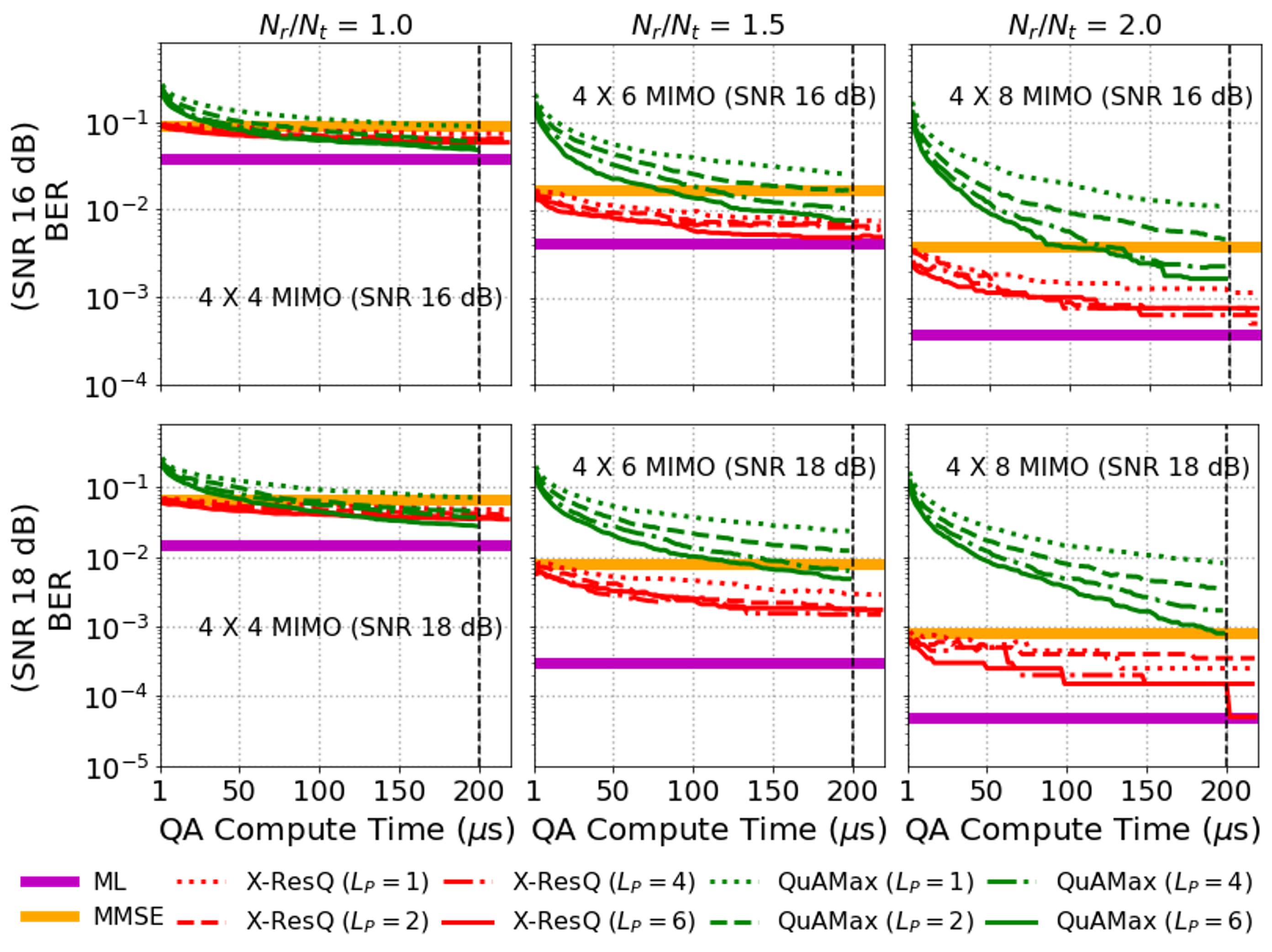}
    \caption{\small BER across QA time at SNR 16 dB (\emph{upper}) and 18 dB (\emph{lower}).
    % \caption{\small BER across SNRs at QA time 110~$\mu$s (\emph{upper}) \& 220~$\mu$s (\emph{lower}).
    % Chance of leveraging the trade-off between exploitation and exploration in optimization existing in RA.
    }
    \label{f:BER_time}
    \end{subfigure}

  \begin{subfigure}[b]{\linewidth}
   \centering\includegraphics[width=0.76\linewidth]{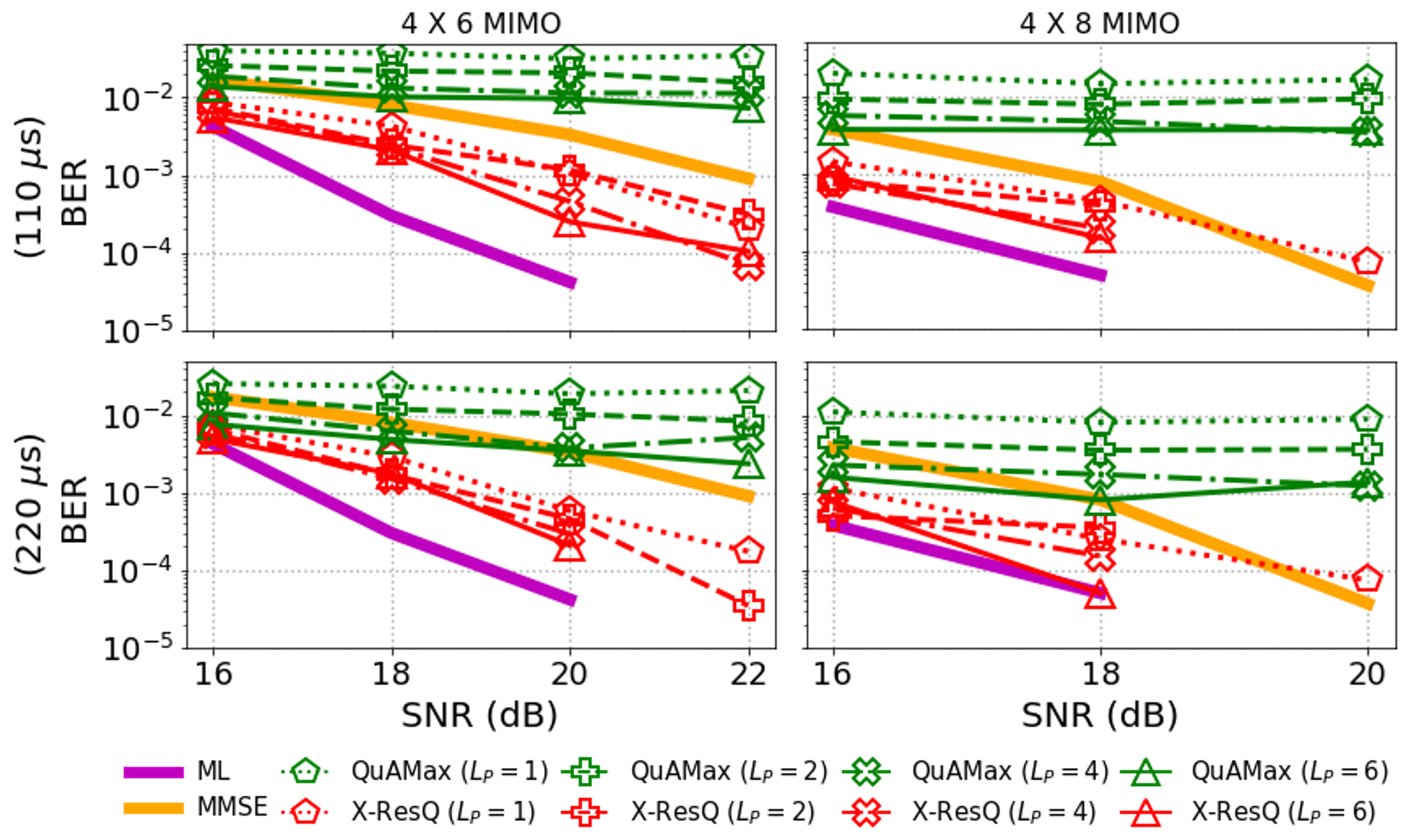}
    \caption{\small BER across SNRs at QA time 110~$\mu$s (\emph{upper}) \& 220~$\mu$s (\emph{lower}).
    % Chance of leveraging the trade-off between exploitation and exploration in optimization existing in RA.
    }
    \label{f:ber_vs_snr}
    \end{subfigure}

    \vspace{-0.1cm}
    \caption{\small
    \textbf{BER performance of \emph{fully parallel} \systemname{} (requiring approx. $40\cdot L_P$ qubits) for 4-user MIMO varying $L_P$, BS antenna counts $N_r$, and SNRs. IoT-ResQ that requires $768$ qubits even for its minimum parallelism ($L_P=16$) cannot be programmed on the state-of-the-art Zephyr-topology QA machine. \vspace{-0.15cm}
    }} 
\label{f:fully-XresQ}
\end{figure}

% Perhaps, the larger $N$ (like 20) might be able to enable 8-user MIMO, we leave the experiments as our future work (with the coming full-sized Advantage2 machine). 

For QA MIMO detection evaluations, we test Ising ML detection instances with 100 anneals to make compute time at most few hundred microseconds (up to 220~$\mu$s). Due to the limited QA hardware, we consider a 4-user MIMO \systemname{} for WLAN scenarios ($N_r$ = $4,6,8$) with convolutional FEC coding, while for cellular network scenarios ($N_r=16,64$), we use cumulative sequential QA to estimate fully parallel performance.\footnote{In Appendix~\ref{s:cumul_QA}, we empirically verify that cumulative QA runs can estimate the performance of a fully parallel QA run (as an upper bound performance) without significant performance gaps for these levels of parallelism.} We report BER of MIMO detection (fundamental metric) and \emph{system throughput} in bits/s/Hz (or empirically obtained spectral efficiency) using an adaptive coding scheme that selects the best code rate among $\frac{1}{3}, \frac{1}{2}$ and $\frac{2}{3}$, given SNRs with the convolutional code for 1500-byte packets. In this subsection, IoT-ResQ is not reported, since IoT-ResQ even with its minimum $L_P$ cannot be programmed on the hardware. 
% For the tested 4-user instances, IoT-ResQ achieves the optimal performance, not because of the QA part, but because of its initial classical non-linear FSD.
More comprehensive evaluations (including direct comparisons against IoT-ResQ) are in the next subsection.
% available in the next subsection. 

\begin{figure}
%\vspace*{4ex}
    \centering
\includegraphics[width=0.73\linewidth]{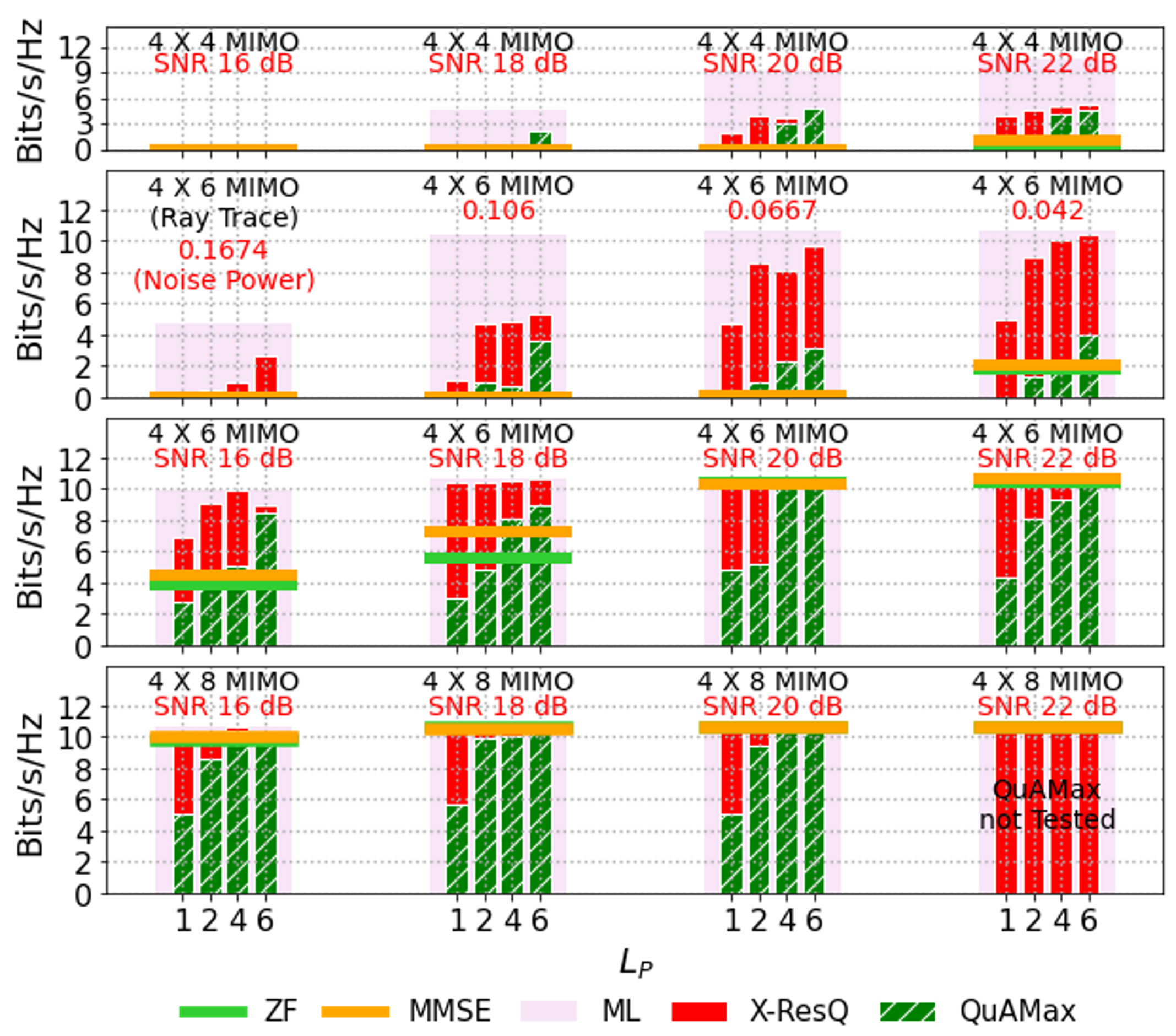}\vspace{-0.1cm}
\caption{\small\textbf{Throughput performance (bits/s/Hz) of \emph{fully parallel} \systemname{} (220~$\mu$s) with adaptive convolutional FEC (w/ 1/3, 1/2, 2/3 rate) for 1500-byte packets. QA detectors are with $N_a=100$.
% (QuAMax: 200~$\mu$s, X-ResQ: 220~$\mu$s).
}}
    % Pink dotted lines report the best cost/rank.
\label{f:throughput}
\end{figure}

First, we compare \emph{fully parallel} \systemname{} (without the split detection method) against QuAMax. For fair comparisons, we reimplement QuAMax on the same Zephyr-topology prototype machine used for \systemname{}. Figure~\ref{f:fully-XresQ} shows BER for 4-user MIMO with different receiver antenna counts ($N_r$) and varying applied levels of parallelism ($L_P: 1, 2, 4, 6$).\footnote{Rarely, BER becomes higher in \emph{better} scenarios (\emph{e.g.,} higher SNRs, $L_P$, and/or compute times). This is because QA is a probabilistic heuristic solver and lower-energy states do not always correspond to lower-bit-error states, particularly when the energies of the best solutions are high (\emph{i.e.,} bad quality).} Figure~\ref{f:BER_time} plots BER as a function of compute time (by translating $N_a$ into time using each anneal duration) at SNR 16 dB (\emph{upper}) and 18 dB (\emph{lower}). Both QuAMax and \systemname{} have generally better BER as higher $L_P$ is applied. As $N_r$ increases, it is clearly observed that \systemname{} can outperform QuAMax and MMSE, reporting over an order of magnitude better BER for the same $L_P$. For example, for $4\times8$ MIMO at SNR 18 dB, while \systemname{} reports below $10^{-4}$ (optimal) BER with $L_P=6$ around 200~$\mu$s QA compute time, while QuAMax and MMSE obtain about $10^{-3}$ BER. Note that \systemname{} with $N_a=50$ (110~$\mu$s) works better than QuAMax with $N_a=100$ (200~$\mu$s) in most scenarios.
% In Figure~\ref{f:ber_vs_snr}, we show more scenarios varying SNRs. 
Figure~\ref{f:ber_vs_snr} plots BER across SNRs at fixed computing time with $N_a=50$ and $100$, \emph{i.e.,} 110 and 220 $\mu$s for \systemname{}, while 100 and 200 $\mu$s for QuAMax. We also observe that \systemname{} outperforms QuAMax (and MMSE) and tends to achieve lower BER as more qubits are assigned; \systemname{} with $L_P \geq 4$ obtain \emph{no bit errors} for $4\times6$ MIMO at SNR 22 dB (out of 28,800 tested bits) and $4\times8$ MIMO at SNR 20 dB (out of 27,040 tested bits).

\begin{figure}
\centering
\includegraphics[width=\linewidth]{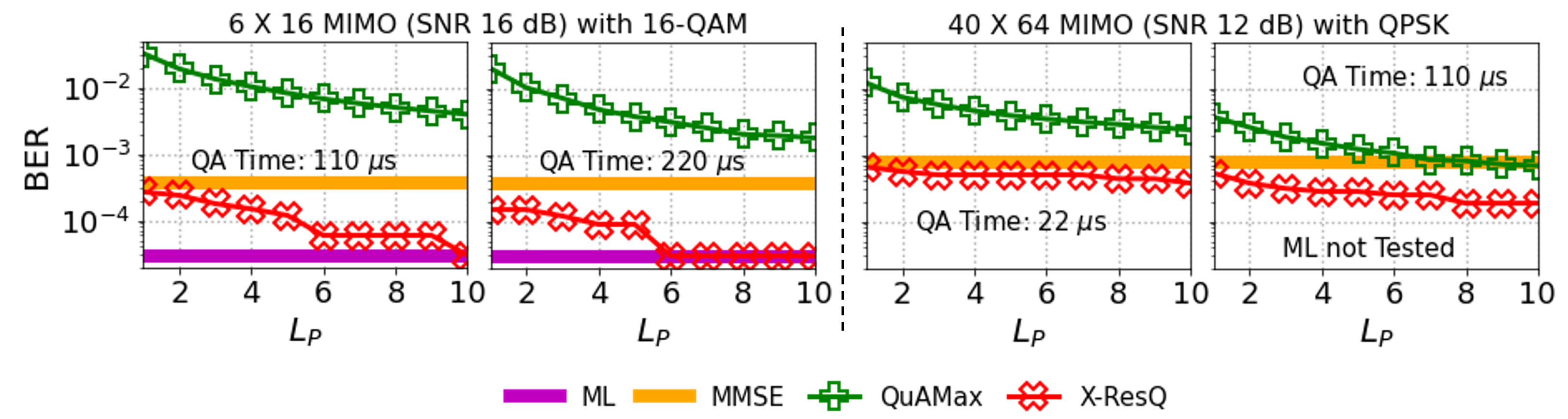}\vspace{-0.1cm}
    \caption{\small \textbf{BER of cumulative sequential \systemname{} (cf. fully parallel \systemname{}) across $L_P$ for $6\times 16$ MIMO w/ 16-QAM and $40\times 64$ MIMO w/ QPSK, whose required qubit counts for fully parallel QA would be $96\cdot L_P$ (Zephyr) and $564\cdot L_P$ (Pegasus), respectively.}}\vspace{-0.2cm} 
\label{f:cumul_seq_qpsk}
\end{figure}

In Figure~\ref{f:throughput}, we show system throughput performance with 16-QAM, using 100 random packets varying $L_P$, SNRs, and $N_r$. Note that linear detectors (ZF, MMSE) are not designed for parallelism, so their detection performance is the same, regardless of $L_P$. As expected, they rely heavily on both $N_r/N_t$ and SNRs. As $N_r/N_t$ or SNR decreases (from \emph{bottom} to \emph{top}, from \emph{right} to \emph{left}, respectively), their performance rapidly degrades; for $4\times 4$ MIMO ($N_r/N_t=1$), their throughput becomes almost zero.
% \systemname{} obtains higher throughput as it increases $L_P$.
In the case of the QA MIMO detectors (\systemname{}, QuAMax), while they both generally achieve higher throughput with higher $L_P$ (\emph{i.e.,} more qubits) for the tested scenarios, \systemname{} shows more efficient parallelism, requiring less $L_P$ to reach the (near-)optimal throughput. Interestingly, \systemname{} greatly outperforms the others, particularly with the Ray Trace channel model. For example, with noise power of $0.042$ (far right on the second from top), \systemname{} with $L_P=6$ reaches near-optimal performance (over 10 bits/s/Hz), achieving 2.5--5x throughput compared to the other comparison schemes including QuAMax with the same $L_P$.

\begin{figure}
\centering
    \includegraphics[width=0.7\linewidth]{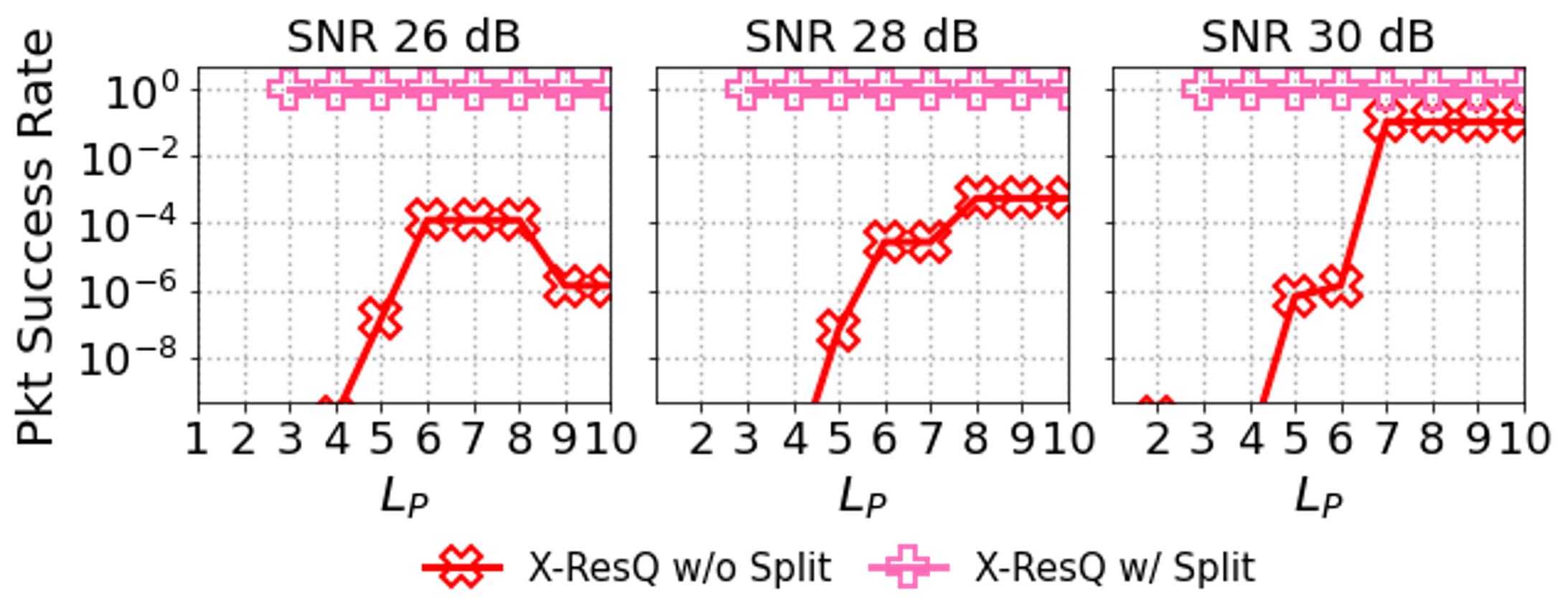}\vspace{-0.1cm}
    \caption{\small
    \textbf{Impact of the split-detection method on the packet success rate of \systemname{} (220~$\mu$s) at high SNRs for $4\times4$ MIMO with 16-QAM, where $L_P=2$ is assigned to the split scheme.
    % $K$ is the levels of parallelism assigned to the method in addition to $L_P$.
    }}
%     When $N+K < \text{input } K$, all active qubits are assigned to the method.}} 
\label{f:split_detect}
\end{figure}

Next, we test the detectors for scenarios with relatively larger MIMO sizes. For this, we estimate parallel \systemname{} performance through cumulative sequential results due to the problem sizes. 
% for 40-user MIMO scenarios, where 
Figure~\ref{f:cumul_seq_qpsk} shows the BER performance as a function of $L_P$, where \systemname{} keeps outperforming QuAMax for the same $L_P$ for both 16-QAM (\emph{two on the left}) and QPSK (\emph{two on the right}) experiments. For 40-user QPSK MIMO, we implement \systemname{} and QuAMax on the Pegasus-based Advantage system (cf. Zephyr-based Advantage2). For $6\times 16$ MIMO with 16-QAM, \systemname{} requires $L_P=10$ (960 qubits) to achieve optimal BER with 110~$\mu$s QA time ($N_a=50$), while with 220~$\mu$s ($N_a=100$) it requires $L_P=6$ (574 qubits), showing trade-off between $L_P$ (or qubit usage) and QA compute times. Similar patterns are observed for QPSK, though ML detection is not reported due to its large $N_t$.

\begin{figure}
%\vspace*{4ex}
    \centering
    % \begin{subfigure}[b]{\linewidth}
    % \centering
    % %\raisebox{1ex}{\mbox{\includegraphics[width=\linewidth]{figures/DWH2.png}}}
    % \includegraphics[width=\linewidth]{Figures/anneal_schedule.png}
    % \caption{Annealing signals across $\tau$.}
    % \label{f:anneal_scheudle}
    % \end{subfigure}
    % \qquad
    
    % \hfill
    % \begin{subfigure}[b]{\linewidth}
    % \includegraphics[width=0.9\linewidth]{Figures/BER_vs_time.png}
    % \caption{BER as a function of compute time (16-QAM at SNR 16 dB).}
    % \label{f:ber_vs_time}
    % \end{subfigure}

    % \begin{subfigure}[b]{\linewidth}\centering
    % \includegraphics[width=0.85\linewidth]{Figures/FA_RA.png}
    % \caption{Cost Landscape View.}
    % \label{f:rank_hdist}
    % \end{subfigure}
    
    \begin{subfigure}[b]{\linewidth}    \centering\includegraphics[width=0.75\linewidth]{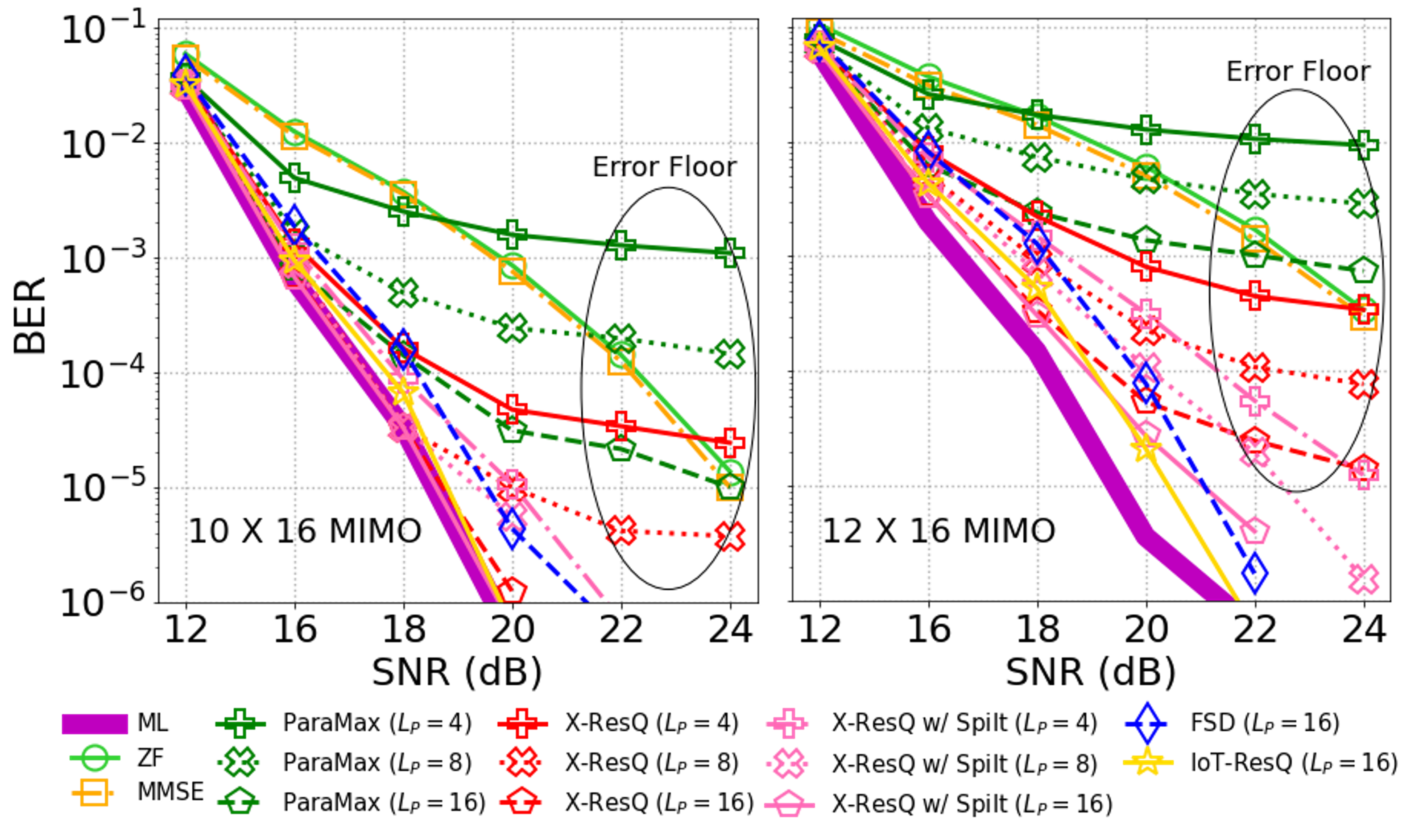}
    \caption{\small Across SNRs for 16-antenna BS MIMO ($N_r=16$).
    % Chance of leveraging the trade-off between exploitation and exploration in optimization existing in RA.
    }
    \label{f:classical_ber_vs_snr}
    \end{subfigure}

            \begin{subfigure}[b]{\linewidth}
    \centering\includegraphics[width=0.78\linewidth]{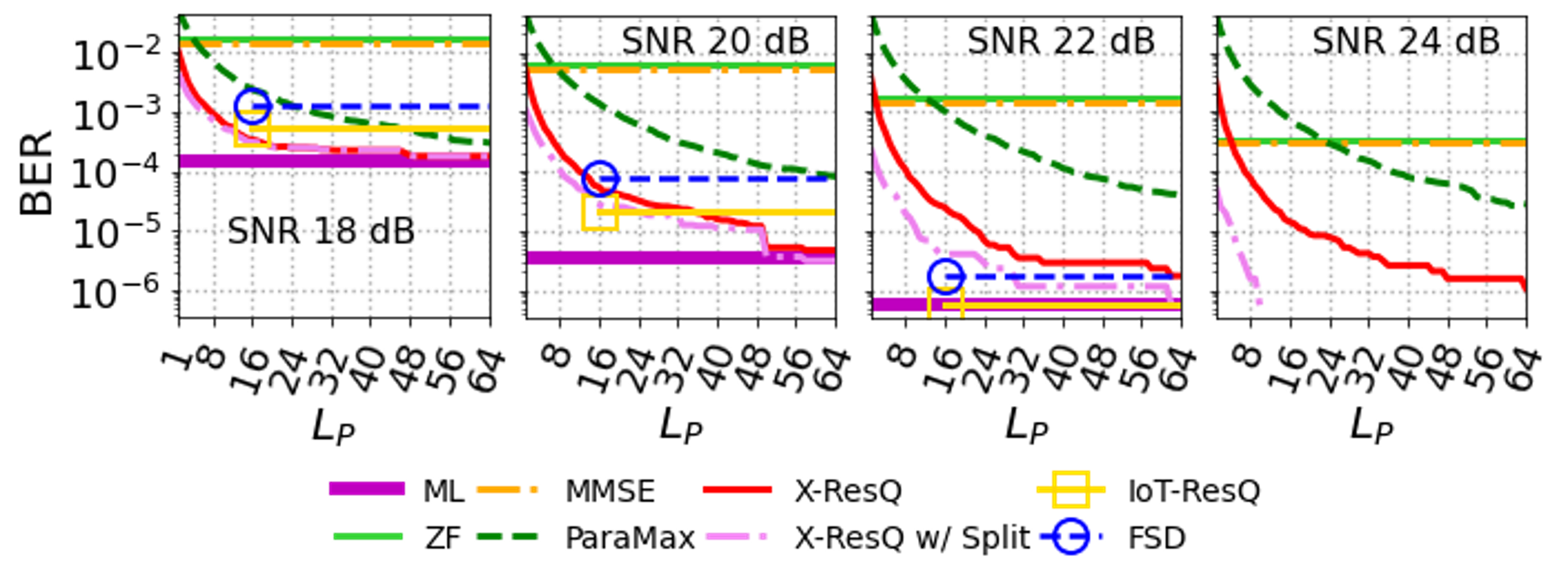}
    \caption{\small As a function of $L_P$ varying SNRs ($12\times16$ MIMO).
    % Chance of leveraging the trade-off between exploitation and exploration in optimization existing in RA.
    }
    \label{f:classical_ber_vs_n}
    \end{subfigure}

    \begin{subfigure}[b]{\linewidth}
    \centering\includegraphics[width=0.70\linewidth]{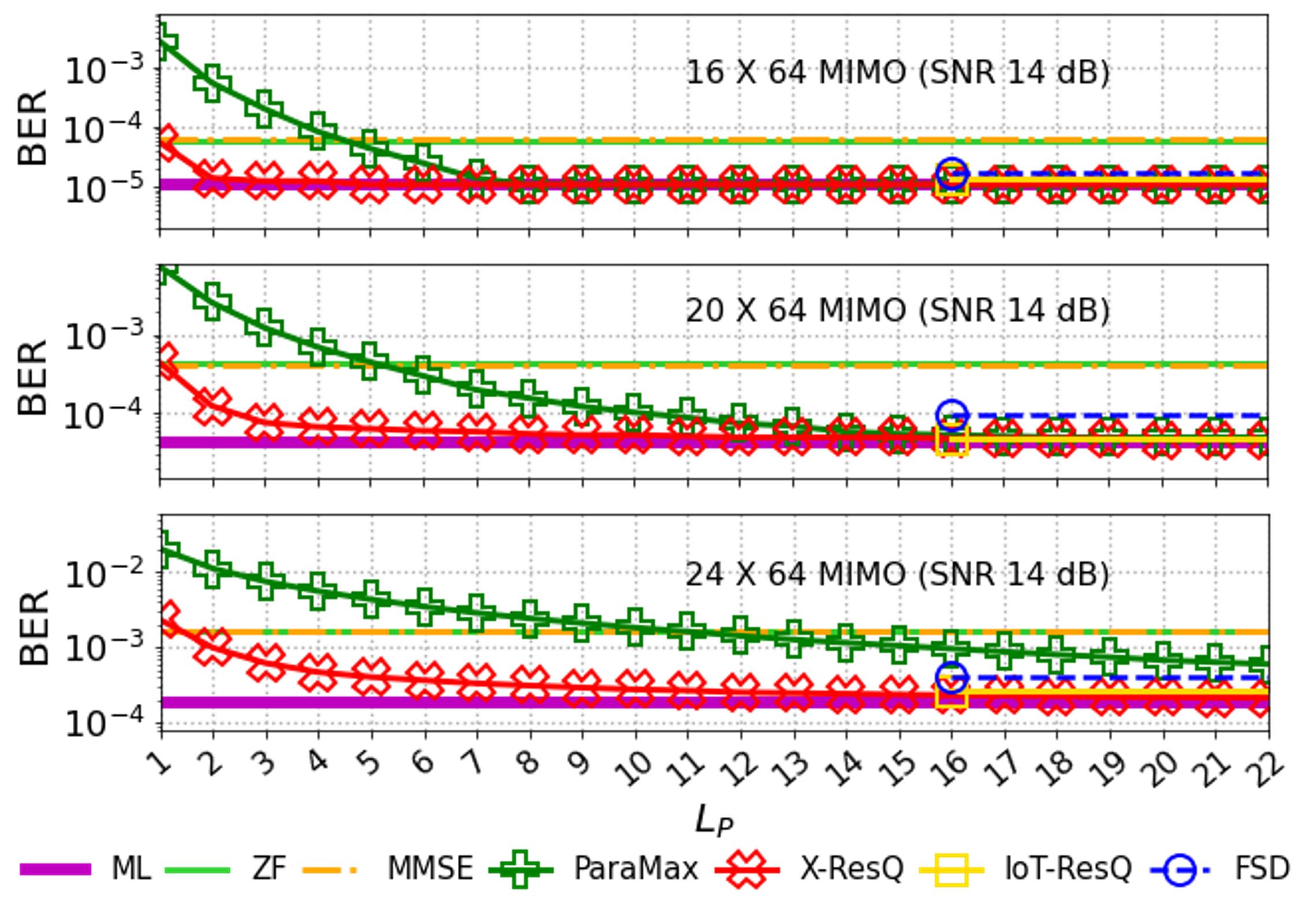}
% {Figures/20X64_mimo.png}
    \caption{Varying users ($N_t$) for MIMO w/ $N_r=64$ at SNR 14 dB.
    % Chance of leveraging the trade-off between exploitation and exploration in optimization existing in RA.
    }
    \label{f:classical_ber_vs_n_64nr}
    \end{subfigure}

        % \vspace{-0.1cm}
    \caption{\small\textbf{BER performance of various MIMO detectors including classically-implemented \systemname{} and IoT-ResQ for Massive MIMO with 16-QAM. Recall that FSD and IoT-ResQ require the minimum $L_P=16$ for fully parallel processing, and current MIMO systems use linear detectors (ZF, MMSE) (\S\ref{s:massive_mimo}). 
    % (and IoT-ResQ) compared to other detectors including ParaMax. Massive parallelism that cannot be tested on the current QA hardware is considered for task parallelism benchmark.
    }}
    % Pink dotted lines report the best cost/rank.
\label{f:classical_imple}
\end{figure}

% \vspace{-0.1cm}
% \subsection{\systemname{} with Split-Detection Method}
% \label{s:split_eval}

% \parahead{Impact of the Split-Detection Method.}
% The split-detection method ($\S$\ref{s:mmse_quadrant}) is designed to resolve this error-floor issue.

To experimentally evaluate the split-detection scheme (\S\ref{s:mmse_quadrant}) on QA hardware, we show the packet success rate (based on the uncoded BER for 1500-byte packets) of sequential cumulative \systemname{} \emph{with} and \emph{without} the method at arbitrarily high SNRs in Figure~\ref{f:split_detect}
(note that \emph{high} SNRs where the error floor occurs are dependent on $L_P$, $N_t$, and $N_r/N_t$). At the tested SNRs, method-assisted \systemname{} effectively improves the performance, resulting in 1.0 packet success rates without errors; \systemname{} without the scheme has several orders of magnitude lower rates, and the MMSE detector achieves $\approx$ 0.0 success rates. However, recall that QA experiments in this work are tested with a limited number of instances, and every optimal solution happens to obtain no errors for this experiment. In general, high-SNR detection experiments need testing more instances to capture the detectors' performance precisely, but conducting QA experiments with a great number of instances is costly today. For this reason, the evaluation is somewhat limited.
, and thus the method will be further discussed in the next subsection, relying on classical experiments.

% We control the levels of parallelism for the split method ($K$) along with the original $N$. Here, $N+K$ becomes the total levels of parallelism. When $N+K$ is less than input $K$, we just apply the split-detection. In Figure~\ref{f:split1}, the error floor issues occur for $K=0,1$, and interestingly at $K=2$ no bit errors are detected for most of the cases. To understand this situation better, we plot BER as a function of $N+K$ for different $K$ at SNR 26 dB in Figure~\ref{f:split2}. When with $K=2$, even 3 levels of parallelism make \systemnames{} BER performance error-free, while without the split-detection \systemname{} suffers from high BER despite increasing parallelism. 
% We also test the method in the next subsection.  

% \begin{figure}
% \centering
%     \includegraphics[width=0.88\linewidth]{example-image-a}
%     % \vspace{-0.1cm}
%     \caption{\small
%     \textbf{Extremely Large MIMO (512x512 1024x1024) with BPSK.
%     }} 
% \label{f:BER_parallelism}
% \end{figure}

% \begin{figure}
% \centering
%     \includegraphics[width=\linewidth]{Figures/QPSK_LARGE.png}
%     % {example-image-a}
%     % \vspace{-0.1cm}
%     \caption{\small
%     \textbf{128X128 AND 256X256 W/ QPSK
%     }} 
% \label{f:BER_parallelism}
% \end{figure}

% \begin{figure}
% \centering
%     \includegraphics[width=0.88\linewidth]{example-image-a}
%     % \vspace{-0.1cm}
%     \caption{\small
%     \textbf{128X128 AND 256X256 W/ QPSK
%     }} 
% \label{f:BER_parallelism}
% \end{figure}

\begin{figure}
%\vspace*{4ex}
    \centering
    % \begin{subfigure}[b]{\linewidth}
    % \centering
    % %\raisebox{1ex}{\mbox{\includegraphics[width=\linewidth]{figures/DWH2.png}}}
    % \includegraphics[width=\linewidth]{Figures/anneal_schedule.png}
    % \caption{Annealing signals across $\tau$.}
    % \label{f:anneal_scheudle}
    % \end{subfigure}
    % \qquad

    % \hfill
    \begin{subfigure}[b]{\linewidth}
    \centering \includegraphics[width=0.83\linewidth]{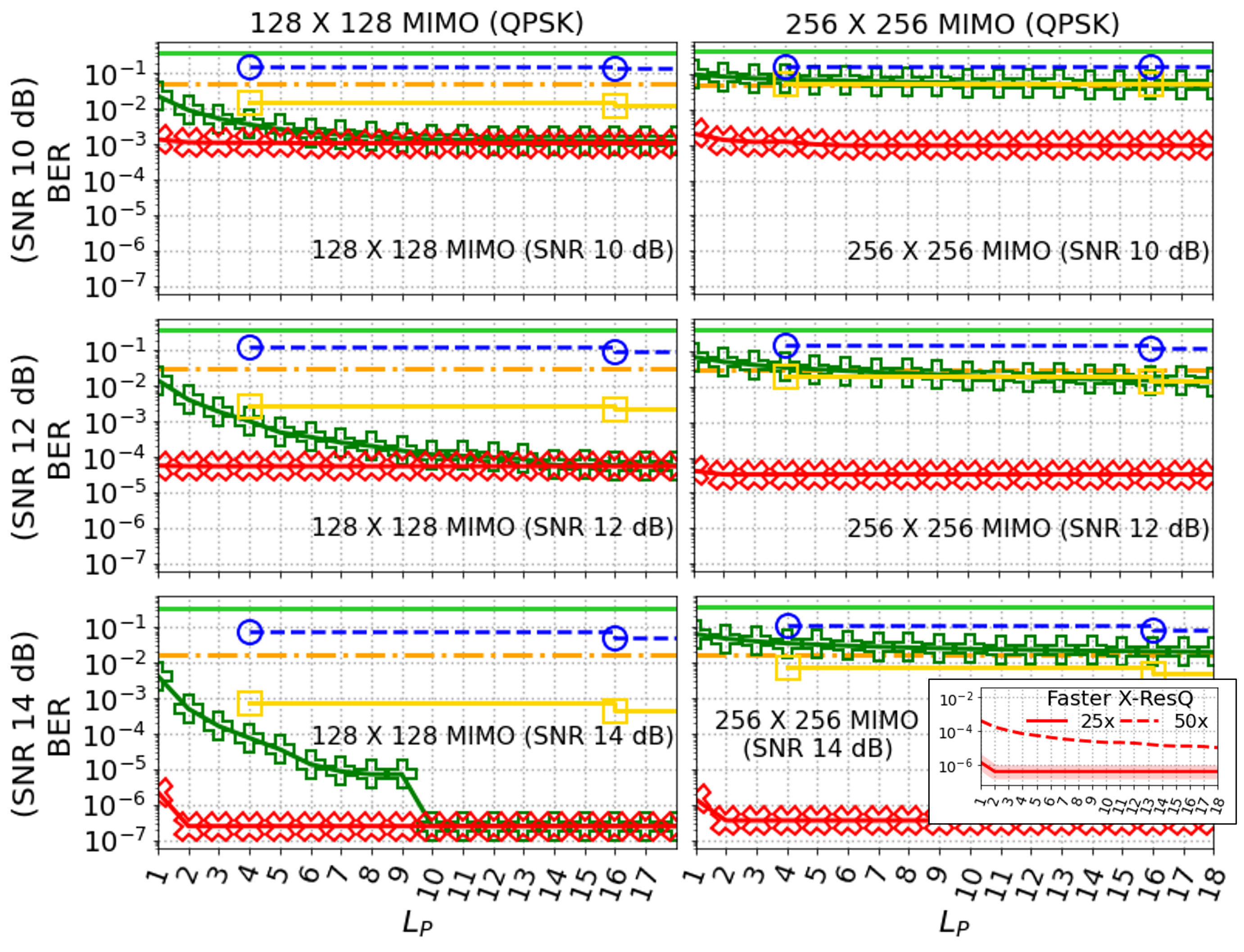}\vspace{-0.1cm}
    \caption{\small Large MIMO ($128\times 128$ and $256\times256$) with QPSK ($|\mathcal{O}|=4$).}
    \label{f:largemimoqpsk}
    \end{subfigure}

    % \begin{subfigure}[b]{\linewidth}\centering
    % \includegraphics[width=0.85\linewidth]{Figures/FA_RA.png}
    % \caption{Cost Landscape View.}
    % \label{f:rank_hdist}
    % \end{subfigure}
    
    \begin{subfigure}[b]{\linewidth}
    \centering\includegraphics[width=0.69\linewidth]{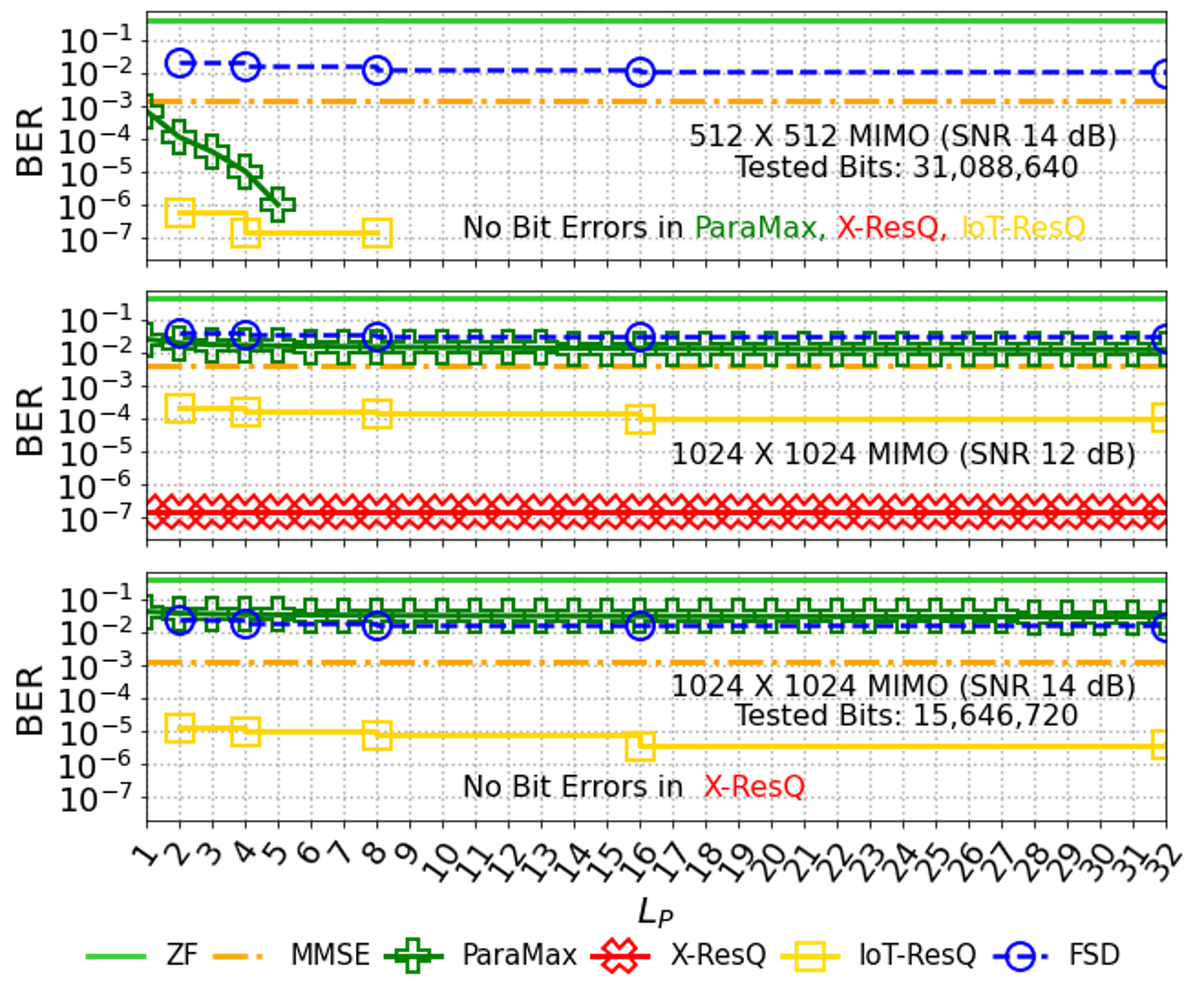}
    \caption{\small Large MIMO ($512\times 512$ and $1024\times1024$) with BPSK ($|\mathcal{O}|=2$).
    % Chance of leveragiang the trade-off between exploitation and exploration in optimization existing in RA.
    }
    \label{f:largemimobpsk}
    \end{subfigure}
        \begin{subfigure}[b]{\linewidth}
    \centering \includegraphics[width=0.78\linewidth]{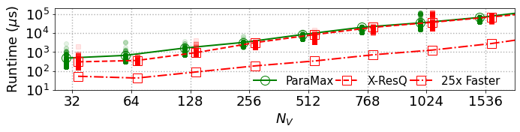}
    \caption{\small Compute latency of classical \systemname{} across spin counts ($N_V$).}
    \label{f:compute_time}
    \end{subfigure}
    \caption{\small\textbf{BER performance for \emph{Ultra-Large} MIMO ($N_r/N_t=1)$ with low-order modulations (BPSK and QPSK) and 99.5-th percentile compute latency of classical \systemname{} across $N_V$ for $N_V/log_2(|\mathcal{O}|)$-user MIMO (\emph{e.g.,} $N_V/2$-user MIMO with QPSK).}}
\label{f:ExtremeLargeMIMO}
\end{figure}

% \begin{figure}
% \centering\vspace{-0.2cm}
% \includegraphics[width=0.8\linewidth]{Figures/compute_time.png}\vspace{-0.2cm}
%     \caption{\small
%     \textbf{Compute latency of classical \systemname{} across $N_V$ for $N_V/log_2(|\mathcal{O}|)$-user MIMO (\emph{e.g.,} $N_V/2$-user MIMO with QPSK).}}
% %     When $N+K < \text{input } K$, all active qubits are assigned to the method.}} 
% \label{f:compute_time}
% \end{figure}

% \vspace{-0.3cm} 
\vspace{-0.1cm}
\subsection{More Comprehensive Evaluations with Classical Implementations}
\label{s:classical_eval}

With classical implementations of \systemname{} and IoT-ResQ ($\S$\ref{s:implementation}), we benchmark their performance with more various MIMO scenarios that current QA hardware cannot support.

Figure~\ref{f:classical_imple} compares the BER performance of various detectors for MIMO scenarios with 16-QAM varying SNRs, $N_t$, $N_r$, and $L_P$. 
Figure~\ref{f:classical_ber_vs_snr} shows 12-user (\emph{left}) and 16-user (\emph{right}) MIMO scenarios with $N_r=16$. As one can see, the performance of linear detectors (ZF, MMSE) is poor in general, showing relatively high BER compared to the other detectors. In the case of ParaMax and \systemname{}, they achieve better BER as $L_P$ increases,
but the error floor phenomenon is observed for them. So, we test \systemname{} with the split-detection method, where half of $L_P$ is used for the split-Ising form. We observe \systemname{} with the split-detection method mitigates the error floor, achieving nearly two orders of magnitude better BER with $L_P=8$ against the baseline one at SNR 24 dB. Interestingly, FSD and IoT-ResQ with their minimum $L_P=16$ outperform the others including \systemname{} with the same $L_P$.

To further analyze this, we plot BER as a function of $L_P$ in Figure~\ref{f:classical_ber_vs_n}. It is observed that even though IoT-ResQ works best at the point of $L_P=16$, \systemname{} can converge to ML performance with less $L_P$, since IoT-ResQ (and FSD) cannot achieve gains until $L_P=256$. Precisely, \systemname{} reaches the optimal BER ($\approx$ $5\cdot10^{-6}$) at SNR 20 dB with around $L_P=50$, flexibly improving the performance with fine-grained parallelism. However, the BER of IoT-ResQ and FSD remains over $10^{-5}$, despite increasing $L_P$ due to their inflexibility. In this figure, we also observe that the split-detection \systemname{} gradually works better as SNRs increase, leading to bigger gaps against the baseline one. Next, we present BER as a function of $L_P$ for different $N_t$ for 64-$N_r$ MIMO scenarios at SNR 14 dB in Figure~\ref{f:classical_ber_vs_n_64nr}. For all the tested $N_t$, \systemname{} reaches near-optimal BER with less $L_P$, even though both FSD and IoT-ResQ can reach it immediately with $L_P=16$.\footnote{However, we observe \systemname{} performs poorly with ``64-QAM modulation'' even for Massive MIMO scenarios, which remains an important challenge in PIC MIMO detectors. 64-QAM negative results are in Appendix~\ref{s:64qam}.}

% Recall that we mimic RA using the PT process and apply their parallelization strategies, where $L_P$ is proportional to the processing elements (PEs) (cf. qubits in QA). 
% for fully parallel processing to maintain the minimum detection latency.

% \parahead{Compute Time.} ParaMax and (classical) IoT-ResQ and X-ResQ feature the same compute time for their optimization processing due to the same PT core, except for their preprocessing including initial detection (\emph{i.e.,} FSD in IoT-ResQ and MMSE in X-ResQ).

\parahead{Ultra-Large MIMO with Low-Order Modulations.} Now, we show the BER performance of \systemname{} for extremely large MIMO (over $100\times 100$) with $N_r/N_t=1$ and low-order modulations, varying SNRs in Figure~\ref{f:ExtremeLargeMIMO}. This is the regime where PIC MIMO detectors show exceptionally better performance compared to other (conventional) detectors. Here, we cannot report ML performance because of the large $N_t$. 
% \vspace{-0.08cm}

First, Figure~\ref{f:largemimoqpsk} presents $128 \times 128$ (\emph{left}) and $256 \times 256$ (\emph{right}) MIMO with QPSK. \systemname{} outperforms the others, converging to a certain BER with a few $L_P$, which we assume is the optimal performance, given that both the converged BERs of ParaMax and \systemname{} are the same in $128\times128$ MIMO. 
% , ParaMax also outperforms IoT-ResQ, requiring less $L_P$ to achieve nearly optimal performance.
However, for $256\times256$ MIMO, only \systemname{} performs well and even ParaMax obtains poor detection performance. At SNR 14 dB, \systemname{} obtains five orders of magnitude lower BER than the others, requiring only a few $L_P$. This performance gap is very surprising and interesting, considering that there exists only a simple design difference between ParaMax and \systemname{} (\emph{i.e.,} \systemnames{} initialization). Further, \systemname{} can converge to a BER with quite a few $L_P$. Thus, we also test \emph{faster} \systemname{} by reducing its engine \emph{sweep iteration} count (from original $N_{SW}=50$) that largely affects its computation time~\cite{kim2021physics}. While 50$\times$ faster version ($N_{SW}=1$) shows some performance trade-offs for the same $L_P$, 25$\times$ faster \systemname{} ($N_{SW}=2$) is still able to achieve near-optimal BER even with similar $L_P$, which implies the evidence of speedup potential. The compute latency of classical \systemname{} also depends on spin variable counts ($N_V=N_t\cdot\log_2{|\mathcal{O}|}$) as shown in Figure~\ref{f:compute_time}, where $25\times$ faster \systemname{} operates with $N_{SW}=2$ (instead of $N_{SW}=50$).
% , which implies the evidence of speedup potential. 

Lastly, we test $512\times 512$ and $1024\times 1024$ MIMO with BPSK. Similar results as QPSK are observed in Figure~\ref{f:largemimobpsk}, where ParaMax, IoT-ResQ, and \systemname{} perform well for $512\times512$ MIMO (\emph{top}), resulting in no bit errors among 10s of million tested bits after a few $L_P$. For $1024\times1024$ MIMO (\emph{mid, bottom}), \systemname{} significantly outperforms the others (while IoT-ResQ still works decently). Note that \systemname{} even with $L_P=1$ achieves error-free results at SNR 14 dB, showing great promise of enabling ultra-large MIMO. To our best knowledge, this is the largest spatial multiplexing MIMO ever reported with (assumably) near-optimal detection BER. It has been discussed that this extreme MIMO regime is particularly beneficial for massive IoT connectivity~\cite{kim2022warm}.\vspace{-0.35cm}

\parabreak{}

\section{Discussion}
\label{s:discussion}
% \vspace{-0.15cm}
% In this section, we discuss the limitations of the work with future feasibility and potential research projects.
% including controversial ones. 

% 0.5--1 PAGE REQUIRED FOR THIS

% \parahead{QA Feasibility for Cellular Wireless Networks.} 
\parahead{Feasibility of \systemname{} in Cellular Networks.}  
% The envisioned scenario of QA MIMO detectors  C-RAN architecture.
Apart from current insufficient qubit counts, 
% and limited connectivity degrees, 
\systemname{} is not immediately available largely due to the existing milliseconds of overheads in QPU usage other than pure QA time (\S\ref{s:QA_parallel_strategy}). 
% For example, qubit counts and connectivity degrees are not sufficient enough to support QA parallelization in cellular networks, overheads in QA are currently several milliseconds, and  
However, many studies have already identified specific techniques to reduce them effectively and their expected improvements~\cite{weber2021high}. Collectively, for 10~M-qubit quantum annealers (ca. 2040)~\cite{kasi2021challenge}, the total projected QPU time of \systemname{} with 100 anneals including all the overheads is around 250--500~$\mu$s, consisting of approx. 50~$\mu$s for programming and thermalization~\cite{reed2010fast}, $100\times2$~$\mu$s for readout time and delay~\cite{grover2020fast,walter2017rapid}, and $100 \times T_a$ for pure QA time, \emph{e.g.,} $100$$\times88$~ns assuming the same RA schedule but with the potential minimum 40~ns\footnote{A recent study has identified that QA with the nanosecond-scale anneal duration can follow coherent quantum theory~\cite{king2023quantum}, and indeed, the minimum anneal duration ($T_a$) that real-world quantum annealers support also tends to lower (currently 0.5--1~$\mu$s). 
% (toward 40~ns~\cite{king2022coherent}).
% This makes the direction of QA-accelerated MIMO detection even more compelling. 
\systemname{} already utilizes the minimum $T_a$ $\approx$ 2~$\mu$s that the current machine supports for the RA schedule with $\tau_p=0.4$ (cf. $\approx$ 15~$\mu$s w/ FA~\cite{ducoing2022quantum, tabi2021evaluation}), and thus is expected to keep following the trend. } (48~ns back-and-forth anneal w/ $\tau_p=0.4$ and 40~ns pause). This implies the stringent 5/6G latency requirements can be potentially satisfied, supporting 1000s of subcarriers with data and task parallelism. Such annealers also bring both cost and power advantages over 1.5~nm CMOS hardware to C-RAN~\cite{kasi2021challenge}.

\parahead{Gate-Model Quantum Processors for MIMO Detection.}
It should be noted that the porting of MU-MIMO detection to superconducting gate-model paradigms with \emph{Quantum Approximate Optimization Algorithm} (QAOA) is an active field of research~\cite{cui2022quantum,gulbahar2023maximum}. However, current gate-model processors can support the optimization algorithm targeting fully-connected Ising models with only a few tens of variables. Momentous progress is being made, but current relevant experiments~\cite{maciejewski2023design,mazumder2023benchmarking} still exhibit lower performance with respect to annealers. Furthermore, it has been discussed that QAOA even on future \emph{fault-tolerant} gate processors has several critical challenges to overcome that do not apply to QA~\cite{sanders2020compilation,farhi2020quantum,wang2021noise}. 
% , and some studies state that QAOA even on fault-tolerant gate-model processors might not outperform QA in terms of optimization~\cite{mazumder2023benchmarking} due to challenges in QAOA and/or gate-model processors (\emph{e.g.,} error correction overheads~\cite{}).

% based on quantitative analysis
% For readers who are interested in the expected timeline of the practical QA-enabled wireless networks further, we refer to the  study~\cite{kasi2021challenge}.

% \parahead{Softwarization.}

% \parahead{Large MIMO with QA.}

% \systemnames{} design aims to improve overall QA detection performance. For readers who are interested in the feasibility and expected timeline of the practical QA-enabled wireless networks in C-RAN architectures, we refer to the  study~\cite{kasi2021challenge}.

\parahead{Rethinking Forward Error Correction.} 
% Many large and massive MIMO studies blindly apply standard error correction coding schemes (\emph{e.g.,} LDPC) to detection results. 
% However, 
When (near-)optimal performance becomes achievable in MIMO detection, we should rethink forward error correction (FEC), since much simpler schemes than current LDPC (or very high code rate) could be applied instead. For example, some \systemnames{} detection (uncoded) BERs observed are getting close to the target BER that FEC decoding aims to achieve (\emph{e.g.,} Figure~\ref{f:ExtremeLargeMIMO} w/ $N_r/N_t=1$), which implies slightly increasing $N_r/N_t$ can make detection nearly error-free.
% , even for low SNRs.  
% For low SNRs where even the optimal performance is not quite good, slightly increasing $N_r/N_t$ can make detection BER much lower. 
Currently, LDPC decoding has been identified as the most computationally heavy part of the physical layer whose computational amounts also scale with users~\cite{ding2020agora}. Leveraging a sophisticated MIMO detector for ML performance (\emph{e.g.,} \systemname{}) while exploiting a simple 
% (or even no)\footnote{Although very unlikely, this (no FEC) might be possible for some certain scenarios, where channel characterization and propagation environments become precisely controllable (\emph{e.g.,} with reconfigurable intelligent surface).} 
FEC decoder might be a more efficient design, which is the opposite of the current architecture. 
Moreover, ultra-large MIMO potentially enabled by (near-)optimal detectors (\emph{e.g.,} $1024\times1024$ MIMO with \systemname{}) could not only improve spectral efficiency and device connectivity but also potentially simplify MAC layer designs. For instance, the expected increased capacity can make resource scheduling much simpler and faster.
Overall, we believe realizing scalable optimal MIMO detectors could open up a research area that allows us to explore new system architectures in both mobile devices and base station (or access point in WLAN) systems. 

% \vspace{-0.05cm}
\parahead{End-to-End Systems.} 
% Despite the great promise of accomplishing the
% next levels of wireless performance,
Despite the impressive performance,
the current QA and PIC-based MIMO detectors have met with only limited success in that the studied systems, including this work, are still far from end-to-end (E2E) system implementations and evaluations, leaving their full potential and practicality as part of real wireless systems questionable. 
% The systems do not consider the rest modules of the physical layer such as forward error correction and subcarrier processing. 
Due to the immaturity of technology, building quantum-based E2E wireless systems is many years away. However, classical generic PIC detectors (\emph{e.g.,} classical \systemname{}) can be embedded into an E2E system and evaluated.
Integrating them into the recent softwarization outputs of the MIMO physical layer such as Agora~\cite{ding2020agora} and Hydra~\cite{gong2023scalable} would be a good initiative for the direction. To do so, significant system research efforts will be required for the efficient pipeline and parallelism design for a real-time E2E system with the newly added dimensions of parallelism. 
% which 
We leave this as our future work, where \emph{system automation} will be also considered for intelligence and elasticity in the physical layer (\emph{e.g.,} opportunistic RA, \emph{adaptive} split-detection method, $N_r/N_t$, and level of parallelism).
% $L_P$).
% taking advantage of virtualization

% \parahead{QA Feasibility for Cellular Wireless Networks.}

% throughput and/or spectral efficiency performance. 

% \parahead{New Communications System Design.}

% Since the shapes of the qubit clusters in clique embedding can be triangular, it is an efficient way of using qubits for parallel QA as well, forming a rectangular shape with two triangular clusters and thus supporting maximum parallelism. The embedding can be prepared in a hash table based on the input size ($N_V$). However,
% since the current machine is a prototype and faulty qubits on the hardware need to be considered~\cite{klymko2014adiabatic}, we leave the maximum-parallel clique embedding and its hash table as our future work. Our parallel clique embedding leaves some qubits idle.
% (\emph{e.g.,} Figure~\ref{f:quantum_parallel}). 
% \vspace{-0.2cm}
\section{Conclusion}
\label{s:conclusion}
% \vspace{-0.1cm}
% \vspace{-0.2cm}
The paper introduces \systemname{}, a QA-based MIMO detector that uses multi-seed ensemble RA as its parallelization strategy. 
% We demonstrate the rationale of \systemnames{} parallel QA design, pointing out the limitations of other parallel QA designs. 
We show that
\systemnames{} design is simple but effective, trading off between qubits and compute time.
% , outperforming other (QA) MIMO detectors. 
Given that nearly all QA components such as qubit counts, overheads, and noise control are being improved drastically,
%  % such as exponentially increasing qubit counts, higher precision, reduced overheads, and better noise control
the significantly enhanced MIMO detection performance achieved by \systemname{} over prior designs on the current machine 
% % compared to QuAMax 
is a good indication
of the potential general use of QA MIMO detectors with a long-term vision. Furthermore,
the paper shows the viability of enabling ultra-large MIMO by classical \systemname{}, which itself is another PIC MIMO detector candidate.

\section*{\large Acknowledgement}
We thank 
% the anonymous shepherd and reviewers of this paper, 
the Princeton Advanced Wireless Systems (PAWS) Group and the Yale Efficient Computing Lab (ECL) 
% Lin Zhong, and Anurag Khandelwal 
for their extensive technical feedback and helpful discussion.
This research is based upon work supported by InterDigital Communications, and National Science Foundation (NSF) Award No. CNS-1824357 and CCF-1918549. QA experiments on the D-Wave quantum annealers have been supported by InterDigital. This work does not raise any ethical issues.

\clearpage

\let\oldbibliography\thebibliography
\renewcommand{\thebibliography}[1]{%
  \oldbibliography{#1}%
  \setlength{\parskip}{0pt}%
  \setlength{\itemsep}{0pt}%
}

\begin{raggedright}
\bibliographystyle{concise2}
\bibliography{reference}
\end{raggedright}

\clearpage
\appendix
\section{Linear MIMO Detectors}
\label{s:primer_linear}

\emph{Zero-Forcing} (ZF) linear method makes use of the pseudo-inverse of $\mathbf{H}$, $\mathbf{H}^\dagger = (\mathbf{H}^*\mathbf{H})^{-1}\mathbf{H}^*$ with $H^*$ Hermitian transpose, for detection, by multiplying $\mathbf{H}^\dagger$ with $\mathbf{y}$:
\begin{equation}   
\begin{small}
\label{eqn:zf}
\mathbf{H}^\dagger\mathbf{y} = (\mathbf{H}^*\mathbf{H})^{-1}\mathbf{H}^*\mathbf{H}\mathbf{\bar{v}}+(\mathbf{H}^*\mathbf{H})^{-1}\mathbf{H}^*\mathbf{n} = \mathbf{\bar{v}+\mathbf{n'}}
\end{small}
\end{equation}
with $\mathbf{n'}$ the noise vector affected by the amplification factor. For each $t^{th}$ user, 
the symbol $\hat{v}_t$ among $v\in \mathcal{O}$ 
% (cf. $\mathcal{O}^{N_t}$ in ML) 
with the minimum Euclidean distance against $\bar{v}_t+n'_t$ (\emph{i.e.,} the closest member of constellation to $t^{th}$ element in $\mathbf{H}^{\dagger}\mathbf{y}$) 
is detected, forming detected symbol vector $\mathbf{\hat{v}}$. As one can see, ZF is sensitive to both noise vector $\mathbf{n}$ and channel $\mathbf{H}$ that decide $\mathbf{n'}$. When the noise and/or amplification factor becomes large (\emph{i.e.,} low SNRs and/or low $N_r/N_t$), the detection performance is severely degraded. 

\parabreak{}\emph{Minimum Mean Square Error} (MMSE) linear detector multiplies $\mathbf{G}$ with $\mathbf{y}$ (instead of $\mathbf{H}^\dagger$ in ZF) where $\mathbf{G}= \text{SNR}\cdot(\mathbf{I}+\text{SNR}\cdot\mathbf{H}^*\mathbf{H})^{-1}\mathbf{H}^*$, with $\mathbf{I}$ being an identity matrix and SNR a signal-to-noise ratio, which satisfies the minimum $ \left\lVert \mathbf{\bar{v}}-\mathbf{G}\mathbf{y}\right\rVert^2$. Unlike ZF, MMSE considers $\mathbf{n}$ in addition to $\mathbf{H}$ for regularization, thus outperforming ZF.
\section{Split-Detection: Theoretical Analysis}
\label{s:noise_analysis}

As discussed in the main paper, the split-detection method transforms a 16-QAM MIMO ML problem into solving two independent QPSK problems. This method can be generalized for any modulations. 
% 16-QAM ML optimization variables can expressed as $\mathbf{v} = 2\mathbf{{q}}_2 + \mathbf{{q}}_1$, where $\mathbf{{q}}_i \in \{-1-j, -1+j, 1-j, 1+j\}^{N_t}$ consists of QPSK symbols. Note that $\mathbf{{q}}_2$ represents the \emph{quadrant}, while $\mathbf{{q}}_1$ represents the \emph{position of the symbol} within the quadrant, and that $\mathbf{q}$ can be expressed with two spin variables ($\mathbf{q_i}=s_{2i-1}+js_{2i}$).
For a square QAM modulation, the optimization variable $\mathbf{v}$ in ML problem (\S\ref{s:primer_mimo}) can be expressed using $n_{q} = \log_2(\lceil\sqrt{|\mathcal{O}|}\rceil)$ QPSK variables, as $\mathbf{v} = \sum_{i = 1}^{n_{q}} 2^{i-1}\mathbf{{q}}_i$, where $\mathbf{{q}}_i \in \{-1-j, -1+j, 1-j, 1+j\}^{N_t}$ consists of QPSK symbols. We fix the values of ($n_{q}-1$) variables in this representation using the MMSE solution and reduce the problem to effectively have an effective search space of $\{-1-j, -1+j, 1-j, 1+j\}^{N_t}$, which is equivalent to searching with a QPSK modulation.  
% Note that, this is equivalent to searching with a QPSK modulation. 
% For non-square modulations, similarly, we can express the MMSE solution as $\mathbf{x}^{MMSE} = \sum_{i = 1}^{n_{sp}} 2^{i-1}\mathbf{{q}}^{MMSE}_i$. 
If the modulation is not a square constellation, the procedure remains the same, except that $\mathbf{{b}}_{n_{b}} \in \{-1,1\}^{N_t}$ and the corresponding reduced problem is equivalent to BPSK, instead of QPSK.

\begin{figure}
\centering
\includegraphics[width=0.85\linewidth]{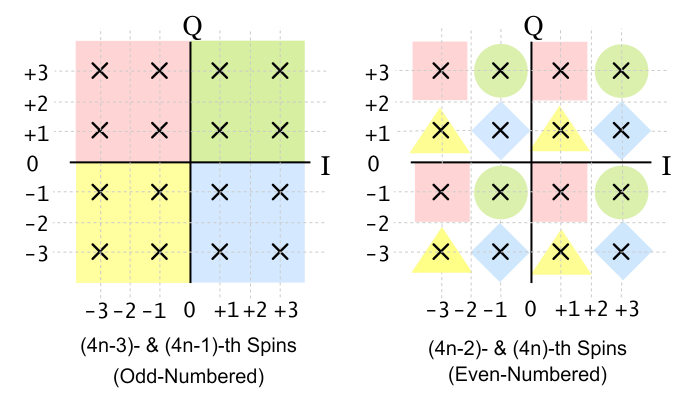}\vspace{-0.1cm}
    \caption{\small
    \textbf{Decision impact of spin variables in ML Ising models. In 16-QAM ($\mathcal{O}=16$), for $n$-th user's symbol, $(4n-3)-$ and $(4n-1)-$th (odd-numbered) spins are related to the quadrant decision (\emph{i.e.,} $2\mathbf{q}_1$ in $\mathbf{v}$), while  $(4n-2)-$ and $(4n)-$th spins (even-numbered spins) to the position decision  (\emph{i.e.,} $\mathbf{q}_2$ in $\mathbf{v}$). This can be generalized with higher-order modulations.}} 
\label{f:sym_quad}
\end{figure}

\parahead{Noise Analysis.} Let us analytically look at the noise in the split-detection with 16-QAM to understand it better, considering an $N \times N$ MIMO system (\emph{i.e.,} $N_t=N_r$):
\begin{equation}
    \mathbf{y} = \mathbf{H}\mathbf{\bar{v}} + \mathbf{n},
\end{equation}
where $\mathbf{n}$ is white Gaussian noise with $E\left[||n||\right] = 0$ and $E\left[||n||^2\right] = \sigma^2$. 
Note that the channel is assumed to be Rayleigh fading (\textit{i.e.}, each element $H_{ij}$ is drawn from a complex normal distribution). 
For 16-QAM, the transmit vector can be expressed as $\mathbf{v} = 2\mathbf{q}_1 + \mathbf{q}_2 \in \mathcal{O}^{N}$, where $\mathbf{q}_1,\mathbf{q}_2 \in \{-1-i,-1+i,1-i,1+i\}^N$, each of which can be expressed with two spin variables: $q = s_1 +js_2$. Note that, $\mathbf{q}_1$ is the quadrant of the transmit symbol and $\mathbf{q}_2$ expresses the position within the quadrant as shown in Figure~\ref{f:sym_quad}.

When we perform split detection assisted by the MMSE solution $\mathbf{v}_m = 2\mathbf{q}_{m1} + \mathbf{q}_{m2}$, and we try to decode $\mathbf{q}_1$, after subtracting $\mathbf{Hm}_2$ to cancel interference of $\mathbf{q}_2$, the equivalent system is 
\begin{equation}
    \dfrac{\mathbf{y}}{2} = \mathbf{H}\mathbf{q}_1 + 0.5(\mathbf{n} + \mathbf{H}(\mathbf{q}_2 - \mathbf{q}_{m2}))
\end{equation}
let us define $\mathbf{\delta} = \mathbf{q}_2 - \mathbf{q}_{m2}$, where for user $i$, $|\delta_i|^2 = 0$ if MMSE estimate is correct, $|\delta_i|^2 = 4$ if only real or imaginary part of MMSE estimate is correct, and $|\delta_i|^2 = 8$ if MMSE estimate is completely wrong.

Then the \emph{effective noise power} is given by,
\begin{equation}
  0.25(E\left[||\mathbf{n} + \mathbf{H}\mathbf{\delta}||^2\right])
\label{eqn:noise_power}
\end{equation}

\begin{figure}
\centering
    \includegraphics[width=0.9\linewidth]{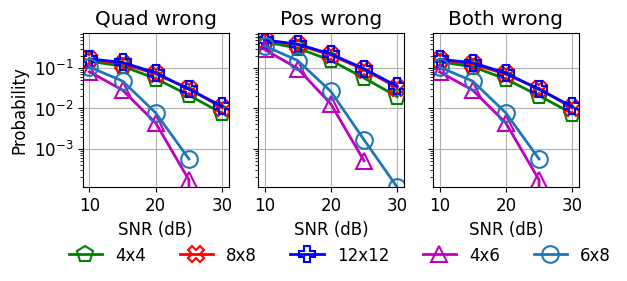}\vspace{-0.1cm}
    \caption{\small
    \textbf{Probability of MMSE solution wrongly predicting the quadrant and/or position of the ML solution. We see that the MMSE solution becomes increasingly accurate with an increase in SNR; hence, the split-detection-based Ising form tends to be equivalent to the original ML problem as SNR increases.}} 
\label{f:mmse_wrong}
\end{figure}

and therefore by triangle inequality 
\begin{equation}
      \leq 0.25(E\left[||\mathbf{n}||^2\right] + E\left[||\mathbf{H}\mathbf{\delta}||^2\right] + 2E\left[||\mathbf{n}||\cdot||\mathbf{H}\mathbf{\delta}||\right]).
\end{equation}
The Cauchy-Swartz inequality for two random variables $X$ and $Y$ states,
\begin{equation}
    |E[XY]| \leq \sqrt{E[X^2]E[Y^2]}
\end{equation}
by Cauchy-Schwartz inequality,
\begin{small}
\begin{equation}
    \leq 0.25(E\left[||\mathbf{n}||^2\right] + E\left[||\mathbf{H}\mathbf{\delta}||^2\right] + 2\sqrt{E\left[||\mathbf{n}||^2\right]\cdot E\left||\mathbf{H}\mathbf{\delta}||^2\right])}
\end{equation}
\end{small}
\begin{equation}
    = 0.25(\sigma^2 + E\left[||\mathbf{H}\mathbf{\delta}||^2\right] + 2\sigma\sqrt{ E\left[||\mathbf{H}\mathbf{\delta}||^2\right])}.
    \label{eq:noisePowerBound}
\end{equation}
Let us look at the random variable $\mathbf{H\delta}$,
\begin{equation}
    E[||\mathbf{H\delta}||^2] = E\left[\delta^\dag\mathbf{H}^\dag\mathbf{H}\delta\right].
\end{equation}
Let $\mathbf{D} = \mathbf{H}^\dag\mathbf{H}$,  
\begin{small}
\begin{equation}
    E[||\mathbf{H\delta}||^2] = E[\delta^\dag\mathbf{D}\delta] = E\left[\sum_i\sum_jD_{ij}\delta_i\delta_j^*\right]
\end{equation}
\end{small}
As $||\mathbf{H\delta}||^2$ is real,
\begin{small}
\begin{equation}
    =  E\left[\sum_i\sum_j\Re{D_{ij}\delta_i\delta_j^*}\right]
\end{equation}
\begin{align*}
    =  \sum_i\sum_jE\left[\Re{D_{ij}}\Re{\delta_i}\Re{\delta_j}\right] \\ -E\left[\Re{D_{ij}}\Im{\delta_i}\Im{\delta_j}\right] \\
    -E\left[\Im{D_{ij}}\Re{\delta_i}\Im{\delta_j}\right]\\
    -E\left[\Im{D_{ij}}\Im{\delta_i}\Re{\delta_j}\right] \nonumber.
\end{align*}
\end{small}
Let us look at the term $\left|E\left[\Re{D_{ij}}\Re{\delta_i}\Re{\delta_j}\right]\right|$,  by Cauchy-Schwartz inequality,
\begin{small}
\begin{equation}
     \leq \sum_i\sum_j\sqrt{E[\Re{D_{ij}}^2]}\sqrt{E[\Re{\delta_i}^2\Re{\delta_j}^2]}.
\end{equation}
\end{small}
Now, 
\begin{small}
\begin{equation}
    E[\Re{\delta_i}^2\Re{\delta_j}^2] \leq \sqrt{E[\Re{\delta_i}^4]}\sqrt{E[\Re{\delta_j}^4]}.
\end{equation}
\end{small}
Note that $\Re{\delta_i}$ can only take values $\{0,-2,2\}$
\begin{equation}
    E[\Re{\delta_i}^4] = \mathcal{P}(\Re{\delta_i}\neq 0)\cdot 16
\end{equation}
which implies
\begin{equation*}
   E[\Re{\delta_i}^2\Re{\delta_j}^2] \leq 16 \cdot \sqrt{\mathcal{P}(\Re{\delta_i}\neq 0)\mathcal{P}(\Re{\delta_j\neq 0)}}.
\end{equation*}
Similarly,
\begin{small}
\begin{align*}
    E[\Im{\delta_i}^2\Im{\delta_j}^2] \leq 16 \cdot \sqrt{\mathcal{P}(\Im{\delta_i}\neq 0)\mathcal{P}(\Im{\delta_j\neq 0)}} \\
     E[\Re{\delta_i}^2\Im{\delta_j}^2] \leq 16 \cdot \sqrt{\mathcal{P}(\Re{\delta_i}\neq 0)\mathcal{P}(\Im{\delta_j\neq 0)}}\\ 
     E[\Im{\delta_i}^2\Re{\delta_j}^2] \leq 16 \cdot \sqrt{\mathcal{P}(\Im{\delta_i}\neq 0)\mathcal{P}(\Re{\delta_j\neq 0)}}.
\end{align*}
\end{small}
Let us look at $E\left[\Re{D_{ij}}^2\right]$
\begin{small}
\begin{equation*}
     = E\left[\left(\sum_k \Re{H_{ki}}\Re{H_{kj}} + \Im{H_{ki}}\Im{H_{kj}}\right)^2\right].
\end{equation*}
\end{small}
Note that the real and imaginary parts of channel coefficients are I.I.D Gaussian random variables with zero mean and variance 0.5.\\
For $i \neq j$,
\begin{small}
 \begin{align*}
    E\left[\Re{D_{ij}}^2\right]= \sum_k E[\Re{H_{ki}}^2]E[\Re{H_{ki}}^2] +\\
    E[\Im{H_{ki}}^2]E[\Im{H_{ki}}^2]  \\
    = 0.5N.
\end{align*}
\end{small}
For $i = j$,
\begin{small}
 \begin{align*}
     E\left[\Re{D_{ii}}^2\right] = E\left[\left(\sum_k \Re{H_{ki}}^2+
    \Im{H_{ki}}^2\right)^2\right] \\
    = \sum_k E[\Re{H_{ki}}^4]+E[\Im{H_{ki}}^4] + \\ 
    2 \cdot\sum_{j,k,j\neq k}E[\Re{H_{ki}}^2\Im{H_{ki}}^2] \\
    = 6N\cdot (1/2)^4 + (1/2)^4 \cdot N(N-1).
\end{align*}
\end{small}
Next, $E\left[\Im{D_{ij}}^2\right]$
\begin{small}
\begin{equation*}
   = E\left[\left(\sum_k \Re{H_{ki}}\Im{H_{kj}} -\Im{H_{ki}}\Re{H_{kj}}\right)^2\right]
\end{equation*}
for $i \neq j$,
 \begin{align*}
     = \sum_k E[\Re{H_{ki}}^2]E[\Im{H_{kj}}^2] \\ +
    E[\Im{H_{ki}}^2]E[\Re{H_{kj}}^2]  = 0.5N 
\end{align*}
\end{small}
     
and zero otherwise.

Therefore, for $i\neq j$, $E\left[(\Re{D_{ij}}\Re{\delta_i}\Re{\delta_j}\right]$
\begin{equation*}
     \leq \sqrt{0.5N} \cdot \sqrt{16 \sqrt{\mathcal{P}(\Re{\delta_i}\neq 0)\mathcal{P}(\Re{\delta_j\neq 0)}}}
\end{equation*}
and for $i = j$, 
\begin{small}
\begin{equation*}
    \leq 4\sqrt{\left[6N\cdot (1/2)^4 + (1/2)^4 \cdot N(N-1)\right]}\sqrt{\mathcal{P}(\Re{\delta_i}\neq 0)}.
\end{equation*}
\end{small}

Putting everything together, $E[||\mathbf{H\delta}||^2]$
\begin{small}
\begin{align*}
     \leq \sum_i 4\sqrt{\left[6N\cdot (1/2)^4 + (1/2)^4 \cdot N(N-1)\right]} \\
     \times(\sqrt{\mathcal{P}(\Re{\delta_i}\neq 0)} + \sqrt{\mathcal{P}(\Im{\delta_i}\neq 0)}) \\
      + \sum_{i,j,i\neq j}\sqrt{8N}\cdot[(\mathcal{P}(\Re{\delta_i}\neq 0)\mathcal{P}(\Re{\delta_j}\neq 0))^{\frac{1}{4}} \\
      + (\mathcal{P}(\Im{\delta_i}\neq 0)\mathcal{P}(\Im{\delta_j}\neq 0))^{\frac{1}{4}} \\
      + (\mathcal{P}(\Re{\delta_i}\neq 0)\mathcal{P}(\Im{\delta_j}\neq 0))^{\frac{1}{4}} \\
      + (\mathcal{P}(\Im{\delta_i}\neq 0)\mathcal{P}(\Re{\delta_j}\neq 0))^{\frac{1}{4}}]
\end{align*}
\end{small}
We assume $\mathcal{P}(\Re{\delta_j} \neq 0) = \mathcal{P}(\Re{\delta_i}\neq 0) = \mathcal{P}(\Im{\delta_j}\neq 0) = \mathcal{P}(\Im{\delta_i}\neq 0) = P^2$, given the symmetry of the problem. Therefore, $E[||\mathbf{H\delta}||^2]$
\begin{equation}
     \leq 2P \cdot [N\sqrt{\left[6N+N(N-1)\right]} + 
    N^{\frac{3}{2}}(N-1)].
    \label{eq:bound17}
\end{equation}

As wireless noise reduces \textit{i.e.},  $\sigma \rightarrow 0$ (as SNRs increase), MMSE becomes increasingly accurate and will have lesser bit errors $\implies$ $P \rightarrow 0$. Therefore, according to Eq.~\ref{eq:bound17}, $E[||\mathbf{H\delta}||^2] \rightarrow 0$, and the effective noise given by Eq.~\ref{eq:noisePowerBound} goes to zero, leading to good performance at high SNR. Note that a similar analysis can be performed when we try to decode $\mathbf{q}_2$, resulting in the same conclusion. Further, in Figure~\ref{f:mmse_wrong} we illustrate the accuracy of MMSE solution in predicting the quadrant and position of the ML solution. Similar pre-decision schemes based on collected samples in the preceding run have been discussed in \cite{karimi2017boosting,kim2021physics}, but the methods require iterative runs of PIC optimization which are not allowed in QA MIMO detection due to the programming time (\S\ref{s:QA_parallel_strategy}).

\parahead{Limitations of a Na\"ive Alternative Method.} A natural alternative to our proposed strategy could be a method that treats either the quadrant or the position as part of noise, rather than fixing them to the MMSE solution. However, such a strategy will not be suitable for X-ResQ as it will significantly increase the effective noise in the problem.

If we try to perform split detection and try to detect $\mathbf{q}_1$ first then the equivalent system is given by, 
\begin{equation}
    \dfrac{\mathbf{y}}{2} = \mathbf{H}\mathbf{q}_1 + 0.5(\mathbf{n} + \mathbf{H}\mathbf{q}_2)
\end{equation}
and effective noise power is given by,
\begin{small}
\begin{align*}
     0.25\cdot (E\left[||\mathbf{n}||^2\right] + E\left[||\mathbf{H}\mathbf{q}_2||^2\right]) \\
     = 0.25 \cdot (\sigma^2 + E\left[\mathbf{q}_2^\dag\mathbf{H^\dag H}\mathbf{q}_2\right])
\end{align*}
\end{small}
We know that for any two random variables $X$ and $Y$, $E[XY] = E[X \cdot E[Y|X]]$. Further. given $\mathbf{n}$, $\mathbf{H}$, and $\mathbf{q}_2$ are independent and $E[H^\dag H] = N \cdot \mathbf{I}$ (Rayleigh fading channels)
\begin{align*}
    = 0.25 \cdot (\sigma^2 + E\left[\mathbf{q}_2^\dag E[\mathbf{H^\dag H}\mathbf{q}_2|\mathbf{q}_2^\dag]\right]) \\
    = 0.25 \cdot(\sigma^2 + E\left[\mathbf{q}_2^\dag E[\mathbf{H^\dag H}]\mathbf{q}_2\right]) \\
    = 0.25 \cdot (\sigma^2 + N \cdot E[||\mathbf{q}_2||^2]).
\end{align*}
Now, $E[||\mathbf{q}_2||^2] = 2N$
\begin{equation}
    = 0.25\cdot(\sigma^2 + 2N^2)
\end{equation}

If we try to decode $\mathbf{q}_2$ instead then the equivalent system is given by, 
\begin{equation}
    \mathbf{y} = \mathbf{H}\mathbf{q}_2 + (\mathbf{n} + 2\mathbf{H}\mathbf{q}_1)
\end{equation}
and effective noise power is given by,
\begin{small}
\begin{equation}
(E\left[||\mathbf{n}||^2\right] + E\left[||2\mathbf{H}\mathbf{q}_2||^2\right])
\end{equation}
\end{small}
\begin{small}
\begin{equation}
    = (\sigma^2 + 8N^2) 
\end{equation}
\end{small}

and hence decoding $\mathbf{q}_1$ will experience lower noise. Unlike before (Eq.~\ref{eqn:noise_power}), $\delta$ is now independent of $\mathbf{H}$ and $\mathbf{n}$. Note that for both these scenarios, there is a significant interference; even when $\sigma \rightarrow 0$, effective noise power doesn't go to zero, leading to a bad performance even at high SNRs. 

% but unlike before $\delta$ is not independent of $\mathbf{H}$ and $\mathbf{n}$, 

\parahead{Summary.} It is observed that, unlike the na\"ive method, the MMSE-assisted split method used in X-ResQ is effective, as MMSE becomes increasingly accurate with an increase in SNR, and the probability that it wrongly predicts both quadrant and position of the ML solution, is very low. Based on the analytical and empirical analysis, we draw the following conclusions:
\begin{itemize}
    \item At high SNRs, the likelihood of both the quadrant and position of the MMSE solution being wrong is very low. As MMSE more accurately predicts at least one of quadrant or position correctly, the effective noise in the split Ising forms becomes very low. Therefore, the MMSE-assisted split detection is an effective method of transforming the original 16-QAM ML problem into an equivalent QPSK ML problem that is a much easier problem for QA and other PIC methods. 
    \item We demonstrate through our experimental analysis (\S\ref{s:evaluation}) that the split-detection formulation successfully mitigates the error floor phenomenon observed in QA and PIC MIMO detection with 16-QAM at high SNRs.
    \item Our split method has much less noise compared to the alternate splitting strategy that considers either the quadrant or position as part of the noise. Also, note that the method can be used along with any PIC algorithm.

\end{itemize}

% \input{QA_Appendix}

% \section{A Fully-Parallel QA Run versus Cumulative Sequential QA Runs}
\section{Parallel QA Hardware Programming}
\label{s:cumul_QA}
\vspace{-0.1cm}

\begin{figure}[!ht]
\centering
    \includegraphics[width=0.62\linewidth]{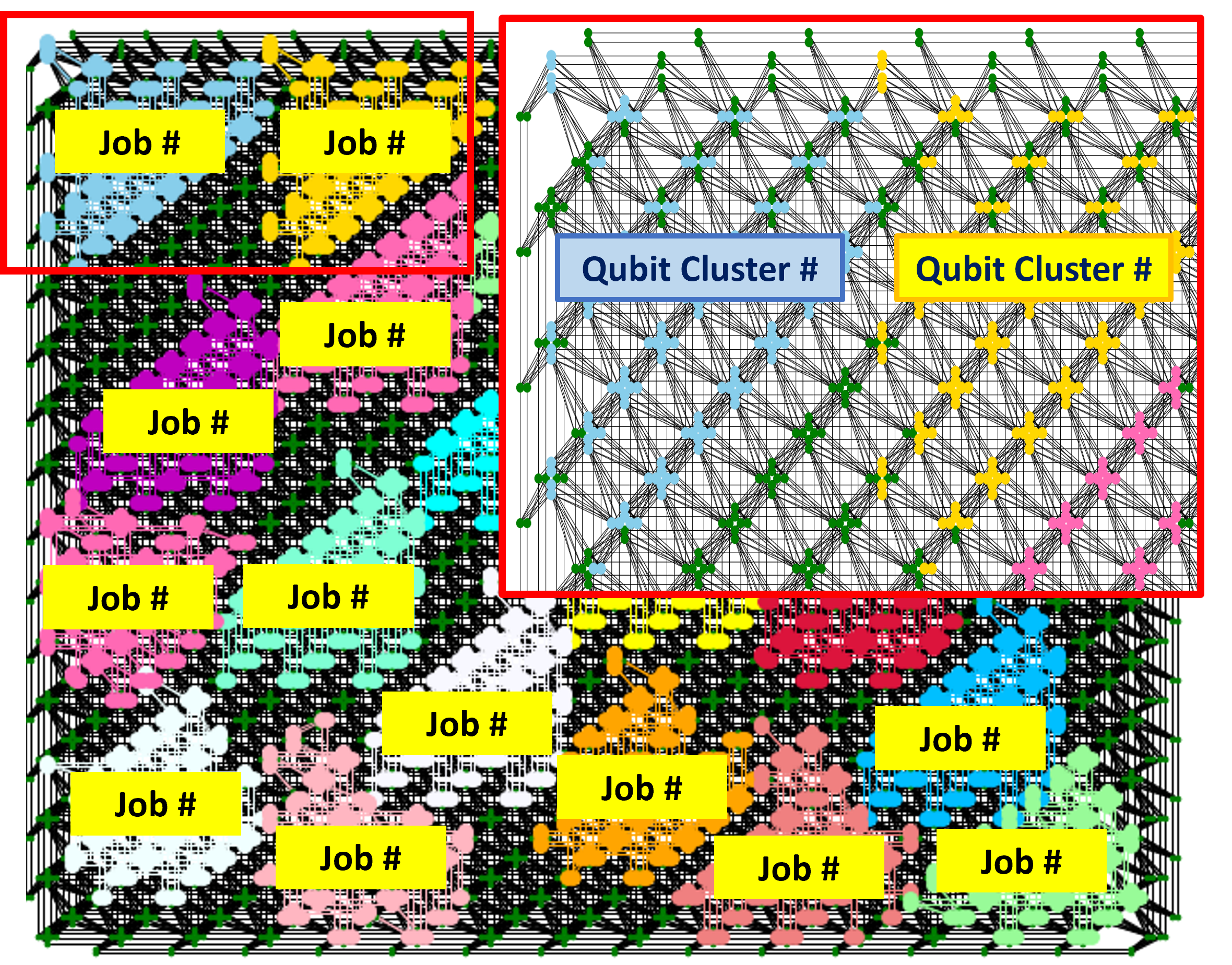}
    \caption{\small
    \textbf{Clique embedding for parallel quantum processing on Pegasus-topology QA hardware using 16-user QPSK MIMO detection. Color coding represents different jobs in parallelism.}} 
\label{f:quantum_parallel}
\end{figure}

% \begin{figure}
% \centering
%     \includegraphics[width=0.80\linewidth]{Figures/ber_vs_N.png}
%     \caption{\small
%     \textbf{BER performance of \systemname{} (cumulative sequential runs to mimic parallel runs) as a function of levels of parallelism ($N$) with 110~$\mu$s and 220~$\mu$s compute time. 
%     % Here, $N\times96$ physical qubits are required for parallelism, implying over 50 parallel jobs available in the full-sized Advantage2 (expected in 2023-2024).}
%     }} 
% \label{f:BER_N}
% \end{figure}

\vspace{-0.2cm}
\parahead{Embedding.} Unlike Ising spin models, physical qubits on real hardware are rather sparsely connected. Thus, we need extra qubits to compile the Ising problem on the machine. This procedure is called \emph{hardware minor embedding}~\cite{choi2008minor}. We use the \emph{clique embedding} technique that is designed to embed fully connected graph models into sparse graphs since our ML models are (nearly-)fully connected.
% For each problem in the combined Ising model ($\S$\ref{s:parallel_QA}, we apply the clique embedding (Figure~\ref{f:quantum_parallel}). 
Since the shape of each qubit cluster in clique embedding can be formed as a triangle (\emph{e.g.,} Figure~\ref{f:quantum_parallel} on Pegasus-connectivity hardware and similarly on Zephyr hardware), it is an efficient way of using qubits for parallel QA as well, forming a rectangular shape with two triangular clusters and thus promisingly supporting maximum parallelism. The embedding information can be prepared in a hash table based on the input size ($N_V$) and level of parallelism ($L_P$) without causing high overheads. However,
since the machine we use for the X-ResQ implementation is a prototype machine with small qubit counts and also faulty qubits on the hardware need to be considered~\cite{klymko2014adiabatic},
we leave the customized maximum-parallel clique embedding and its hash table as our future work (with the full-size machine). Our parallel embedding used in this work leaves some qubits idle. Some weird shapes of the clique embedding are observed due to the (small-size) hardware geometry and faulty qubits.

\begin{figure}
\centering
\includegraphics[width=0.68\linewidth]{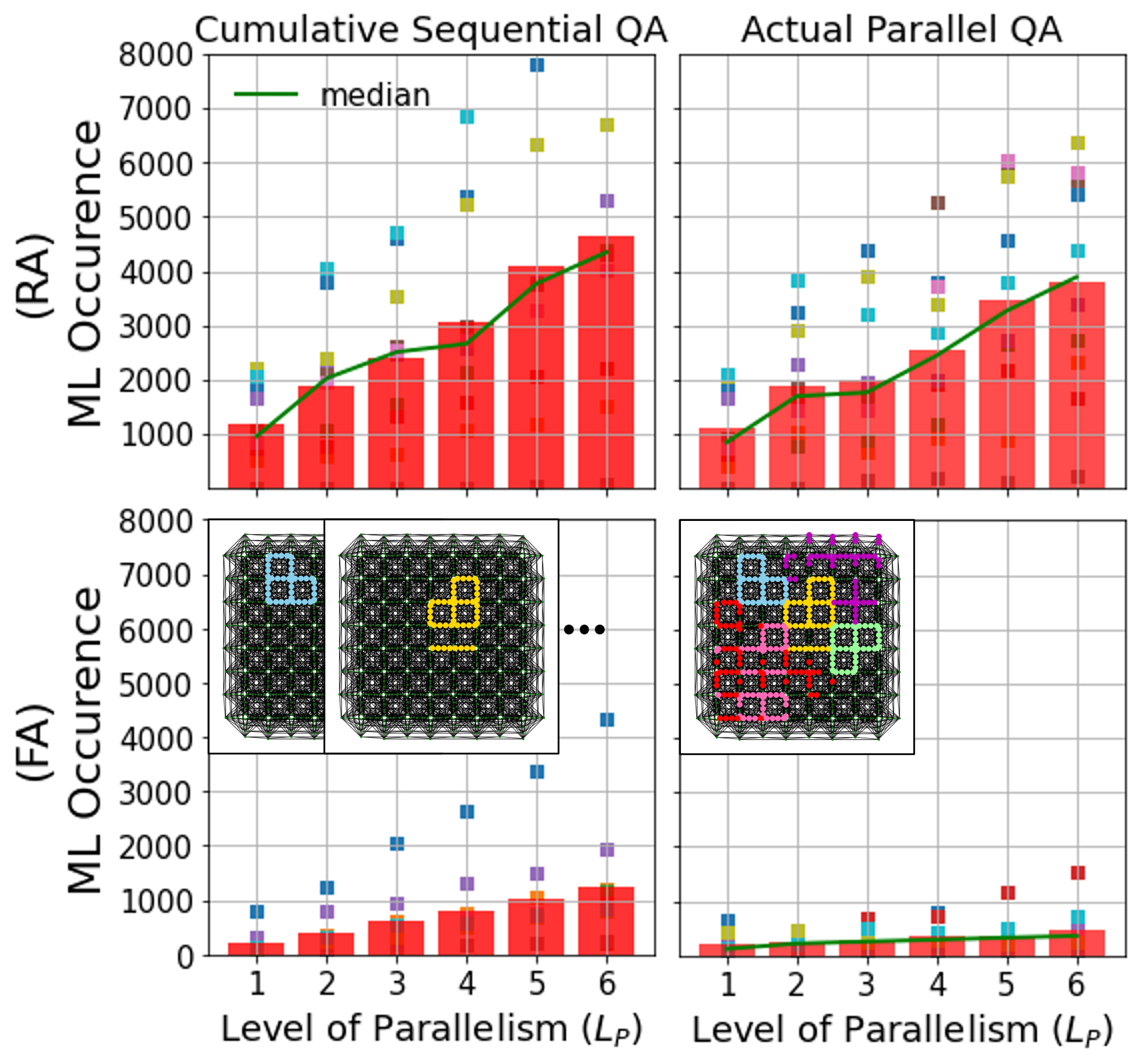}\vspace{-0.1cm}
    \caption{\small
    \textbf{Average ML occurrence across instances (out of $L_P\cdot 5000$ QA samples per instance), comparing cumulative sequential QA runs to mimic parallel QA (\emph{left}) against actual parallel runs (\emph{right}). Both X-ResQ w/ RA (\emph{upper}) and QuAMax w/ FA (\emph{lower}) are tested. $4\times 4$ 16-QAM MIMO instances are used, where each symbol reports each detection instance (color: inst. index).}} 
\label{f:cumul_seq_vs_parallel}
\end{figure}

\parahead{Selection of QA Parameters.} The choice of QA parameters in \systemname{} is decided by the brute-force exploration, following the same steps as the previous work~\cite{kim2019leveraging,kim2020towards}. Ideally, QA parameter settings can be optimized per instance (also, different initialization states should be considered in RA). However, this is not practical due to the lack of theoretical means to acquire the best setting per instance. 
Thus, we use the empirically-obtained best-median setting obtained from a few tens of instances through the brute-force exploration: $2.2~\mu$s anneal duration ($T_a$) and $\tau_p = 0.4$ switching point (\emph{i.e.,} $1.2~\mu$s back-and-forth anneal with $1.0~\mu$s pause). The precise QA schedules used in QuAMax and \systemname{} (and IoT-ResQ) as a form of [$T_a$ ($\mu$s), $\tau$] follows:

\begin{small}
\begin{itemize}
    \item QuAMax (FA with 1~$\mu$s pause at $\tau_p=0.3$):  

$[0.0,0.0] \xrightarrow{F} [0.3,0.3] \xrightarrow{P} [1.3, 0.3] \xrightarrow{F} [2.0, 1.0]$,

    \item \systemname{} (RA with 1~$\mu$s pause at $\tau_p=0.4$): 

$[0.0,1.0] \xrightarrow{R} [0.6, 0.4] \xrightarrow{P} [1.6, 0.4] \xrightarrow{F} [2.2, 1.0]$.

\end{itemize}
\end{small}

For further details about the system implementations (\emph{e.g.,} embedding penalty term, coupling dynamic range, and post-processing), we refer to Ref.~\cite{kim2019leveraging}.

\parahead{Cumulative Sequential QA to Estimate Fully Parallel QA.} Current quantum annealer hardware features a limited number of qubits to test parallel QA, especially on the Advantage2 prototype while it is the state-of-the-art machine with the most advanced topology hardware (see Table~\ref{t:DWAVE}). To test large parallelism beyond one that the current machine can support, we also estimate parallel QA performance by accumulating separate (sequential) QA results. 

\begin{figure}\centering
    \includegraphics[width=0.8\linewidth]{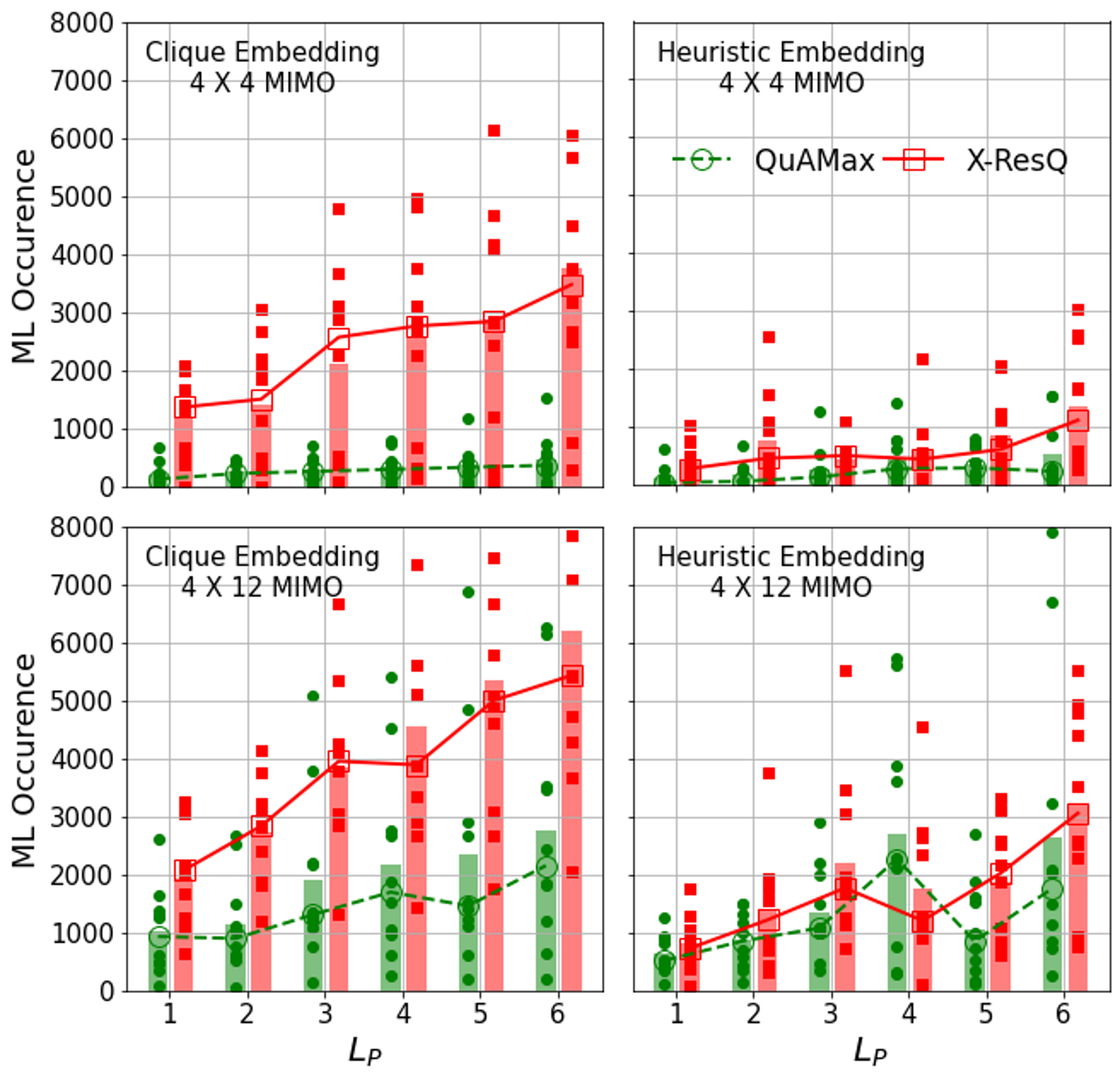}\vspace{-0.15cm}
    \caption{\small
    \textbf{ML occurrence comparison between clique embedding (applied in the implementation) (\emph{left}) versus heuristic embedding (\emph{right}) for QuAMax (FA) and X-ResQ (RA) using fully parallel QA with 4-user MIMO at SNR 20~dB (\emph{upper}: $4\times4$, \emph{lower}: $4\times12$). Bars report the mean while lines report the median.
    % In the case of IoT-ResQ and QFSD, they cannot achieve gains between 1 and 16 parallelism levels}.
    }}
\label{f:clique_vs_heuristic}
\end{figure}

% \parahead{ML Occurrence for Parallel QA benchmark.} 
We empirically validate this
% (\emph{e.g.,} ) 
by comparing the cumulative results against the actual fully parallel QA run with small-size instances.
For the comparisons, we use the \emph{ML occurrence} out of total anneal samples as a benchmark metric, instead of TTS. This is because parallel QA's motivation is to hit the global optimal more at the cost of more qubits, even though the quality of annealing in each task might get slightly degraded due to the extended Hamiltonian and cross-talk among qubits. Thus, we care about absolute anneal counts that hit the ML solution (global optimum) out of the total collected samples. We plot ML occurrence as a function of levels of parallelism in Figure~\ref{f:cumul_seq_vs_parallel} to compare cumulative sequential QA runs that mimic a parallel QA run against actual parallel run performance tested on the Zephyr Adavantange2 machine. As expected, actual parallel runs obtain slightly degraded results than cumulative sequential results for both RA and FA. However, considering the nature of the probabilistic heuristics we believe these gaps are not quite critical (for these levels of parallelism) in that at $L_P=6$ the median difference between them is less than 3\% in both FA and RA, given 30,000 total collected samples ($N_a\cdot L_P$). Furthermore, the gaps are mainly caused by certain instances that work particularly well with sequantial runs.
Thus, we believe the performance of cumulative sequential QA runs can be used to estimate fully parallel QA performance, to some extent.

\parahead{Clique versus Heuristic Embedding for Parallel QA.} We also compare two embedding methods for parallel QA, clique embedding and heuristic embedding.
% Here, the heuristic embedding (offered by D-Wave APIs) accepts the entire combined Ising form for parallel QA (\S\ref{s:parallel_QA}) as an input and then results in a big qubit cluster that contains all parallel task information (cf. different qubit clusters for different tasks in the clique embedding). 
Figure~\ref{f:clique_vs_heuristic} shows the comparison with 4-user MIMO detection instances. It plots the average ML occurrence out of $L_P\cdot5000$ samples as a function of $L_P$ for QuAMax and \systemname{}. We observe that while the ML occurrences keep increasing gradually as $L_P$ increases with both the clique embedding and the heuristic embedding, the latter results in relatively smaller ML occurrences. Thus, the clique embedding (\systemname{} opts for) is a more efficient hardware embedding technique for parallel QA with MIMO detection instances. 
% Interestingly, heuristic embedding uses much smaller qubits; for $L_P=3$, clique embedding requires 278 qubits, while heuristic embedding requires only 119 qubits on the Zephyr-topology hardware.

% To test the slightly larger levels of parallelism, we also plot the performance of cumulative sequential QA detectors with $6\times 16$ MIMO at SNR 16 dB in Figure~\ref{f:BER_N}. It is shown that with shorter compute time, \systemname{} requires higher levels of parallelism (more qubits) to reach the (near-)ML performance, while less qubits are required at the expense of compute time (trade-off).

% \newpage
% Thus, we use cumulative sequential approaches for scenarios that require high qubits (\emph{e.g.,} high $N_t$ and/or high $N$). The gaps are relatively smaller in RA, which is probably due to the effects of using initial states in RA.

% In our evaluation, we observe that \systemname{} starts performing poorly for 8 users or more, even with $N_r=16$. 

% (as stated in $\S$\ref{s:primer_ra}) across instances.
\vspace{-0.3cm}
\begin{table}[htbp]
\begin{tiny}
% \begin{footnotesize}
\centering
\caption{D-Wave quantum annealers and features: D-Wave 2000Q (DW2Q) \& Advantage (DWAdv) machine.}
% \vspace{-0.3cm}
\begin{tabularx}{\linewidth}{*{5}{Y}}
\toprule
& \textbf{\scriptsize{DW2Q}} & \textbf{\scriptsize{DWAdv}} & \textbf{\scriptsize{DWAdv2}} \scriptsize{(prototype)} & \textbf{\scriptsize{DWAdv2}} \scriptsize{(full-size)}  \\ \midrule
%Access time & 26.647~ms\\
%Access overhead & 5.834~ms\\
\textbf{\scriptsize{Release Year}} &  \scriptsize{2017} & \scriptsize{2020} & \scriptsize{2024} &
\scriptsize{(exp)} \scriptsize{2025} \\
\textbf{\scriptsize{Topology}} & \scriptsize{Chimera} & \scriptsize{Pegasus} & \scriptsize{Zephyr} & \scriptsize{Zephyr} \\
\textbf{\scriptsize{Qubits}} & \scriptsize{2000+} & \scriptsize{5000+} & \scriptsize{1000+} & \scriptsize{7000+} \\
\textbf{\scriptsize{Connectivity}} & \scriptsize{6} & \scriptsize{15} & \scriptsize{20} & \scriptsize{20} \\
% Connectivity & $\simeq$ 9000~$\mu$s\\
 % & $\simeq$ 300~$\mu$s\\
% Post processing & 445~$\mu$s \\
% \midrule
% $(N_t)$\\
% \bottomrule
% \toprule
% \multicolumn{6}{c}{\textbf{Dense Urban}}\\ 
% \multicolumn{3}{c}{\textbf{Downlink}} & \multicolumn{3}{c}{\textbf{Uplink}} \\\midrule
% \text{poor} & \text{medium} & good & \text{poor} & \text{medium} & good\\
% 5~dB & 12.5~dB  & 17.5~dB  & 5~dB  & 7.5~dB & 9~dB\\
\bottomrule
\end{tabularx}
\vspace{-0.3cm}
\label{t:DWAVE}
\end{tiny}
\end{table}

\vspace{-0.2cm}
\section{Adverse Effect of Problem Decomposition in QA Optimization}
\label{s:decom_adverse_effect}
% \vspace{-0.1cm}

\begin{figure}
\centering
\includegraphics[width=0.75\linewidth]{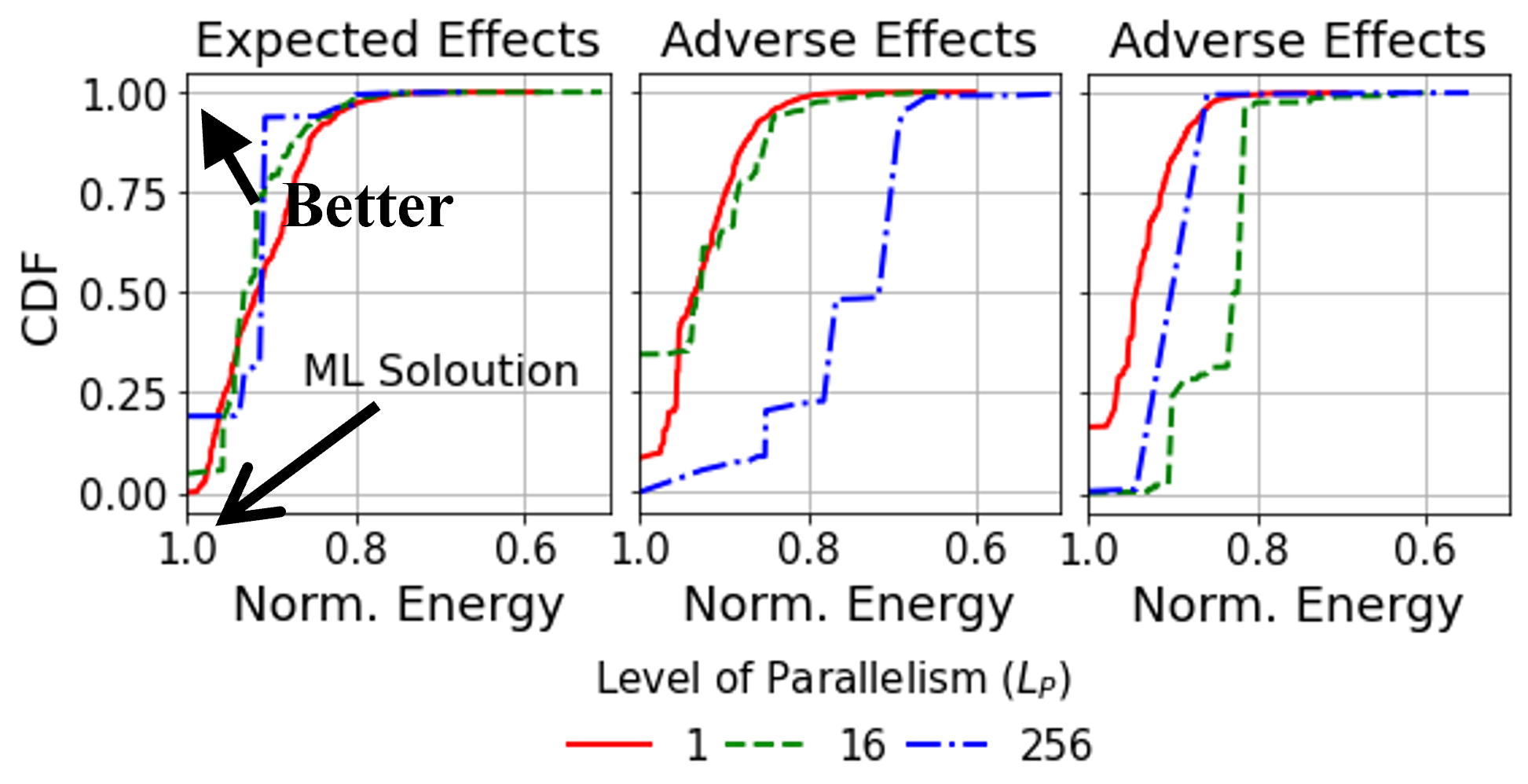}\vspace{-0.2cm}
    \caption{\small
    \textbf{Effects of the decomposition on QA. It is observed that further decomposition could cause harder problems for QA, lowering the probability of finding the ML solution per anneal ($P_G$), despite the reduced search space.}} 
\label{f:decom_limitation}
\end{figure}

\begin{figure}
\centering
\includegraphics[width=0.75\linewidth]{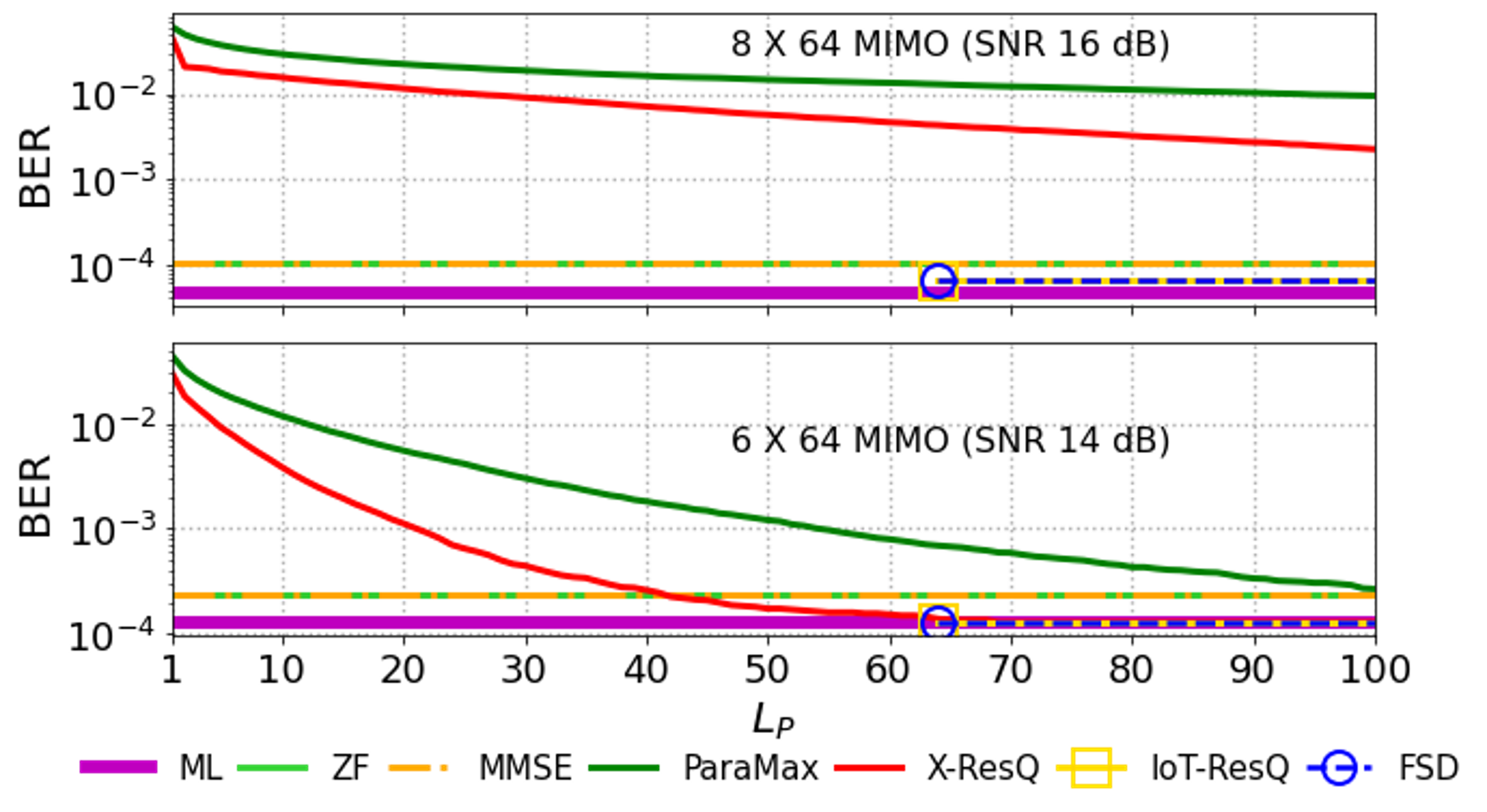}
    \caption{\small
    \textbf{BER performance of classical \systemname{} across $L_P$ for 64-antenna MIMO ($N_r=64$) with 64-QAM ($\mathcal{|O|}=64$).}} 
\label{f:64qam}
\end{figure}

\begin{figure*}
    \centering    \includegraphics[width=0.7\linewidth]{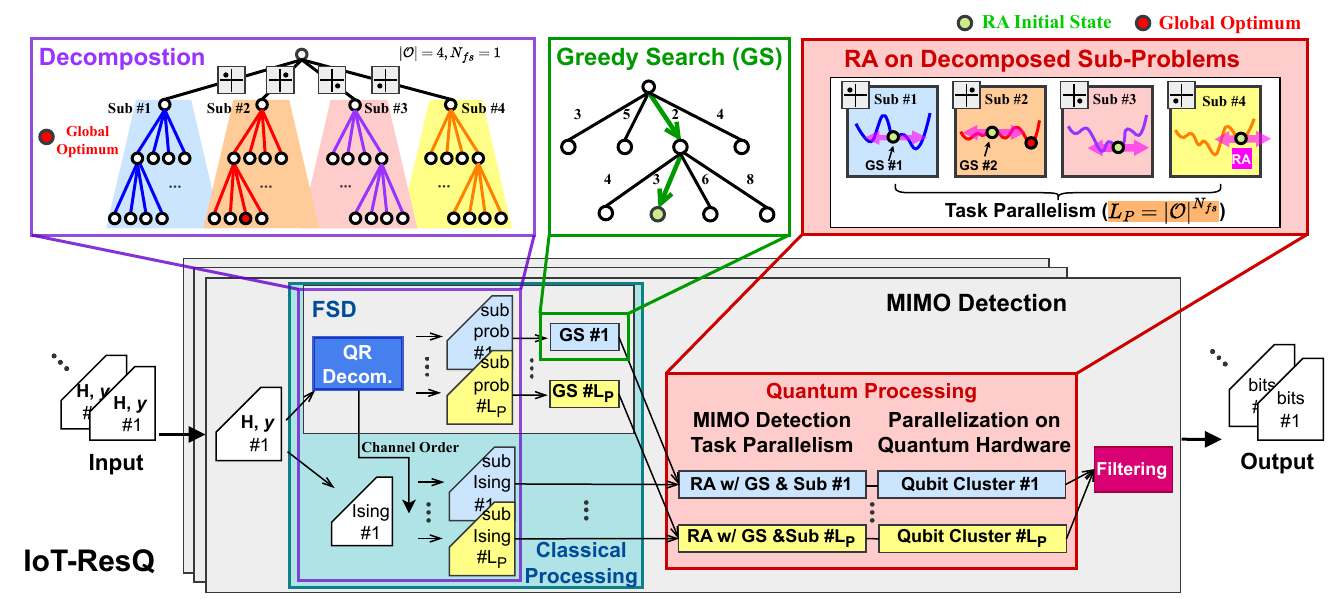}
    \caption{\small\textbf{System architecture of IoT-ResQ~\cite{kim2022warm} consisting of FSD and RA (decomposition-based parallel RA). Since IoT-ResQ relies on the (FSD) decomposition approach, only a \emph{single} sub-problem retains the global optimum (sub-Ising model \#2 in this example). 
    % IoT-ResQ can support only limited parallelism. Further, IoT-ResQ has high overheads due to non-linear classical detector FSD, requiring parallelism in both classical and quantum processing. 
    % When FA is directly applied to sub-Ising problems (instead of applying RA initialized by GS solutions), we call it QFSD (Quantum FSD).     \vspace{-0.2cm}
    }} 
    \label{f:IoT_ResQ}
\end{figure*}

Regarding decomposition-based parallel QA MIMO detectors, we experimentally found that decomposed subproblems could be even harder problems than the original problem for QA, which does not occur in the case of deterministic classical solvers like the greedy search in FSD. We plot the CDF of the normalized Ising energies of the collected samples out of QA (FA) runs with different levels of parallelism in Figure~\ref{f:decom_limitation} for three $4\times 4$ MIMO detection instances at SNR 20 dB. Higher levels of parallelism imply further decomposition ($L_P$ is equal to $16^{N_{fs}}$ with $N_{fs}=0,1,2$) and we focus on only a subproblem that contains the global optimum. While the left panel shows the expected impact of the decomposition approach, where further decomposition improves optimization quality, the others show the adverse effects where further decomposition leads to worse results despite the reduced search space.
Among 50 tested instances, less than 5--10 instances follow the expected effects; similar phenomena have been also observed for different modulations and RA, especially with $N\times N$ MIMO ($N_t=N_r$). Considering the reduced search space in subproblems compared to the one in the original problem, these are unexpected results (further decomposition makes search space size $2^{16-4N_{fs}}$ exponentially smaller). 
While it is difficult to explain the clear reason, it may be due to the 
Ising updates for decomposed subproblems resulting in critical coefficient values to analog-related noise on the machine ($\S$\ref{s:common_challenges}) (\emph{e.g.,} very large coefficient values).

\section{Negative Results with 64-QAM}
\label{s:64qam}
% \vspace{-0.2cm}

In Figure~\ref{f:64qam}, we compare the BER performance of classical \systemname{} against that of other MIMO detectors for 64-QAM massive MIMO systems with 64 receive antennas. While \systemname{} demonstrates promising results for 16-QAM and lower, it is observed that its performance deteriorates for 64-QAM and becomes even worse than linear detectors, ZF and MMSE. We see that in $8 \times 64$ MIMO, \systemname{} would require a very high $L_P$ to achieve similar performance as classical conventional detectors. 
This is surprising in that \systemname{} is based on the MMSE solution and its final solution's Ising energy is always lower than that of MMSE. 
In our future work, we will analyze this further as an effort to enable 64-QAM. 
Nevertheless, recall that \systemname{}'s RA optimization can be opportunistically skipped when MMSE performs well; for these MIMO scenarios, \systemname{} can rely only on the MMSE detector without conducting PIC optimization.

% Currently, this remains a drawback of our methodology and we plan to address it in our future work.

\section{IoT-ResQ Architecture}
\label{s:iot-resq}
\vspace{-0.1cm}
The overall architecture of IoT-ResQ is shown in Figure~\ref{f:IoT_ResQ} (cf. Figure~\ref{f:design_overview} for X-ResQ). 
% For a more detailed explanation, we refer to the study.

% \input{design_challenges}
%\input{reverse_annealing}
% \input{timing_appendix}
%%%%%%%%%%%%%%%%%%%%%%%%%%%%%%%%%%%%%%%%%%%%%%%%%%%%%%%%%%%%%%%%%%%%%%%%%%%%%%%%
\end{document}